\PassOptionsToPackage{square,comma,numbers,sort&compress}{natbib}
\documentclass[onecolumn,12pt,3p,review]{elsarticle}

\usepackage{amsmath,amsfonts}
\usepackage{esvect}
\usepackage{array}
\usepackage{natbib}
\usepackage{appendix}
\usepackage{textcomp}
\usepackage{amssymb}
\usepackage{lscape}
\usepackage{textcomp}
\usepackage{stfloats}
\usepackage{url}
\usepackage{verbatim}
\usepackage{graphicx}
\usepackage{graphbox}
\usepackage{float}
\usepackage[noend]{algpseudocode}
\usepackage[ruled]{algorithm2e}
\usepackage{setspace}
\usepackage{bookmark}
\usepackage{lineno,hyperref}

\usepackage[tight,footnotesize]{subfigure}
\usepackage{picins} 
\usepackage{multirow,bigstrut} 
\usepackage{booktabs}  
\usepackage{etoolbox}
\usepackage{color}

\DeclareMathOperator*{\minimize}{minimize}

\makeatletter
\patchcmd{\@algocf@start}
{-1.5em}
{0pt}
{}{}
\makeatother

\journal{Elsevier}
\begin{document}
	\sloppy
	\begin{frontmatter}
		\title{Multi-step-ahead Stock Price Prediction Using Recurrent Fuzzy Neural Network and Variational Mode Decomposition}
		\author[aut]{Hamid Nasiri}
		\ead{h.nasiri@aut.ac.ir}
		\author[aut]{Mohammad Mehdi Ebadzadeh \corref{cor1}}
		\cortext[cor1]{Corresponding author}
		\address[aut]{ Department
			of Computer Engineering, Amirkabir University of Technology (Tehran Polytechnic), Tehran, Iran}
		\ead{ebadzadeh@aut.ac.ir}

\begin{abstract}
Financial time series prediction, a growing research topic, has attracted considerable interest from scholars, and several approaches have been developed. Among them, decomposition-based methods have achieved promising results. Most decomposition-based methods approximate a single function, which is insufficient for obtaining accurate results. Moreover, most existing researches have concentrated on one-step-ahead forecasting that prevents stock market investors from arriving at the best decisions for the future. This study proposes two novel methods for multi-step-ahead stock price prediction based on the issues outlined. DCT-MFRFNN, a method based on discrete cosine transform (DCT) and multi-functional recurrent fuzzy neural network (MFRFNN), uses DCT to reduce fluctuations in the time series and simplify its structure and MFRFNN to predict the stock price. VMD-MFRFNN, an approach based on variational mode decomposition (VMD) and MFRFNN, brings together their advantages. VMD-MFRFNN consists of two phases. The input signal is decomposed to several IMFs using VMD in the decomposition phase. In the prediction and reconstruction phase, each of the IMFs is given to a separate MFRFNN for prediction, and predicted signals are summed to reconstruct the output. Three financial time series, including Hang Seng Index (HSI), Shanghai Stock Exchange (SSE), and Standard \& Poor's 500 Index (SPX), are used for the evaluation of the proposed methods. Experimental results indicate that VMD-MFRFNN surpasses other state-of-the-art methods. VMD-MFRFNN, on average, shows 35.93\%, 24.88\%, and 34.59\% decreases in RMSE from the second-best model for HSI, SSE, and SPX, respectively. Also, DCT-MFRFNN outperforms MFRFNN in all experiments.
\end{abstract}

\end{frontmatter}


\section{Introduction}

Economic globalization, as one of the three main dimensions of globalization, cause the stock market to be highly fluctuated, polymorphic, and complex. Stock price index prediction is vital for investors and financial analysts to build investment strategies and get steady returns \cite{liu2022stock, deng2022multi}. Hence, extensive studies \cite{farimani2022investigating,huang2021new,seong2022forecasting} have been done on improving financial time series prediction approaches as an active research topic. 
A stock price time series is a series of successive observations collected from the stock market over time. The prediction of these time series can be made in one-step or multi-step ahead forecasting. Obviously, a long-term prediction (i.e., multi-step ahead) is more beneficial for stock market investors to arrive at the best decisions and plans for the future. Nevertheless, most existing stock price prediction researches \cite{cao2019stock,guo2022forecasts,lin2021forecasting,rezaei2021stock} only address one-step ahead forecasting. Therefore, there is a need to investigate multi-step ahead prediction of stock prices \cite{deng2022multi}.

Another issue that requires attention is that financial time series are usually dynamic, non-linear, extremely fluctuated, and complex, which makes their prediction challenging \cite{yujun2021research, rezaei2021stock,xu2022self}. To address this issue, decomposition-based methods have been proposed. The main idea behind decomposition-based methods is to decompose complex time series into a number of components with a simpler structure, i.e., reducing time series complexity. Since these components have simpler structures and more stationary trends, their prediction is easier, and the expected accuracy is higher \cite{nguyen2021ensemble}.

Empirical mode decomposition (EMD), an adaptive and data-driven decomposition method, was proposed by Huang et al. \cite{huang1998empirical} in 1998. EMD decomposes a time series into some intrinsic mode functions (IMFs) and a monotonic residual signal in an iterative process \cite{liu2020robust}. So, it can simply isolate highly fluctuating data into lower frequency components \cite{rezaei2021stock}. Although many researchers used EMD for financial time series prediction, it has many inherent problems such as mode mixing, lack of mathematical foundations, sensitivity to noise and sampling, and choice of the interpolation method \cite{dragomiretskiy2014variational, guo2022forecasts}. To address these shortcomings, variational mode decomposition (VMD) \cite{dragomiretskiy2014variational} was proposed by Dragomiretskiy and Zosso in 2014.
VMD is a completely non-recursive decomposition method based on a solid mathematical framework. It decomposes a time series into several eigenmode components with a specific bandwidth in the spectral domain by solving a convex optimization problem. These band-limited components reconstruct the input time series exactly or in the least squares sense. In other words, VMD is capable of separating tones of similar frequencies and avoiding the mode mixing problem \cite{dragomiretskiy2014variational, liu2022stock, yujun2021research}.

Although in recent years, many researchers have developed decomposition-based methods using VMD instead of EMD, as far as the authors' knowledge goes, all of them approximate a single function, i.e., produce a specific output determined by the current and previous inputs. As described in \cite{nasiri2022mfrfnn}, in challenging nonlinear problems, different outputs might be produced for a certain input depending on the state of the system. Therefore, to obtain accurate results in financial time series prediction, a method is needed that can identify the system's states and approximate a specific function for each of them, i.e., it should be able to simultaneously learn multiple functions. Multi-functional recurrent fuzzy neural network (MFRFNN) \cite{nasiri2022mfrfnn} proposed by Nasiri and Ebadzadeh in 2022 can learn multiple functions simultaneously. In light of the issues outlined, two novel methods for financial time series prediction have been proposed in this study: DCT-MFRFNN and VMD-MFRFNN.

DCT-MFRFNN is based on discrete cosine transform (DCT) and MFRFNN. It applies DCT to the time series and computes DCT coefficients. A sequence of high-frequency DCT coefficients is dropped to eliminate high frequencies, reduce fluctuations in the time series, and simplify its structure. Then the time series is reconstructed using only the remaining coefficients. Finally, the reconstructed signal is given to the MFRFNN model for prediction.

VMD-MFRFNN is an approach based on VMD and MFRFNN and brings together their advantages. It consists of two phases: a) decomposition phase and b) prediction and reconstruction phase. The input time series is decomposed to several IMFs using VMD in the decomposition phase. In the prediction and reconstruction phase, first, each of IMFs  is given to a separate MFRFNN model for prediction. Then the predicted signals, i.e., the output of each MFRFNN, are summed to reconstruct the final output. The following are the main contributions of this research:

\begin{enumerate}
	\item {DCT-MFRFNN and VMD-MFRFNN, two novel financial time series prediction methods, are proposed. DCT-MFRFNN first applies DCT to the input signal and drops a sequence of high-frequency DCT coefficients. Then the input time series is reconstructed using the remaining coefficients. Finally, the MFRFNN model uses the reconstructed signal for prediction. VMD-MFRFNN first decomposes the input signal into several IMFs and then predicts each by a separate MFRFNN model. Finally, the outputs of MFRFNNs are summed to generate the final prediction.}
	\item{Developing new training algorithms for DCT-MFRFNN and VMD-MFRFNN.}
	\item{The prediction performance of DCT-MFRFNN and VMD-MFRFNN is evaluated using three financial time series, including HSI, SSE, and SPX. The experimental results indicate that VMD-MFRFNN shows promising performance in multi-step-ahead prediction compared with other existing methods. VMD-MFRFNN, on average, shows 35.93\%, 24.88\%, and 34.59\% decreases in RMSE from the second-best model for HSI, SSE, and SPX, respectively. Also, DCT-MFRFNN outperforms MFRFNN in all experiments.}
\end{enumerate} 

The remainder of the paper is organized as follows: Section \ref{sec:LiteratureReview} outlines an overview of the related works. In Section \ref{sec:Preliminaries}, DCT, VMD, and MFRFNN are briefly introduced, with Section~\ref{sec:ProposedMethod} presenting the DCT-MFRFNN and VMD-MFRFNN. Section \ref{sec:ResultsAndDiscussion} evaluates the effectiveness of the proposed methods and provides results and discussion. Finally, this study is concluded with a summary in Section \ref{sec:Conclusion}.
    
\section{Literature Review}
\label{sec:LiteratureReview}
This section outlines an overview of the literature that relates to this study. As a vital problem for investors and financial analysts, financial time series prediction has been extensively studied in recent years \cite{zhou2019emd2fnn,yang2022novel,yujun2021research,deng2022multi,tang2020multistep,nguyen2021ensemble,rezaei2021stock,lin2021forecasting,guo2022forecasts,liu2022stock,salimi2022novel,nasiri2022mfrfnn}. Several researchers have developed decomposition-based methods using EMD for time series forecasting. Zhou et al. \cite{zhou2019emd2fnn} proposed EMD2FNN, a hybrid method composed of EMD and factorization machine-based neural networks (FMNN) for predicting stock market trends. EMD2FNN brings together the benefits of EMD and FMNN. Yang et al. \cite{yang2022novel} developed a hybrid method for time series prediction based on recursive EMD and long short-term memory (LSTM). They proposed a novel recursive EMD to solve the mode mixing problem of EMD and used LSTM to predict the IMFs. More recent work by Deng et al. \cite{deng2022multi} introduced a hybrid modeling framework based on multivariate EMD (MEMD) and LSTM for stock price index forecasting. In this framework, first, MEMD was used to decompose relevant features of the time series. Then, the extracted features were used to train LSTM for the multi-step-ahead prediction task. Moreover, the orthogonal array tuning approach was employed to tune the LSTM's hyperparameters. Deng et al. evaluated their proposed method on three real-world financial time series and obtained promising results.

Although many researchers used EMD for time series forecasting, it has many inherent problems, such as mode mixing, marginal effects, lack of mathematical foundations, and sensitivity to noise \cite{liu2020non}. To overcome these issues, ensemble empirical mode decomposition (EEMD) \cite{wu2009ensemble} was proposed by Wu and Huang in 2009. EEMD is a self-adaptive decomposition method that eliminates the mode mixing problem of EMD by adding white noise \cite{tang2020multistep, nguyen2021ensemble}. Rezaei et al. \cite{rezaei2021stock} developed CEEMD-CNN-LSTM, a hybrid method using complete EEMD (CEEMD), convolutional neural network, and LSTM. Their proposed method can extract deep features and time sequences. Although CEEMD-CNN-LSTM achieved encouraging results in the one-step-ahead stock price prediction, it has not been evaluated in multi-step-ahead prediction tasks. In 2021, Lin et al. \cite{lin2021forecasting} proposed a hybrid approach combining CEEMD with adaptive noise (CEEMDAN) and LSTM. They assessed their proposed method on SPX and Chinese Securities 300 Index time series. Despite overcoming the problems associated with EMD, EEMD suffers from incompleteness, reconstruction error, high computational complexity, and predicament of parameters selection \cite{lin2021forecasting,rashida2020intelligent}.

VMD, a completely non-recursive adaptive decomposition method, was proposed by Dragomiretskiy and Zosso \cite{dragomiretskiy2014variational} in 2014 to address EMD's shortcomings. Due to its merits, it has gained increasing popularity in recent years, and there have been several attempts to develop decomposition-based methods using VMD. Lahmiri \cite{lahmiri2016variational} introduced a novel prediction model by integrating VMD and general regression neural network. VMD-LSTM \cite{niu2020hybrid}, a hybrid technique based on VMD and LSTM, was developed by Niu et al. In another paper, Tang et al. \cite{tang2020multistep} proposed a novel hybrid approach by combining VMD, EEMD, and extreme learning machine (ELM). Their proposed method first applies VMD to the time series and then applies EEMD to the residual signal obtained by VMD. They used ELM to predict the stock price index and differential evolution to optimize its weights. Same as \cite{niu2020hybrid}, Yujun et al. \cite{yujun2021research} developed a prediction model based on VMD and LSTM for stock price prediction. A more recent paper by Wang et al. \cite{wang2022asian} proposed a novel hybrid method based on secondary decomposition (SD), multi-factor analysis, and attention-based LSTM. In this method, SD uses improved VMD and improved CEEMDAN to filter the noise effectively and capture additional nonlinear features. Attention layer was added to LSTM to increase the weights of important data instances and improve LSTM's performance. VML \cite{liu2022stock} was developed by Liu et al. in 2022. VML integrates meta-learning and VMD. It uses VMD to decompose time series and applies model-agnostic meta-learning and LSTM to predict each component. Liu et al. used the Shanghai Stock Exchange index and Dow Jones Industrial Average to evaluate the VML. Guo et al. \cite{guo2022forecasts} introduced a hybrid method based on VMD, autoregressive integrated moving average (ARIMA), and Taylor expansion forecasting (TEF). They used VMD to decompose time series, ARIMA to forecast the linear component of each IMF, and TEF to predict the nonlinear component.

\section{Preliminaries }
\label{sec:Preliminaries}
This section gives a concise overview of DCT, VMD, and MFRFNN methods to provide further insight into the proposed method. 
\subsection{Discrete Cosine Transform}
DCT is a time-frequency linear orthogonal transformation that decreases redundant calculations in the discrete Fourier transform \cite{vinciguerra2021discrete}. It represents a finite sequence of samples as a sum of cosine functions with different frequencies. The DCT coefficients ($y(k)$) of a discrete signal ($x(n)$) can be calculated as follows \cite{begum2022hybrid}:

\begin{equation}
	y(k)=\alpha(k) \sum_{n=0}^{N-1} x(n) \cos \left(\frac{\pi(2 n+1) k}{2 N}\right), k=0,1, \cdots, N-1
\end{equation}

\noindent where $N$ is the signal length, and $\alpha(k)$ denotes the scaling factor that is determined by Eq.~\eqref{eq:DCT_alpha}:

\begin{equation}
	\label{eq:DCT_alpha}
	\alpha(k)=\left\{\begin{array}{lc}
		\sqrt{1 / N} & \:\:\:\:  k=0 \\
		\sqrt{2 / N} & \:\:\:\:      1 \leq k \leq N-1
	\end{array}\right.
\end{equation}

\subsection{Variational Mode Decomposition}
VMD, developed by Dragomiretskiy and Zosso \cite{dragomiretskiy2014variational}, is an adaptive decomposition method based on a solid mathematical framework \cite{zhang2021application}. It decomposes a complex signal into $K$ IMFs with a specific bandwidth in the spectral domain by solving the variational problem \cite{he2022fault,yang2021denoising}.
Let $f$ and $t$ denote the input signal and time step, respectively. IMFs ($u_{k}(t)$) and their corresponding center frequency ($\omega_{k}$) can be obtained by solving the following constrained optimization problem:

\begin{equation}
\label{eq:VMD_constrained}
\begin{aligned}
 &\minimize_{\left\{u_{k}\right\},\left\{\omega_{k}\right\}}\:\:\:\:\:\sum_{{k}=1}^{{K}}\left\|\partial_{t}\left[\left(\delta(t)+\frac{j}{\pi t}\right) * u_{k}(t)\right] e^{-j \omega_{k} t}\right\|_{2}^{2}
\\& \text{subject to}\:\:\:\:\:\sum_{{k}=1}^{{K}} u_{k}=f
\end{aligned}
\end{equation}

\noindent where $\delta(\cdot)$ represents the Dirac delta function and $\partial_{t}$ denotes the partial derivative with respect to $t$ \cite{lian2018adaptive,huang2021new}. 

To ensure signal reconstruction's accuracy under Gaussian noise, a quadratic penalty term is added to Eq.\;\eqref{eq:VMD_constrained}, and to convert it into an unconstrained optimization problem, a Lagrange multiplier is applied as follows \cite{lian2018adaptive, yang2021denoising}:
\begin{equation}
\label{eq:VMD_unconstrained}
\begin{split}
	L\left(\left\{u_{k}\right\},\left\{\omega_{k}\right\}, \lambda\right)=\alpha \sum_{k=1}^{K}\left\|\partial_{t}\left[\left(\delta(t)+\frac{j}{\pi t}\right) * u_{k}(t)\right] e^{-j \omega_{k} t}\right\|_{2}^{2}+\left\|f(t)-\sum_{k=1}^{K} u_{k}(t)\right\|_{2}^{2} \\
	+\left\langle\lambda(t), f(t)-\sum_{k=1}^{K} u_{k}(t)\right\rangle
\end{split}
\end{equation}

\noindent where $\lambda$ and $\alpha$ denote the Lagrange multiplier and penalty parameter, respectively. Finally, by solving Eq. \eqref{eq:VMD_unconstrained} iteratively using the Alternating Direction Method of Multipliers, the input signal is decomposed into $K$ IMFs \cite{lian2018adaptive, dragomiretskiy2014variational}.

\subsection{Multi-Functional Recurrent Fuzzy Neural Network}
\begin{figure}[!htbp]
	\centering
	\includegraphics[width=5cm]{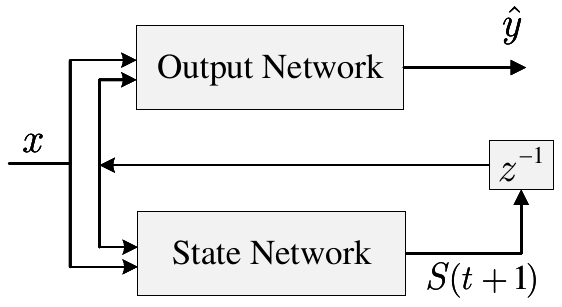}
	\caption{Overview of MFRFNN \cite{nasiri2022mfrfnn}.}
	\label{fig:MFRFNN_Overview}
\end{figure}

MFRFNN, proposed by Nasiri and Ebadzadeh \cite{nasiri2022mfrfnn} in 2022, is a recurrent fuzzy neural network for time series forecasting. It comprises two fuzzy neural networks (FNNs). One of them, i.e., the output network, generates the system's output, whereas the other, i.e., the state network, identifies the state of the system (Fig. \ref{fig:MFRFNN_Overview}). MFRFNN uses the least squares algorithm and particle swarm optimization (PSO) technique to learn the weights of output and state networks, respectively. This network can learn and memorize historical data from previous time series samples and approximate several functions simultaneously by incorporating the states in its structure \cite{nasiri2022mfrfnn}.

\section{Proposed Methods}
\label{sec:ProposedMethod}
This section presents a detailed description of the proposed methods for financial time series prediction.

\subsection{DCT-MFRFNN}
\begin{figure}[!htbp]
	\centering
	\includegraphics[width=0.99\textwidth]{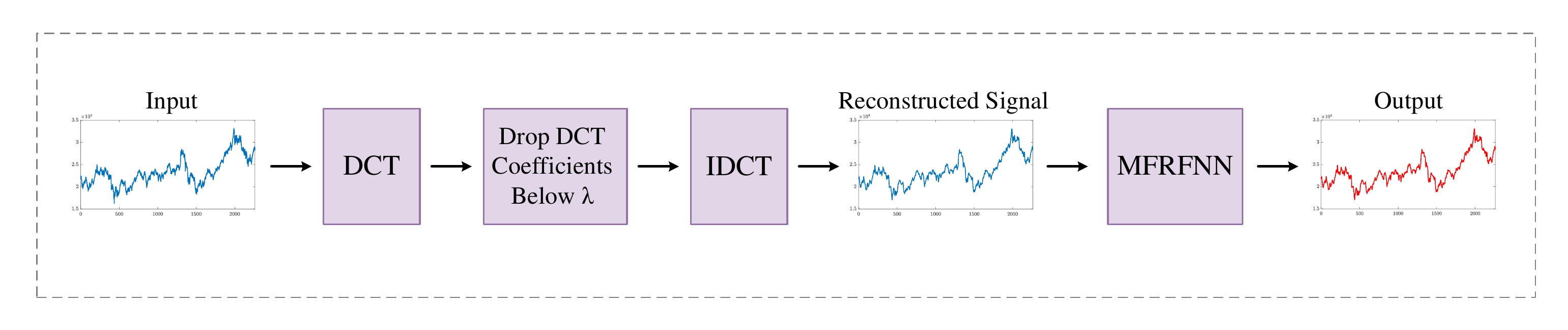}
	\caption{Structure of DCT-MFRFNN.}
	\label{fig:DCT_MFRFNN}
\end{figure}

The proposed DCT-MFRFNN method consists of the following steps: 1) The DCT is applied to the input time series, and DCT coefficients ($\mathbf{c}$) are computed. 2) $\lambda$ percent of high-frequency DCT coefficients are set to zero. 3) The inverse DCT is applied to the remaining coefficients, and the input signal is reconstructed. 4) The reconstructed signal ($\hat{\mathbf{x}}$) is given to the MFRFNN as input for training. 5) The MFRFNN generates the final prediction ($\hat{\mathbf{y}}$). Fig. \ref{fig:DCT_MFRFNN} shows the structure of DCT-MFRFNN.

In DCT-MFRFNN, a sequence of high-frequency DCT coefficients is dropped to eliminate high frequencies and reduce fluctuations in the time series. As a result, a time series with a simpler structure and more stationary trends is given to MFRFNN for prediction. Intuitively, since this time series has a simpler structure, its prediction is easier for the model. DCT-MFRFNN training algorithm is shown in Algorithm \ref{alg:DCT_MFRFNN_Training}.

\begin{algorithm}[!t]
	\setstretch{1.10}
	\small  
	\caption{DCT-MFRFNN training algorithm}
	\label{alg:DCT_MFRFNN_Training}
	\vspace{0.1cm}
	\textbf{Input:} Input time series  $D=\left\{x^{[t]}, x^{[t+h]}\right\}_{t=1}^{p}, \lambda$
	\\
	\textbf{Output:} Predicted signal $\left(\hat{\mathbf{y}}\right)$
	\vspace{0.3cm}
	
	Apply DCT to $\mathbf{x}$ to obtain $\mathbf{c} \in \mathbb{R}^{p \times 1}$
	
	$n \gets \left\lfloor p \times \frac{100-\lambda}{100}\right\rfloor+1 $
	\vspace{0.05cm}	
	
	\For{$i \gets n \:\:\: \KwTo \:\:\: p\:\:$}{
		\texttt{\\}
		$c_i \gets 0$	
		\vspace{0.2cm}	
	
		\vspace{0.2cm}	
	}
	
	Apply IDCT to $\mathbf{c}$ to obtain $\hat{\mathbf{x}}$
	\vspace{0.05cm}	
	
	Train MFRFNN using $\hat{\mathbf{x}}$ as input
	\vspace{0.05cm}	
	
	Predict $\hat{\mathbf{y}}$ using MFRFNN model
	\vspace{0.2cm}	
\end{algorithm}

\subsection{VMD-MFRFNN}
\begin{figure}[!htbp]
	\centering
	\includegraphics[width=0.95\textwidth]{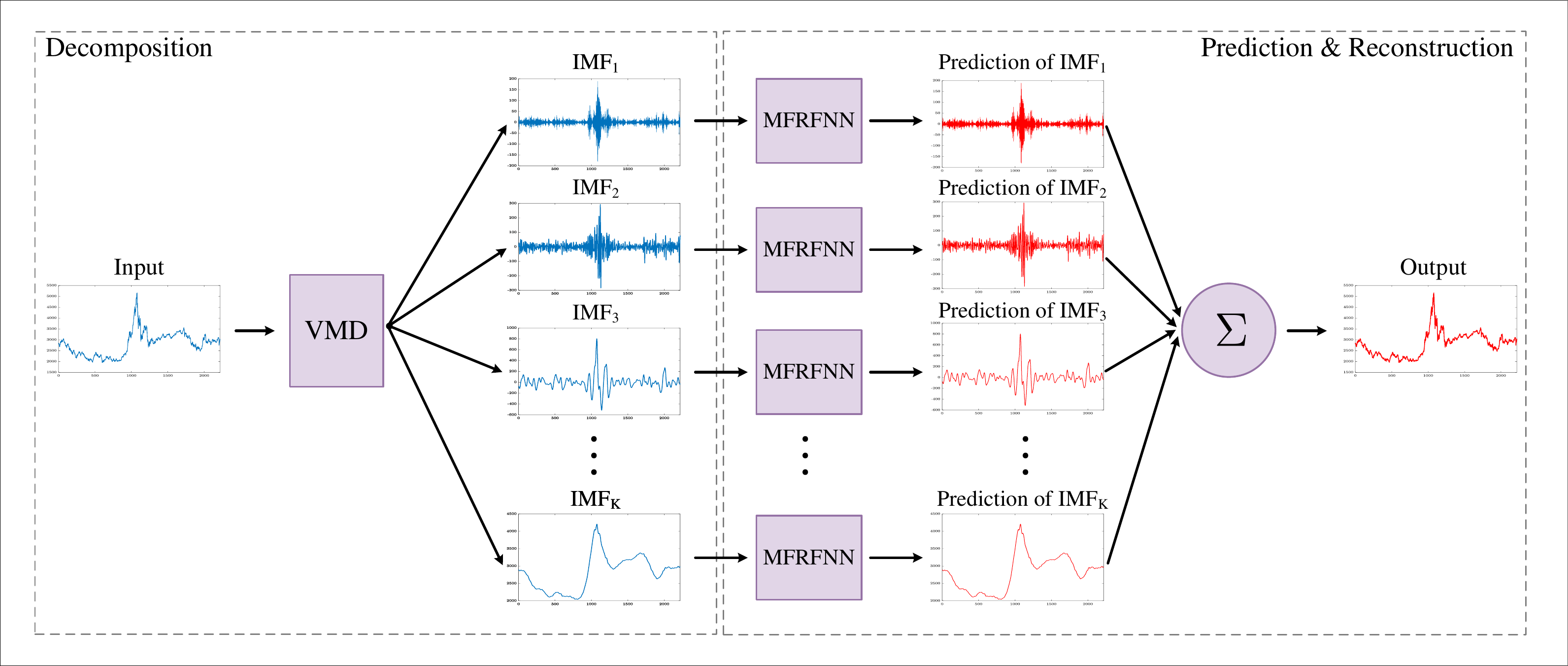}
	\caption{Structure of VMD-MFRFNN.}
	\label{fig:Framework}
\end{figure}

VMD-MFRFNN consists of two phases: a) decomposition phase and b) prediction and reconstruction phase. In the decomposition phase, the input time series is decomposed into several IMFs using the VMD method. In the prediction and reconstruction phase, first, each of IMFs is given to a separate MFRFNN model for training. Then, each model generates its predicted signal, i.e., the output of each MFRFNN. Finally, the predicted signals are summed to reconstruct the final prediction. Fig. \ref{fig:Framework} illustrates the structure of VMD-MFRFNN. A detailed description of VMD-MFRFNN's phases is given below.
     
\begin{figure}[!htbp]
	\centering
	\begin{tabular}{ccc}
		\subfigure[][{\scriptsize SSE Time Series}]{\includegraphics[width=5cm]{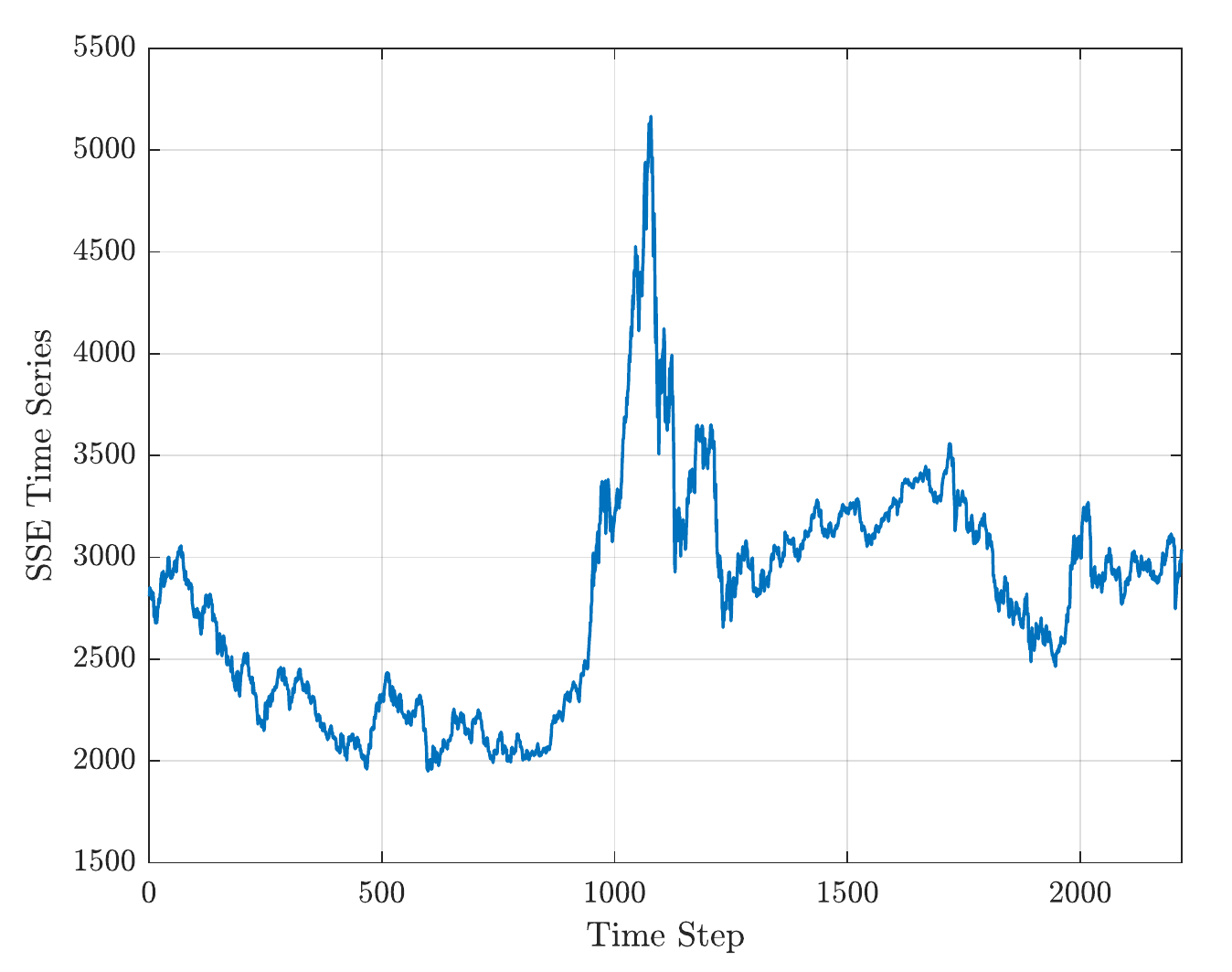}}	&  \subfigure[][{\scriptsize IMF$_1$ Component}]{\includegraphics[width=5cm]{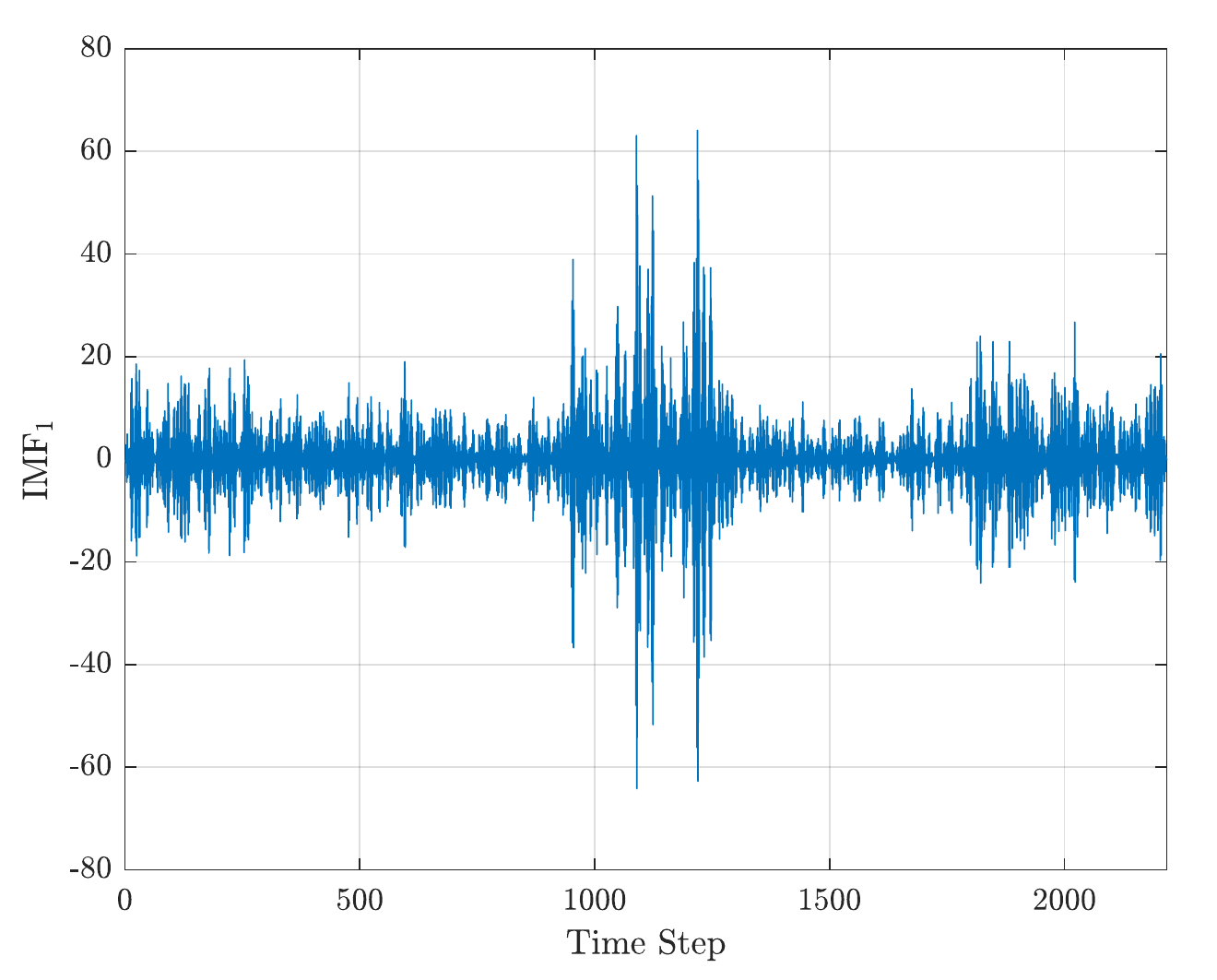}}   &
		\subfigure[][{\scriptsize IMF$_2$ Component}]{\includegraphics[width=5cm]{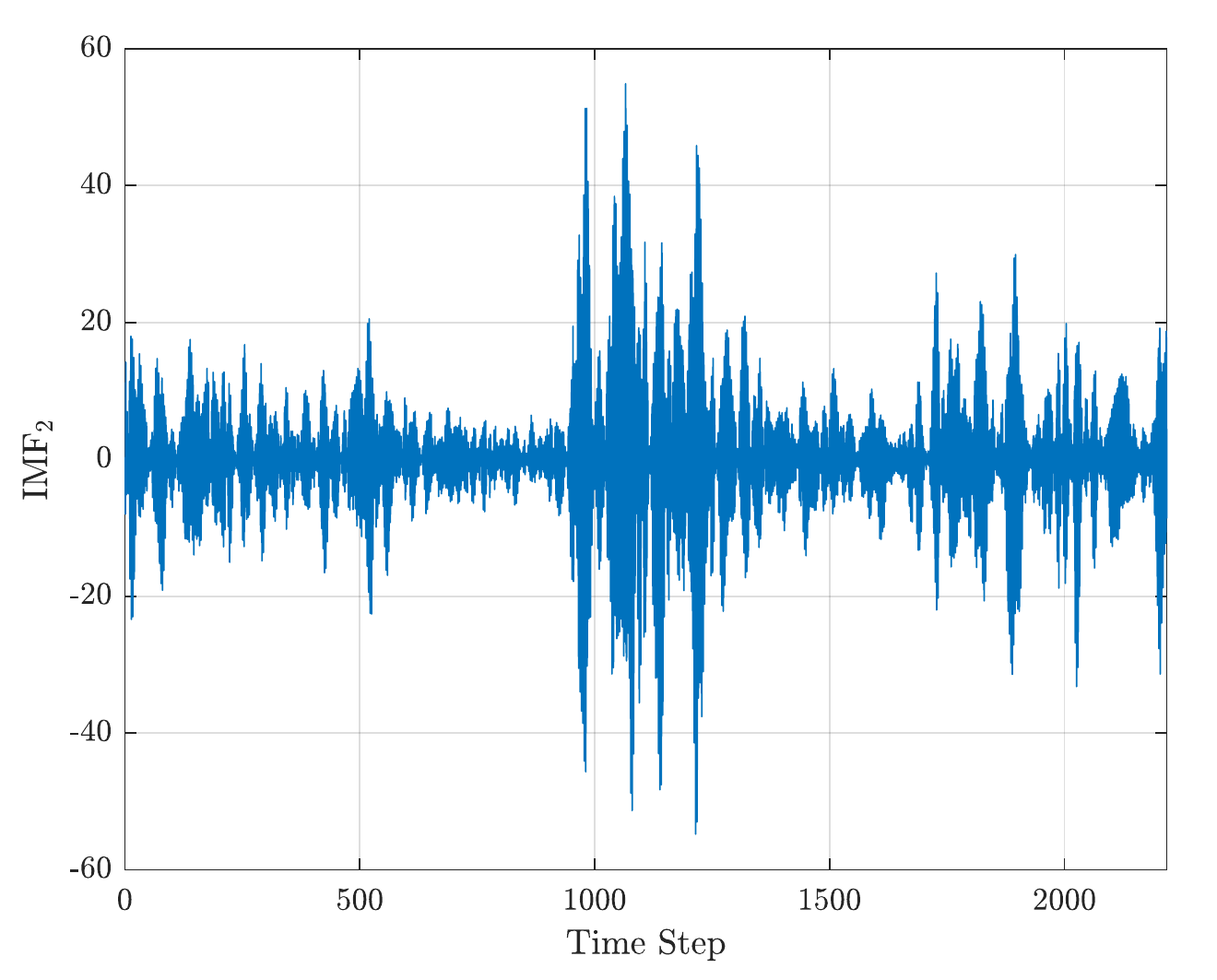}} 	\\   		\subfigure[][{\scriptsize IMF$_3$ Component}]{\includegraphics[width=5cm]{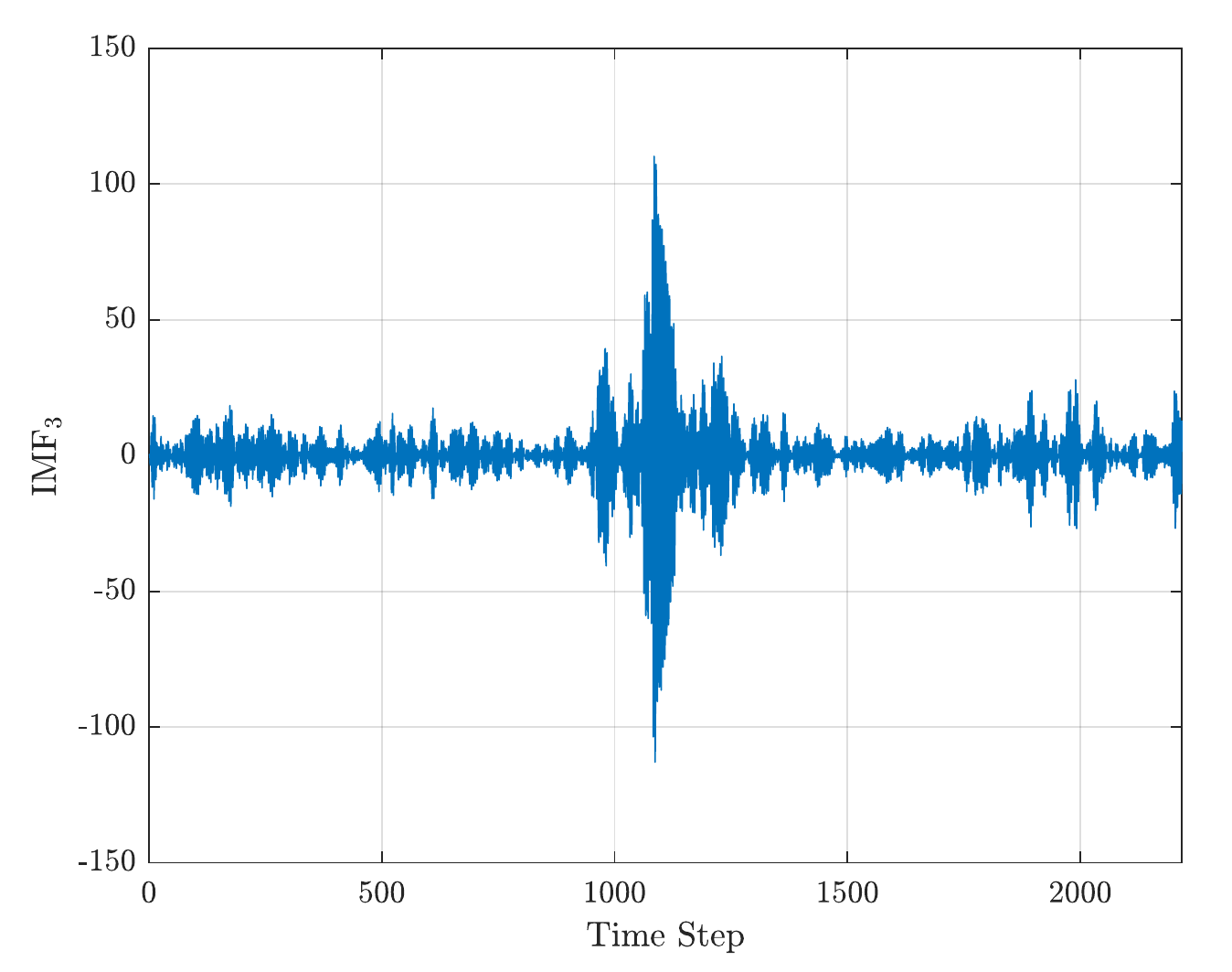}}  &
		\subfigure[][{\scriptsize IMF$_4$ Component}]{\includegraphics[width=5cm]{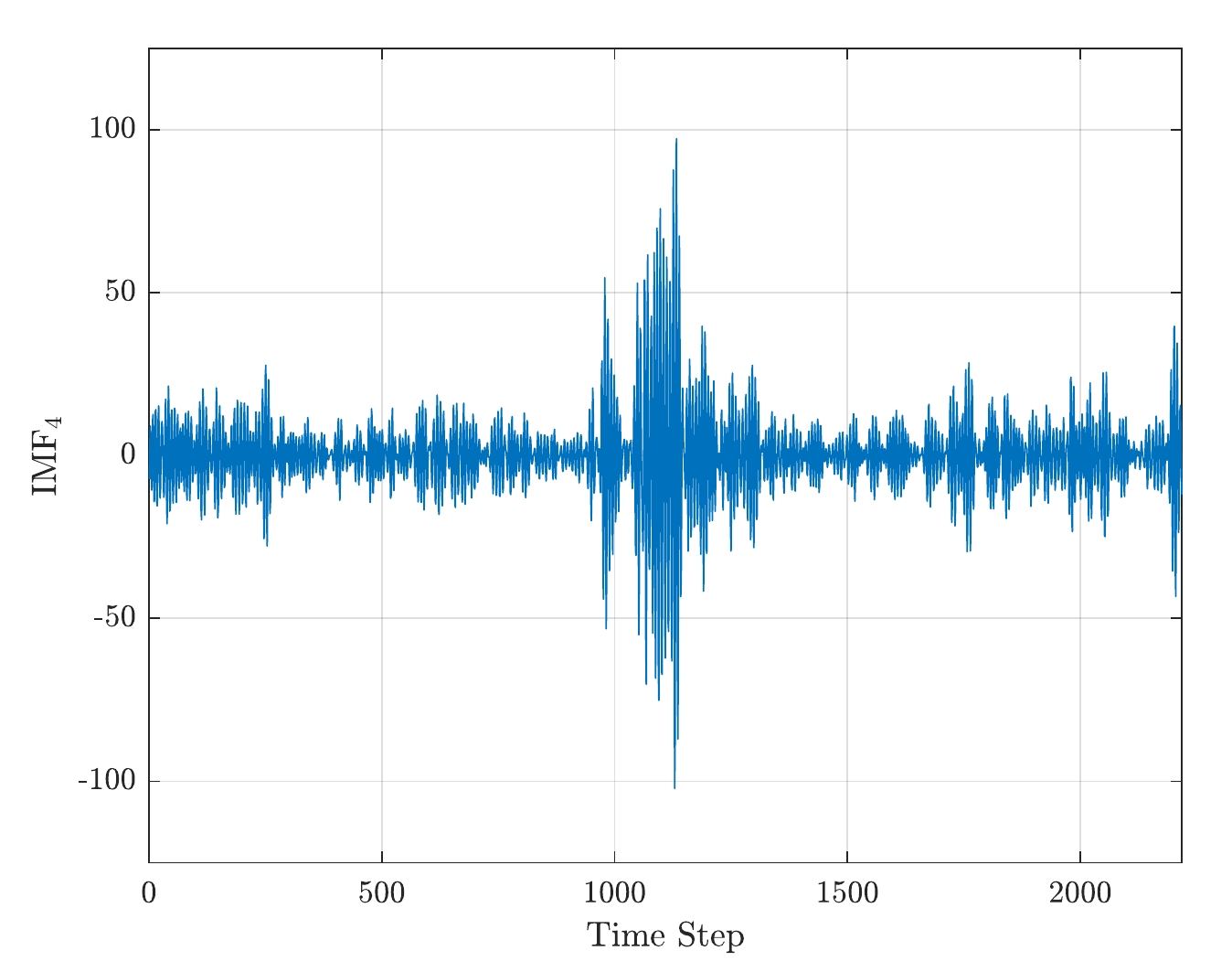}}  &   		\subfigure[][{\scriptsize IMF$_5$ Component}]{\includegraphics[width=5cm]{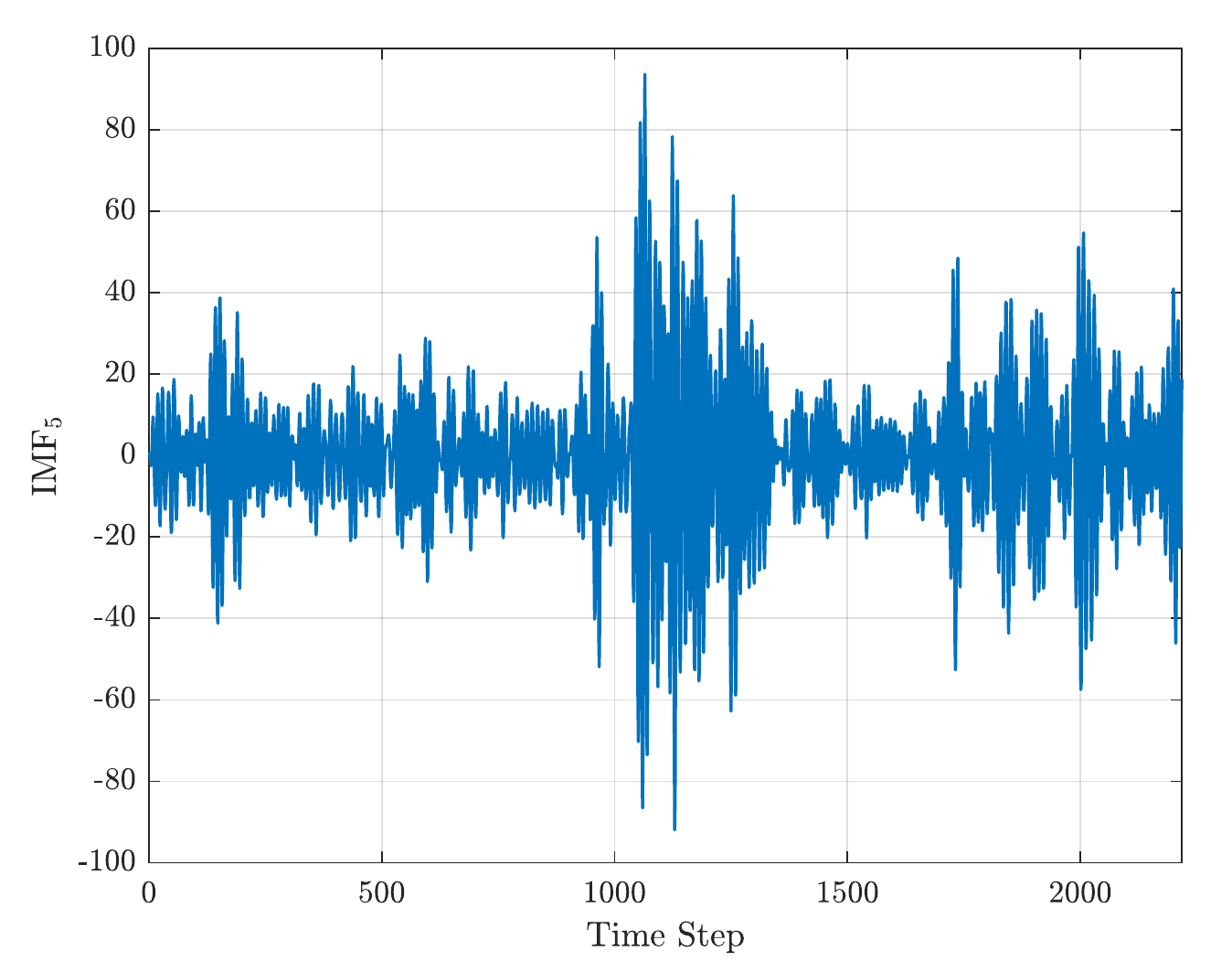}}  \\
		\subfigure[][{\scriptsize IMF$_6$ Component}]{\includegraphics[width=5cm]{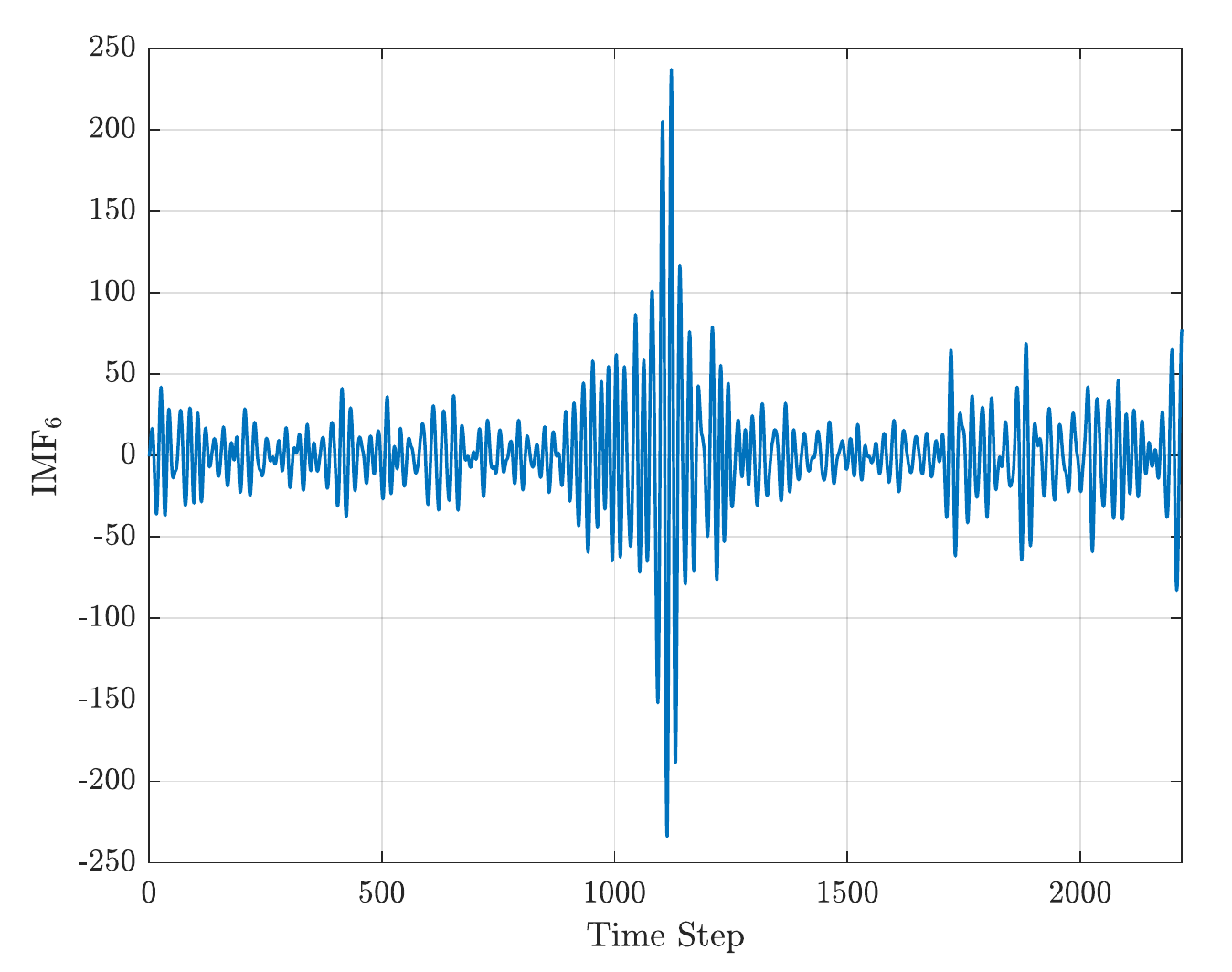}} &   		\subfigure[][{\scriptsize IMF$_7$ Component }]{\includegraphics[width=5cm]{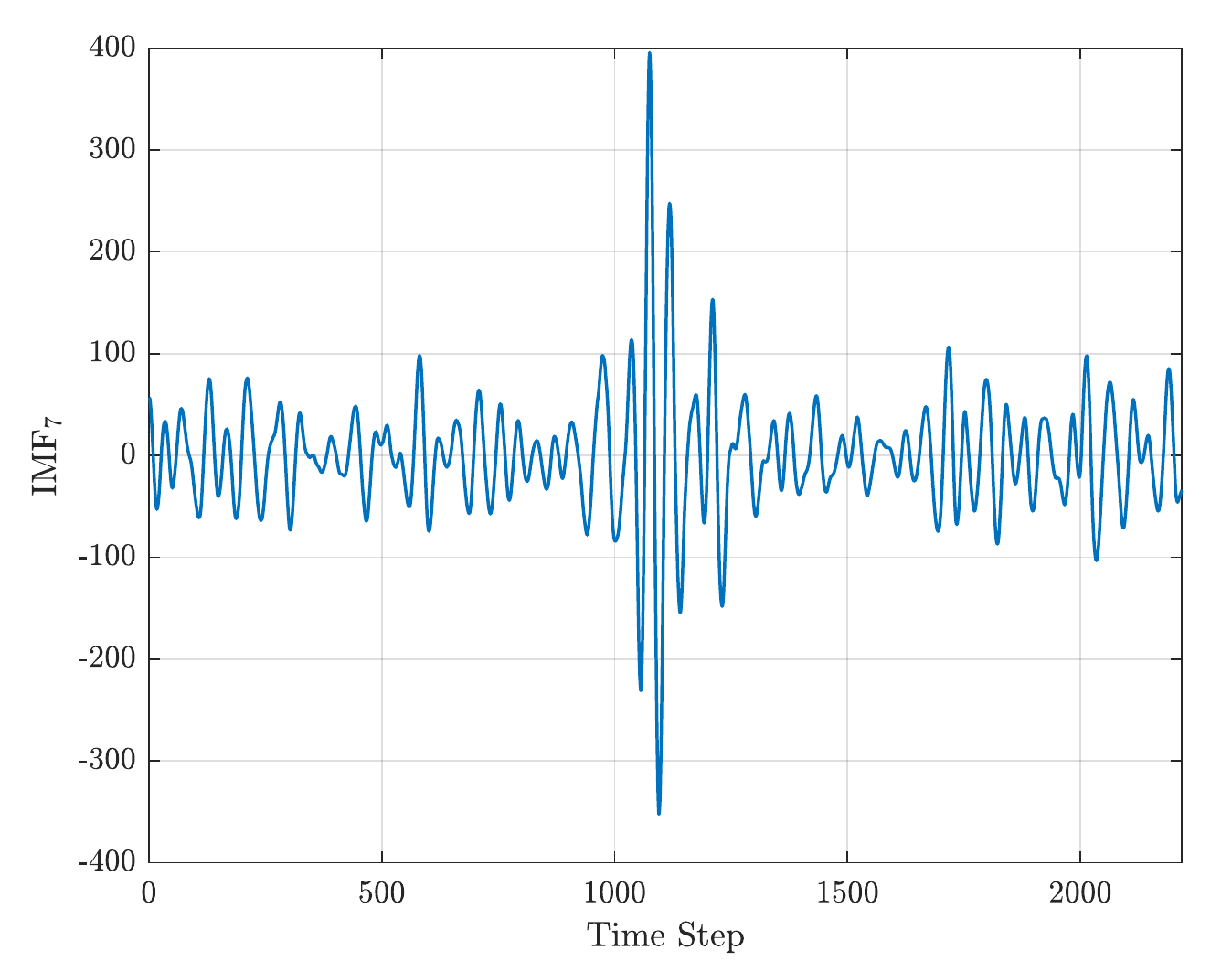}} & 
		\subfigure[][{\scriptsize IMF$_8$ Component}]{\includegraphics[width=5cm]{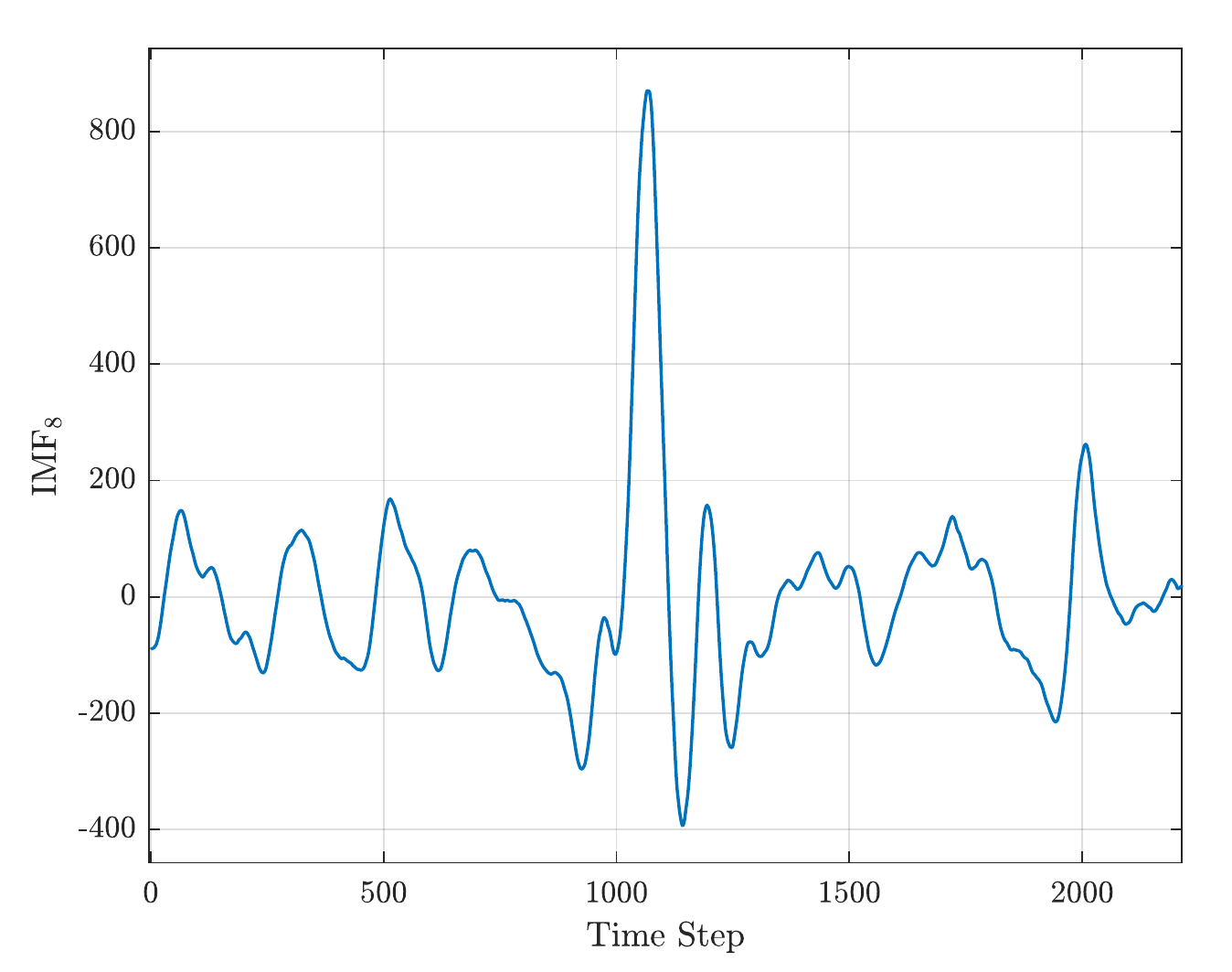}}
	\end{tabular}
	\begin{tabular}{c}
	\subfigure[][{\scriptsize IMF$_9$ Component}]{\includegraphics[width=5cm]{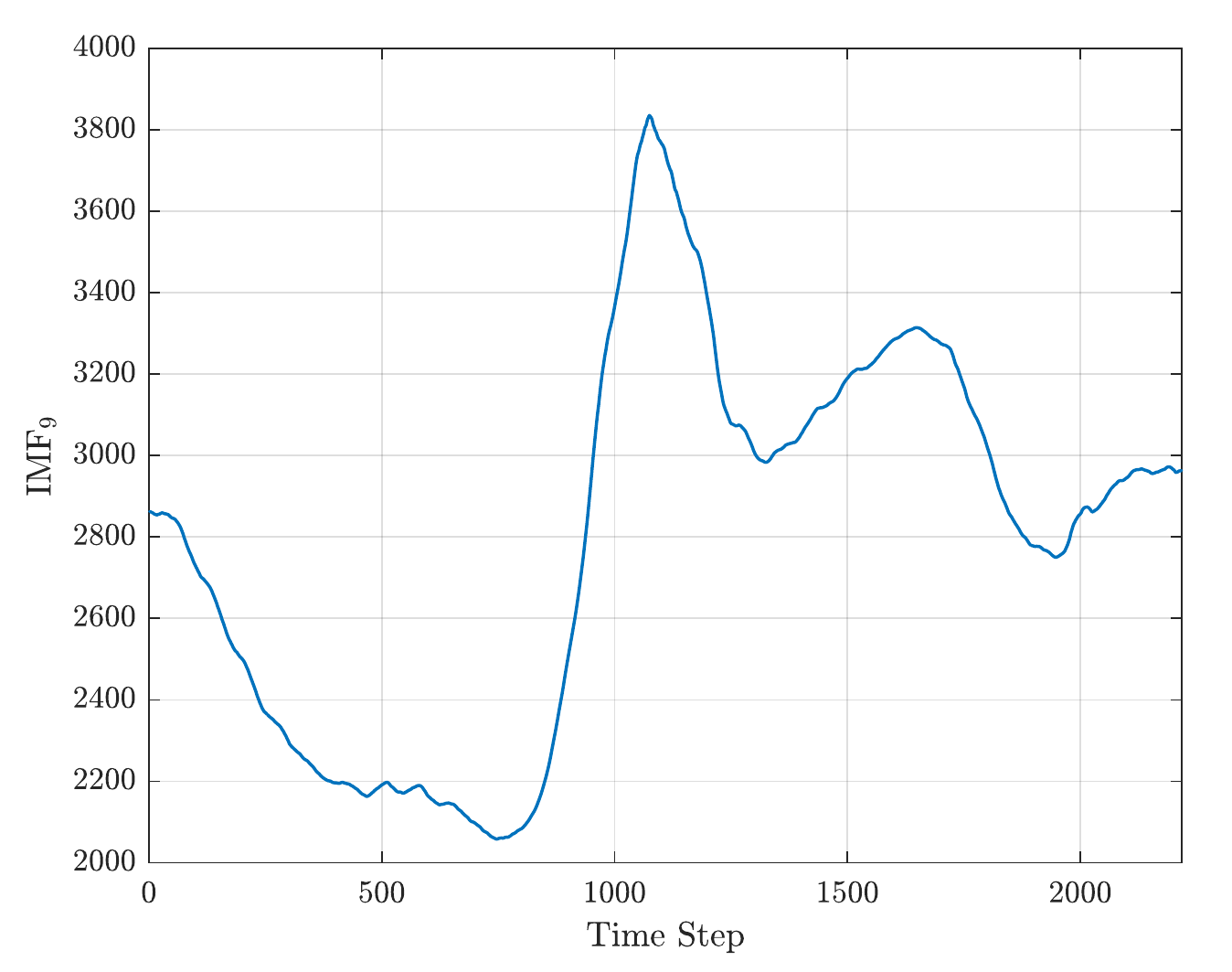}} 	
	\end{tabular}
	\caption{VMD decomposition of the Shanghai Stock Exchange time series.}
	\label{fig:VMD_Decomposition}	
\end{figure}

\textbf{Decomposition Phase:} In this phase, the input signal is decomposed into $K$ IMFs using VMD. Let $D=\left\{x^{[t]}, x^{[t+h]}\right\}_{t=1}^{p}$ represent the input time series, where $h$ and $p$ denote the prediction horizon and number of samples, respectively. VMD decomposes $\mathbf{x}$ to $K$ IMFs by taking $K$ as input. Obviously, these components have simpler structures, so the decomposition phase allows the proposed method to predict simpler structures with more stationary trends and achieve higher accuracy. To clarify the proposed algorithm, in what follows, we denote the $i$th IMF by $\mathbf{imf}_{i}=\left[imf_{i}^{[1]}, imf_{i}^{[2]}, \ldots, imf_{i}^{[p]}\right]$. Fig.  \ref{fig:VMD_Decomposition} shows an example of VMD decomposition of the Shanghai Stock Exchange (SSE) time series. As seen, the VMD extracts IMFs in the order from higher to lower frequencies.

\textbf{Prediction \& Reconstruction Phase:} Each of the IMFs in this phase is given to a separate MFRFNN as input for training the models. In other words, $K$ MFRFNNs are trained in this phase. The output of each MFRFNN is a component of the predicted signal. By summing these components, the final prediction is reconstructed. Employing the MFRFNN models allows the proposed method to learn and memorize historical data from previous observations, capture the dynamic characteristics of the financial time series, and approximate several functions simultaneously. Formally, let $MFRFNN_i$ represent the $i$th MFRFNN model, and $\hat{\mathbf{y}}_{i}=\left[\hat{y}_{i}^{[1]}, \hat{y}_{i}^{[2]}, \ldots, \hat{y}_{i}^{[p]}\right]$ denote the predicted signal by $MFRFNN_i$. The final prediction $\left(\hat{\mathbf{y}}\right)$ can be computed as follows:
\vspace{-0.1cm}
\begin{equation}
	\hat{\mathbf{y}}=\sum_{i=1}^{K}\left[\hat{y}_{i}^{[1]}, \hat{y}_{i}^{[2]}, \ldots, \hat{y}_{i}^{[p]}\right]
\end{equation}

VMD-MFRFNN training algorithm is shown in Algorithm \ref{alg:VMD_MFRFNN_Training}. Assuming that the number of output and state networks' fuzzy rules are equal in all MFRFNNs, the number of VMD-MFRFNN's trainable parameters is $\left(K_{1}+K_{2}\right) \times N \times K$, where $K_1$ and $K_2$ represent the number of output and state networks' fuzzy rules, respectively, and $N$ is the number of states of each MFRFNN.

\begin{algorithm}[!htbp]
	\setstretch{1.10}
	\small  
	\caption{VMD-MFRFNN training algorithm}
	\label{alg:VMD_MFRFNN_Training}
	\vspace{0.1cm}
	\textbf{Input:} Input time series  $D=\left\{x^{[t]}, x^{[t+h]}\right\}_{t=1}^{p}, K$
	\\
	\textbf{Output:} Predicted signal $\left(\hat{\mathbf{y}}\right)$
	\vspace{0.3cm}
	
	Decompose $\mathbf{x}$ to obtain $\left\{\mathbf{imf}_i\right\}_{i=1}^{K}$
	
	\For{$i \gets 1 \:\:\: \KwTo \:\:\: K\:\:$}{
		\texttt{\\}
		Train $MFRFNN_i$ using $\mathbf{imf}_i$ as input
		
		Predict $\hat{\mathbf{y}}_{i}$ using $MFRFNN_i$ model
		
	\vspace{0.2cm}	
	}
	
	\vspace{0.2cm}
	$\hat{\mathbf{y}} \gets \sum_{i=1}^{K}\left[\hat{y}_{i}^{[1]}, \hat{y}_{i}^{[2]}, \ldots, \hat{y}_{i}^{[p]}\right]$
	\vspace{0.2cm}	
\end{algorithm}
\vspace{-0.5cm}
\section{Results and Discussion}
\label{sec:ResultsAndDiscussion}
\vspace{-0.2cm}
In this section, the prediction performance of DCT-MFRFNN and VMD-MFRFNN is evaluated on three financial time series, including Hang Seng Index (HSI), Shanghai Stock Exchange (SSE), and Standard \& Poor's 500 Index (SPX). All time series were collected from Yahoo Finance. Table~\ref{tab:StatisticalDescription} summarizes the statistical description of these time series. The prediction horizons were considered one, three, and five. 

\begin{table}[h]
	\centering
	\caption{The statistical description of financial time series}
	\label{tab:StatisticalDescription}
	\footnotesize
	\begin{tabular}{@{}llll@{}}
		\toprule
		Time Series & HSI      & SSE     & SPX \\ \midrule
		\#Samples   & 2261     & 2218    & 2769    \\
		Min         & 16250.27 & 1950.01 & 1022.58 \\
		Max         & 33154.12 & 5166.35 & 3756.07 \\
		Mean        & 23418.23 & 2798.88 & 2077.30 \\
		STD         & 3097.44  & 550.70  & 674.72  \\ \bottomrule
	\end{tabular}
\end{table}

Two evaluation metrics were used to assess the performance of the proposed methods and compare them with other methods. Mean Absolute Percentage Error (MAPE) and Root Mean Square Error (RMSE). The MAPE measures how accurate the prediction method fits the data and can be computed as follows:

\begin{equation}
	\mathrm{MAPE}=\frac{1}{p} \sum_{t=1}^p\left|\frac{y_t-\hat{y}_t}{y_t}\right|
\end{equation}

\noindent where $y_t$, $\hat{y}_t$, and $p$ represent target output, predicted output, and the number of samples, respectively. The RMSE is able to quantify the derivations between the predicted and target values. Overall results are more important to RMSE. It effectively penalizes significant errors and is expressed by \eqref{eq:RMSE}:

\begin{equation}
	\label{eq:RMSE}
	\mathrm{RMSE}=\sqrt{\frac{1}{p} \sum_{t=1}^p\left(y_t-\hat{y}_t\right)^2}
\end{equation}

The parameters of the proposed methods were selected by a process of trial and error using the validation set in each time series. Table \ref{tab:Parameters} lists the chosen parameters for each time series. As mentioned, the VMD extracts IMFs in the order from higher to lower frequencies. Therefore, intuitively, the number of output network's fuzzy rules for the first several MFRFNNs should be higher than the last several ones. We used triangular membership functions for the proposed methods' input layer. For the PSO algorithm, the inertia weight was selected adaptively in the range $[0.1, 1.1]$, and the cognitive and social acceleration coefficients were set to 1.49 in all experiments. 

\begin{table}[h]
	\centering
	\caption{The parameters of the proposed methods in each financial time series}
	\label{tab:Parameters}
	\footnotesize
	\begin{tabular}{@{}lccc@{}}
		\toprule
		Time Series                                                                      & HSI & SSE & SPX \\ \midrule
		\begin{tabular}[c]{@{}l@{}}Number of Output Network's Fuzzy Rules\\ $[$MFRFNN$_1$, MFRFNN$_2, \cdots,$ MFRFNN$_9]$ \end{tabular} & [6 6 6 5 5 5 4 4 4]  & [4 4 4 3 3 3 2 2 2]   & [4 4 4 3 3 3 2 2 2]   \\
		\begin{tabular}[c]{@{}l@{}}Number of State Network's Fuzzy Rules\\ (State Network)\end{tabular}  & 2   & 4   & 2   \\
		Number of States                                                                 & 2   & 2   & 2   \\
		\begin{tabular}[c]{@{}l@{}}Maximum Number of FES\\ (PSO Method)\end{tabular}     & 600 & 800 & 600 \\
		Number of IMFs                                                                   & 9   & 9   & 9 \\ $\lambda$ & 80 & 85 & 86
		  \\ \bottomrule
	\end{tabular}
\end{table}

To evaluate the effectiveness of DCT-MFRFNN and VMD-MFRFNN in the time series prediction task, we compared their results with six other methods, including ARIMA as a classic statistical model, feedforward neural network (FFNN) as an example of traditional machine learning models, LSTM as a deep learning method, EMD-LSTM as a decomposition-based method, and MEMD-LSTM \cite{deng2022multi} and MFRFNN \cite{nasiri2022mfrfnn} as two state-of-the-art models proposed in 2022. In all experiments, we executed twenty independent runs for MFRFNN, DCT-MFRFNN, and VMD-MFRFNN and reported the averages and standard deviations of the results over these runs.

\subsection{Hang Seng Index (HSI)}
\label{sec:HSI}

The Hang Seng Index (HSI) is a stock market index in Hong Kong. It is the primary indicator of the performance of the Hong Kong stock market as a whole and is utilized to track daily changes in the top companies listed on the Hong Kong stock exchange. This index is an example of relatively developed markets with active transactions \cite{deng2022multi}. The HSI data was collected from Yahoo Finance website during a nine-year period from 4-January-2010 to 8-March-2019. The training, validation, and test set in this experiment consist of 1509, 376, and 376 samples, respectively.

Table \ref{tab:HSI_Results} presents the one, three, and five-step-ahead prediction results of the HSI. The results showed better performance of VMD-MFRFNN compared to other methods in one and three-step-ahead predictions of HSI time series, closely followed by MEMD-LSTM. In five-step-ahead prediction based on MAPE, MEMD-LSTM outperformed other methods, and after that, VMD-MFRFNN had the lowest MAPE. Based on RMSE, VMD-MFRFNN had the highest accuracy in five-step-ahead prediction, closely followed by MEMD-LSTM. Based on this metric, VMD-MFRFNN showed a decrease of 63.27\%, 41.27\%, and 3.25\% from the MEMD-LSTM in one, three, and five-step-ahead forecasting tasks, respectively.

\begin{table}[h]

	\centering
	\caption{One, three, and five-step-ahead prediction results of the HSI}
	\label{tab:HSI_Results}
	\footnotesize
	\begin{tabular}{@{}lccccc@{}}
		\toprule
		Method     & Step & \multicolumn{1}{c}{MAPE (\%)}                                          & $p$-value & \multicolumn{1}{c}{RMSE}                                               & $p$-value \\ \midrule
		ARIMA      & 1    & 3.40E+00                                                               &  2.72E$-$38
		       & 7.30E+02                                                               &  8.78E$-$33
		       
		       \\
		& 3    & 4.24E+00                                                               &   3.67E$-$45
		
		      & 7.84E+02                                                               &  1.72E$-$36
		      
		             \\
		& 5    & 5.20E+00                                                               &  4.59E$-$45
		
		      & 8.16E+02                                                               &  2.70E$-$37
		      
		             \\ \midrule
		FFNN       & 1    & 1.93E+00                                                               &   5.98E$-$33
		
		      & 6.21E+02                                                               &  3.67E$-$31
		      
		             \\
		& 3    & 2.18E+00                                                               &  1.77E$-$38
		
		      & 6.37E+02                                                               &  4.01E$-$34
		      
		            \\
		& 5    & 2.61E+00                                                               &   1.85E$-$37
		
		      & 6.51E+02                                                               &   6.71E$-$34
		      
		            \\ \midrule
		LSTM       & 1    & 1.20E+00                                                               &   5.49E$-$28
		
		      & 4.26E+02                                                               &  4.14E$-$27
		      
		             \\
		& 3    & 1.43E+00                                                               &    1.75E$-$33
		
		     & 4.37E+02                                                               &   3.72E$-$29
		     
		           \\
		& 5    & 1.53E+00                                                               &    3.12E$-$29
		
		     & 4.43E+02                                                               &  1.96E$-$25
		     
		            \\ \midrule
		EMD-LSTM   & 1    & 6.21E$-$01                                                               &   2.98E$-$19
		
		      & 3.51E+02                                                               &  8.36E$-$25
		      
		             \\
		& 3    & 7.93E$-$01                                                               &  2.26E$-$23
		
		       & 3.66E+02                                                               &    2.70E$-$26
		       
		            \\
		& 5    & 8.44E$-$01                                                               &  1.14E$-$09
		
		       & 3.79E+02                                                               &    7.35E$-$19
		       
		            \\ \midrule
		MEMD-LSTM  & 1    & 5.12E$-$01                                                               &   1.12E$-$15
		
		      & 3.24E+02                                                               &  8.70E$-$24
		      
		             \\
		& 3    & 5.87E$-$01                                                               &   3.12E$-$11
		
		      & 3.32E+02                                                               &  1.80E$-$24
		      
		             \\
		& 5    & \textbf{6.27E$-$01}                                                               &    2.62E$-$22
		
		     & 3.38E+02                                                               &  4.04E$-$07
		     
		            \\ \midrule
		MFRFNN     & 1    & \begin{tabular}[c]{@{}c@{}}8.59E$-$01\\(1.79E$-$03)\end{tabular}   &    4.69E$-$24
		
		      & \begin{tabular}[c]{@{}c@{}}3.24E+02\\(4.88E$-$01)\end{tabular}                                                   &   7.95E$-$24
		      
		           \\	& 3    & \begin{tabular}[c]{@{}c@{}}1.57E+00\\(9.94E$-$04)\end{tabular}                                                  &   8.03E$-$35
		           
		                 & \begin{tabular}[c]{@{}c@{}}5.65E+02\\(4.68E$-$01)\end{tabular}                                                   &    8.37E$-$33
		                 
		                      \\
		& 5    & \begin{tabular}[c]{@{}c@{}}2.06E+00\\(1.14E$-$03)\end{tabular}                                                  &   2.13E$-$34
		
		      & \begin{tabular}[c]{@{}c@{}}7.50E+02\\($6.88$E$-$01)\end{tabular}                                                   &   9.84E$-$37
           \\ \midrule
DCT-MFRFNN     & 1    & \begin{tabular}[c]{@{}c@{}}8.14E$-$01\\(1.65E$-$03)\end{tabular}   &    2.49E$-$23

& \begin{tabular}[c]{@{}c@{}}3.18E+02\\(1.79E+00)\end{tabular}                                                   &   4.65E$-$24

\\	& 3    & \begin{tabular}[c]{@{}c@{}}1.31E+00\\(3.31E$-$02)\end{tabular}                                                  &   1.18E$-$38

& \begin{tabular}[c]{@{}c@{}}4.71E+02\\(1.26E+01)\end{tabular}                                                   &    3.80E$-$40

\\
& 5    & \begin{tabular}[c]{@{}c@{}}1.96E+00\\(2.10E$-$02)\end{tabular}                                                  &   1.31E$-$54

& \begin{tabular}[c]{@{}c@{}}6.89E+02\\(1.00E+01)\end{tabular}                                                   &   1.44E$-$46
		           \\ \midrule
		VMD-MFRFNN & 1    & \textbf{\begin{tabular}[c]{@{}c@{}}3.16E$-$01\\ (3.65E$-$02)\end{tabular}} &  $-$       & \textbf{\begin{tabular}[c]{@{}c@{}}1.19E+02\\ (1.41E+01)\end{tabular}} &   $-$      \\
		& 3    & \textbf{\begin{tabular}[c]{@{}c@{}}5.29E$-$01\\ (1.91E$-$02)\end{tabular}} &  $-$       & \textbf{\begin{tabular}[c]{@{}c@{}}1.95E+02\\ (8.67E+00)\end{tabular}} &   $-$      \\
		& 5    & \begin{tabular}[c]{@{}c@{}}8.99E$-$01\\ (2.24E$-$02)\end{tabular} &  $-$       & \textbf{\begin{tabular}[c]{@{}c@{}}3.27E+02\\ (6.53E+00)\end{tabular}} &  $-$       \\ \bottomrule
	\end{tabular}
\end{table}

ARIMA had the lowest accuracy in all experiments. Moreover, the results illustrated in Table \ref{tab:HSI_Results} indicate that DCT-MFRFNN outperformed MFRFNN in all prediction horizons. Based on RMSE, DCT-MFRFNN showed a decrease of 1.85\%, 16.64\%, and 8.13\% from MFRFNN in prediction horizons of one, three, and five, respectively. The good performance of VMD-MFRFNN is due to its structure. A structure with a feedback loop allows this method to memorize past observations. Moreover, employing multiple states and the VMD decomposition method enables the VMD-MFRFNN to store historical information and capture the dynamic characteristics of the time series over longer prediction horizons. 

Comparing the results obtained by LSTM with EMD-LSTM and MFRFNN with~ VMD-MFRFNN revealed that decomposition approaches have a favorable effect on forecasting results. Decomposition-based methods decompose complex time series into several IMFs with simpler structures and more stationary trends. Reducing the complexity will make it easier for the model to predict the time series.
In addition, a two-tailed Welch's t-test with a significance threshold of $\alpha=0.05$ was used to determine if the superiority of a model was statistically significant or not. The statistical test was performed between VMD-MFRFNN and other methods. Welch's t-test, a nonparametric two-sample statistical test, is beneficial when the samples have unequal variances \cite{zojaji2022semantic}. As can be seen, in all experiments, the null hypothesis is rejected (i.e., $p$-value $< 0.05$), confirming that the results are statistically significant. 

\begin{figure}[!htbp]
	\centering
	\begin{tabular}{ccc}
		\subfigure[][{\scriptsize HSI Time Series}]{\includegraphics[width=5cm]{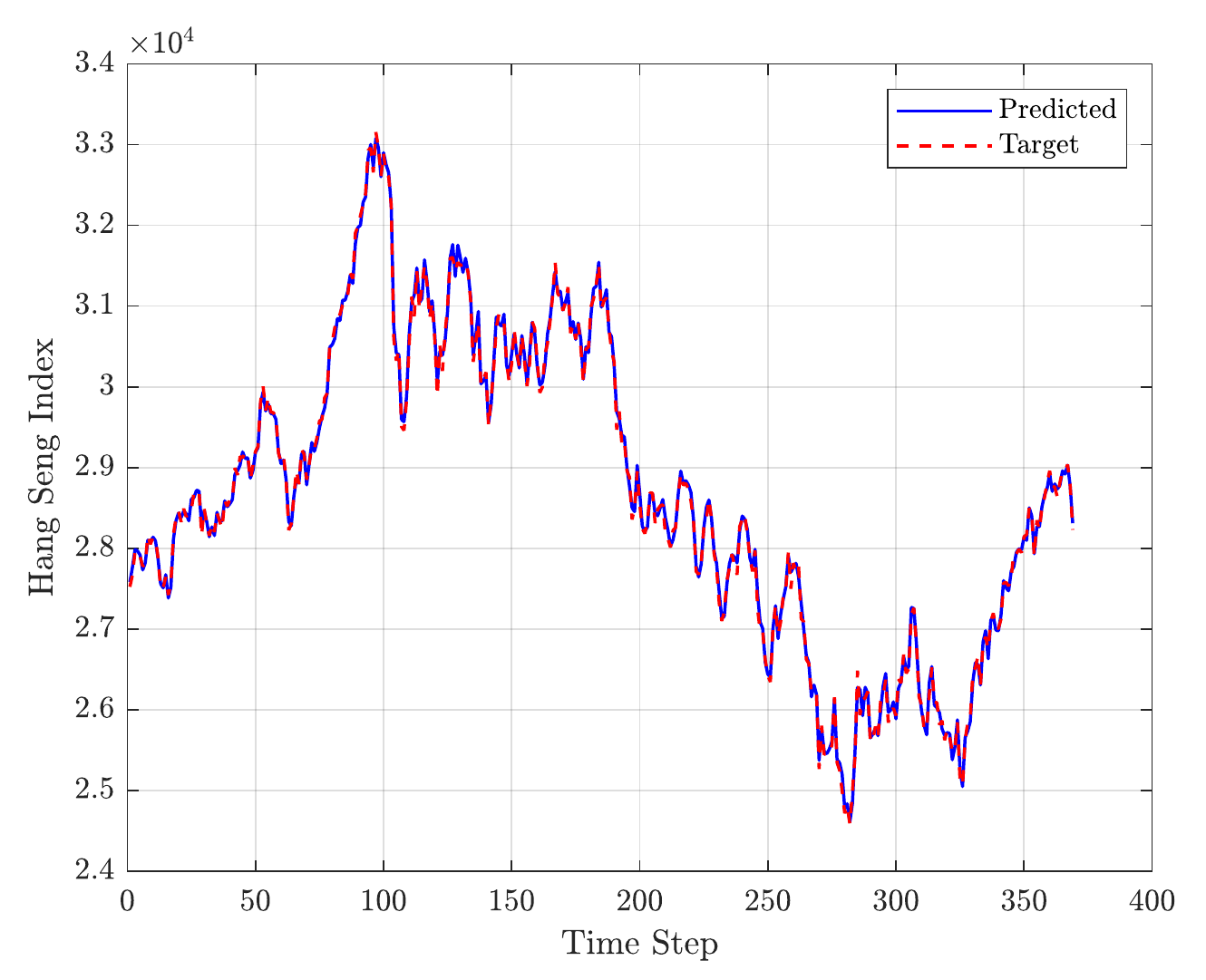}}	&  \subfigure[][{\scriptsize IMF$_1$ Component}]{\includegraphics[width=5cm]{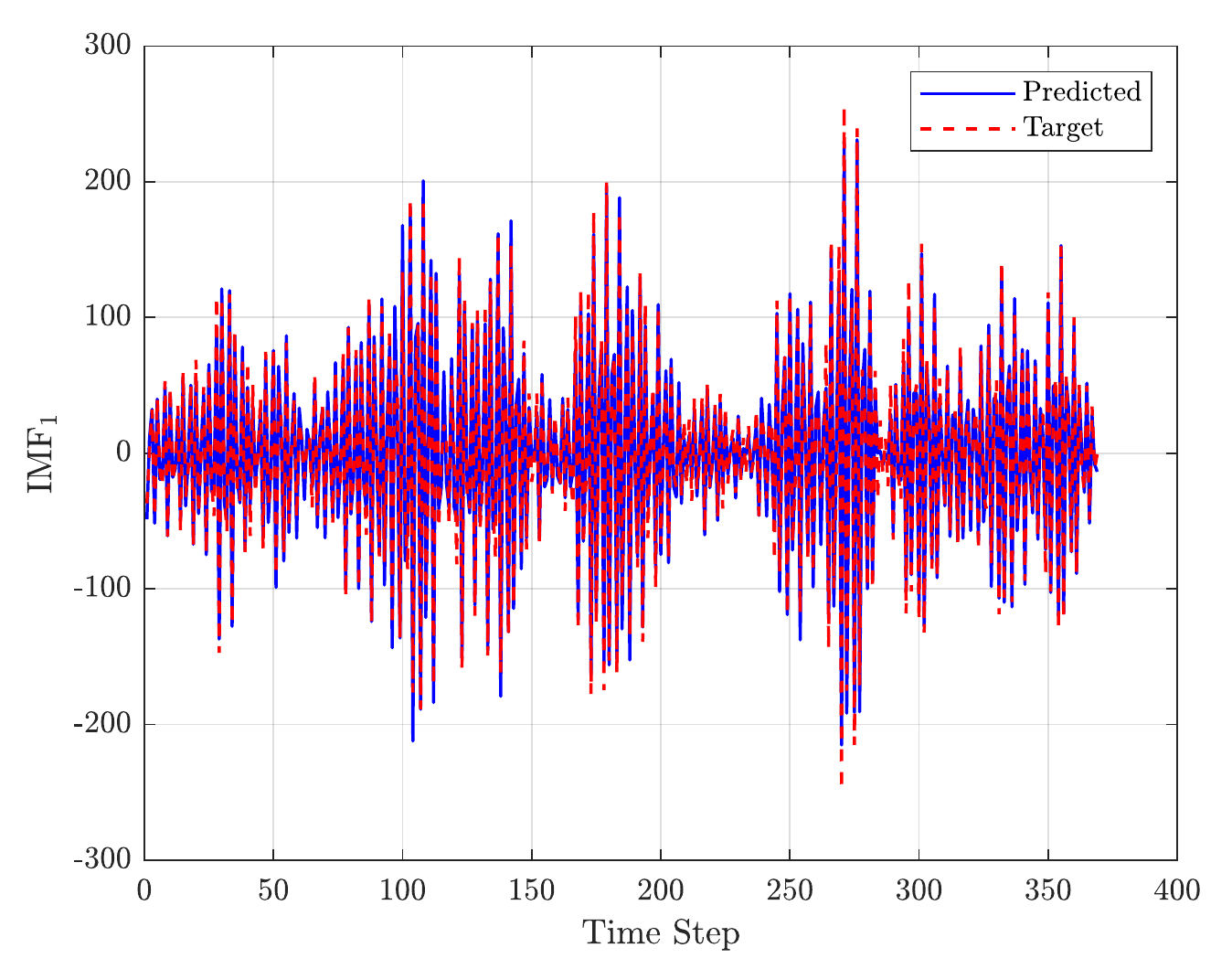}}   &
		\subfigure[][{\scriptsize IMF$_2$ Component}]{\includegraphics[width=5cm]{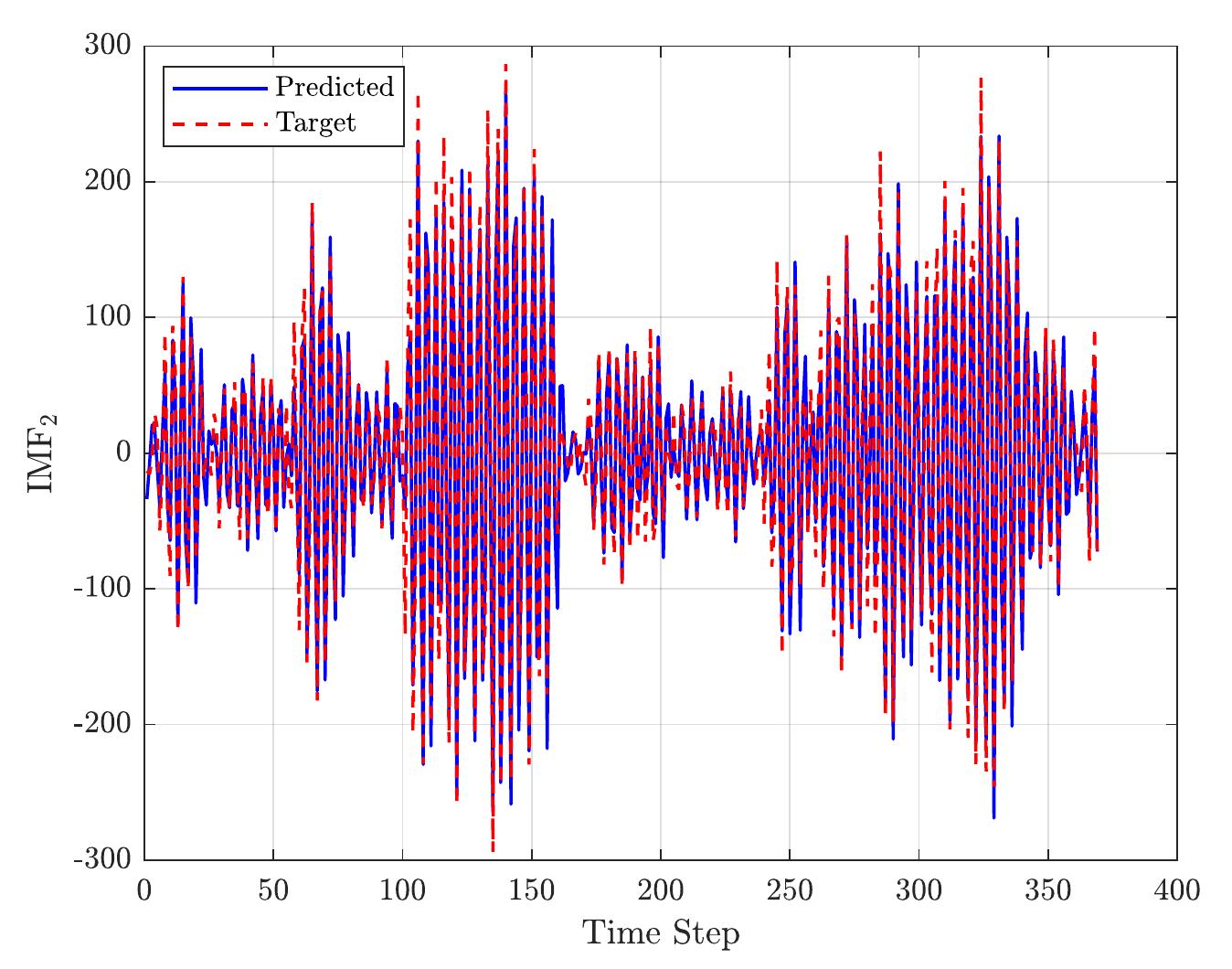}} 	\\   	\subfigure[][{\scriptsize IMF$_3$ Component}]{\includegraphics[width=5cm]{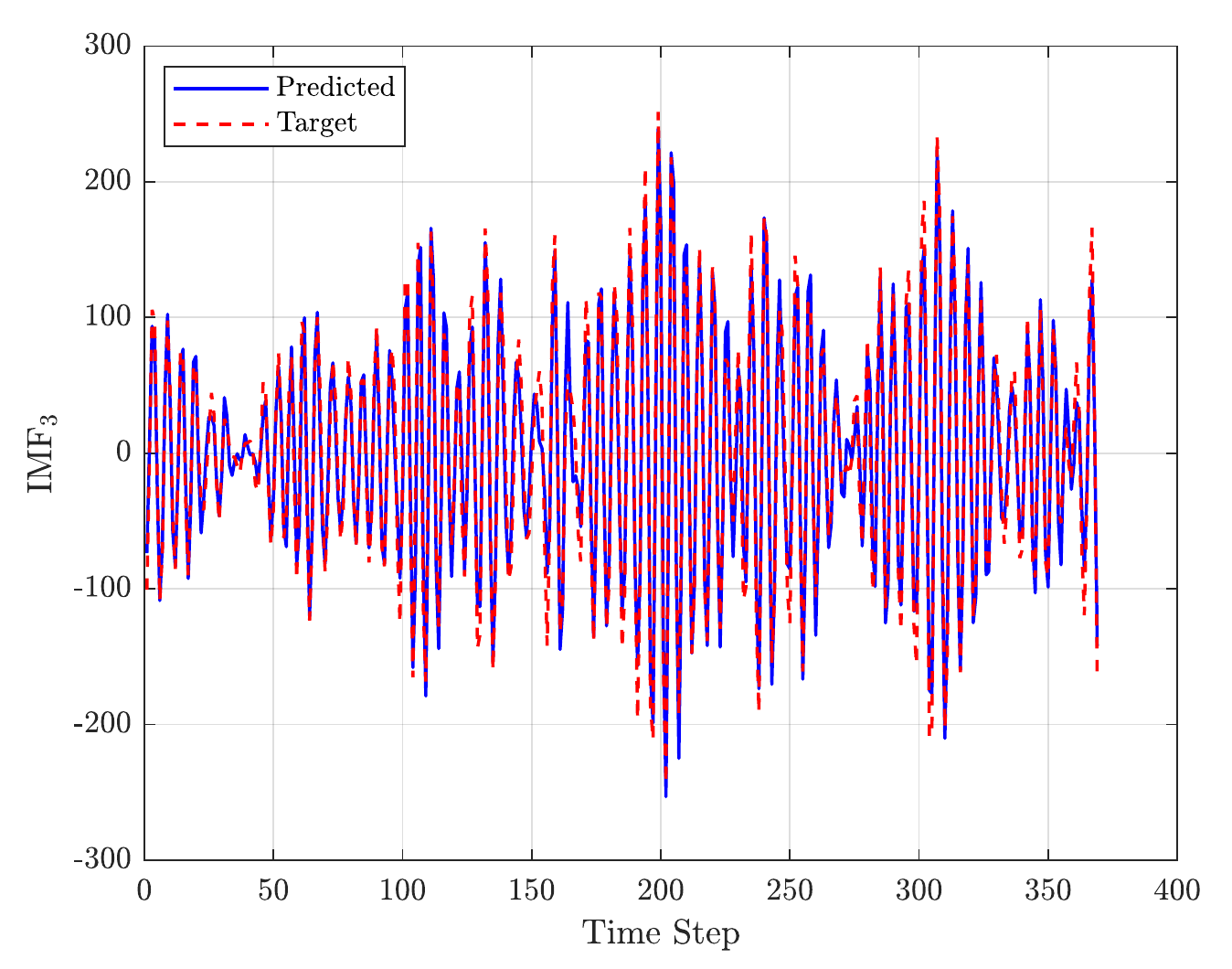}}  &
		\subfigure[][{\scriptsize IMF$_4$ Component}]{\includegraphics[width=5cm]{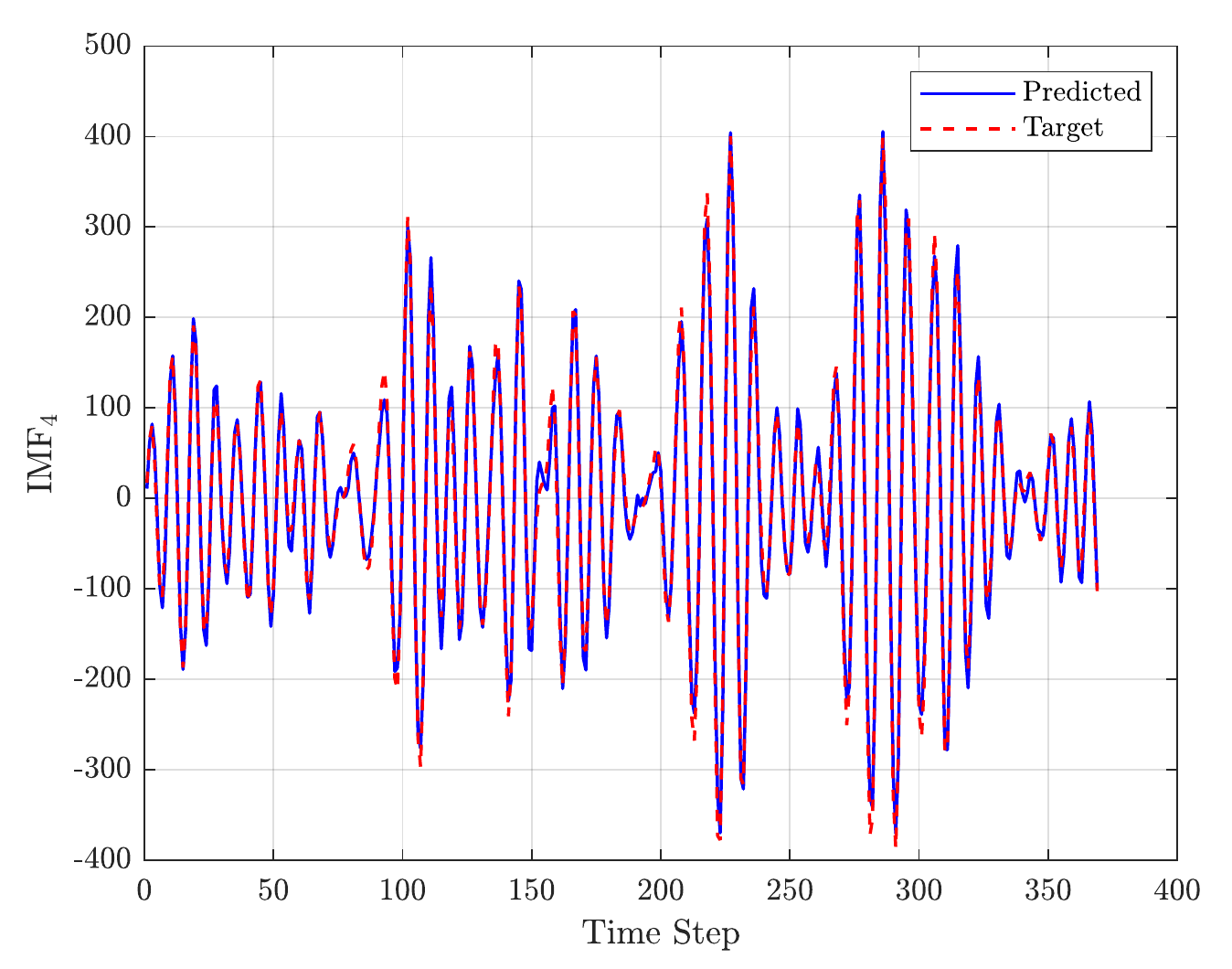}}  &   		\subfigure[][{\scriptsize IMF$_5$ Component}]{\includegraphics[width=5cm]{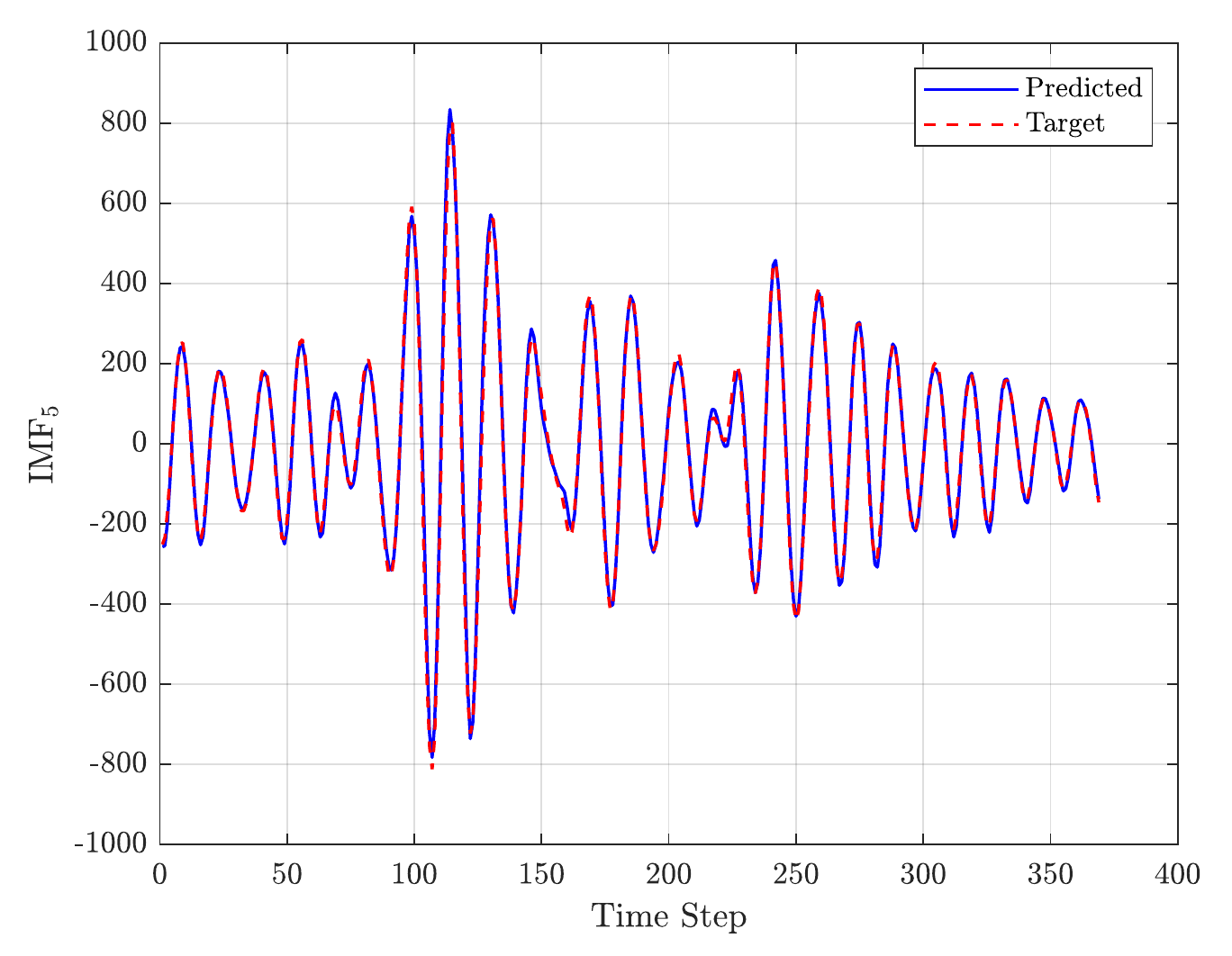}}  \\
		\subfigure[][{\scriptsize IMF$_6$ Component}]{\includegraphics[width=5cm]{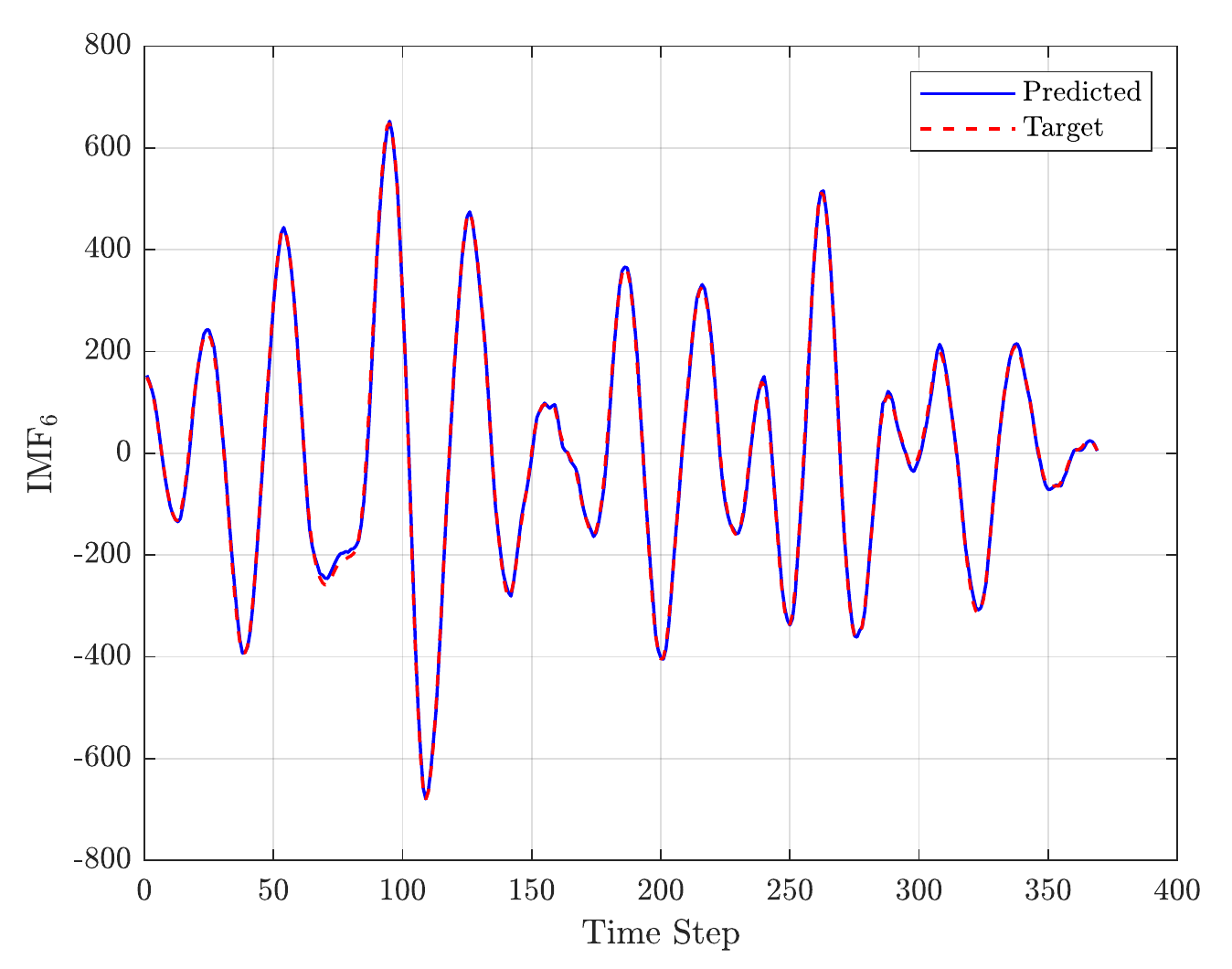}} &   		\subfigure[][{\scriptsize IMF$_7$ Component }]{\includegraphics[width=5cm]{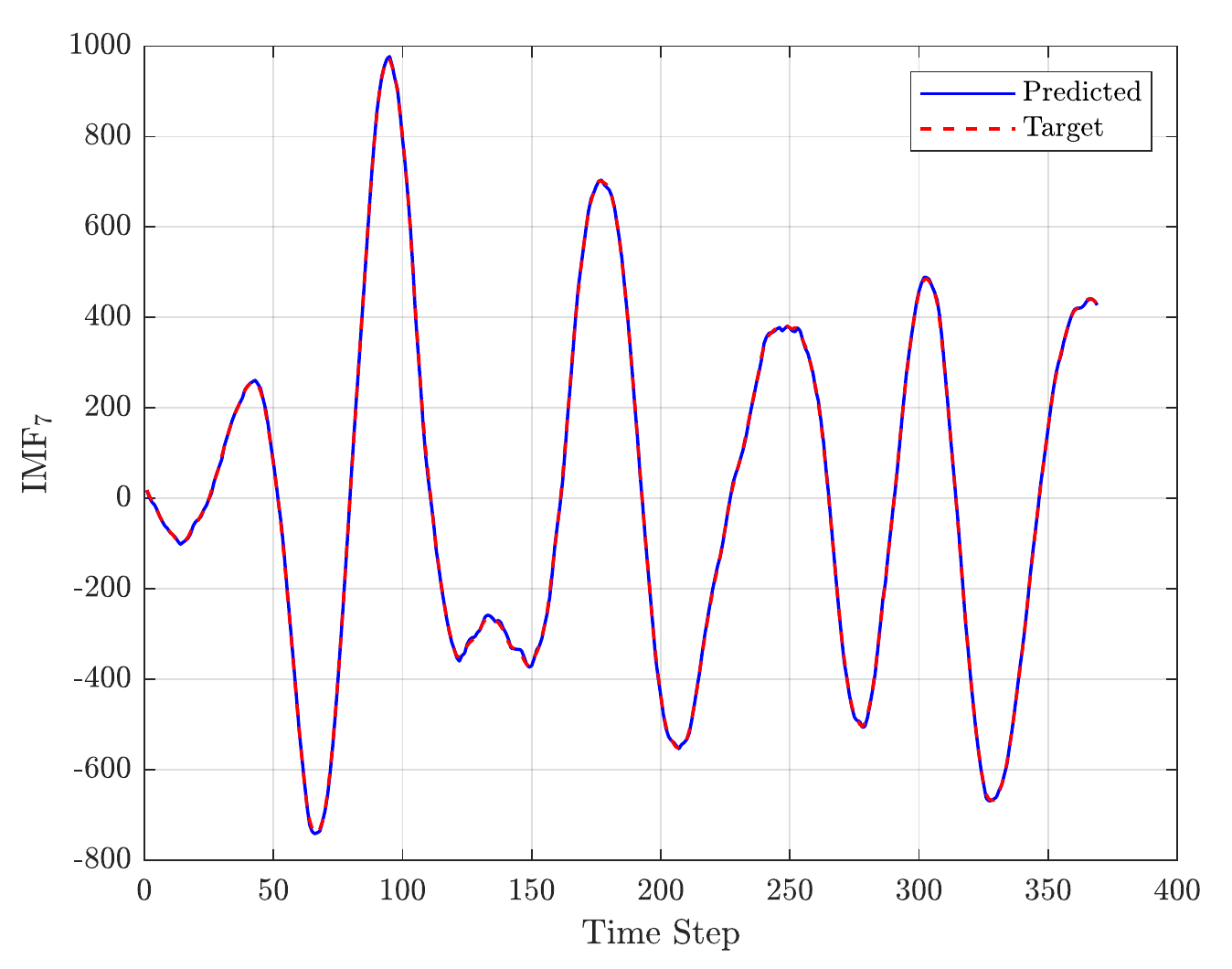}} & 
		\subfigure[][{\scriptsize IMF$_8$ Component}]{\includegraphics[width=5cm]{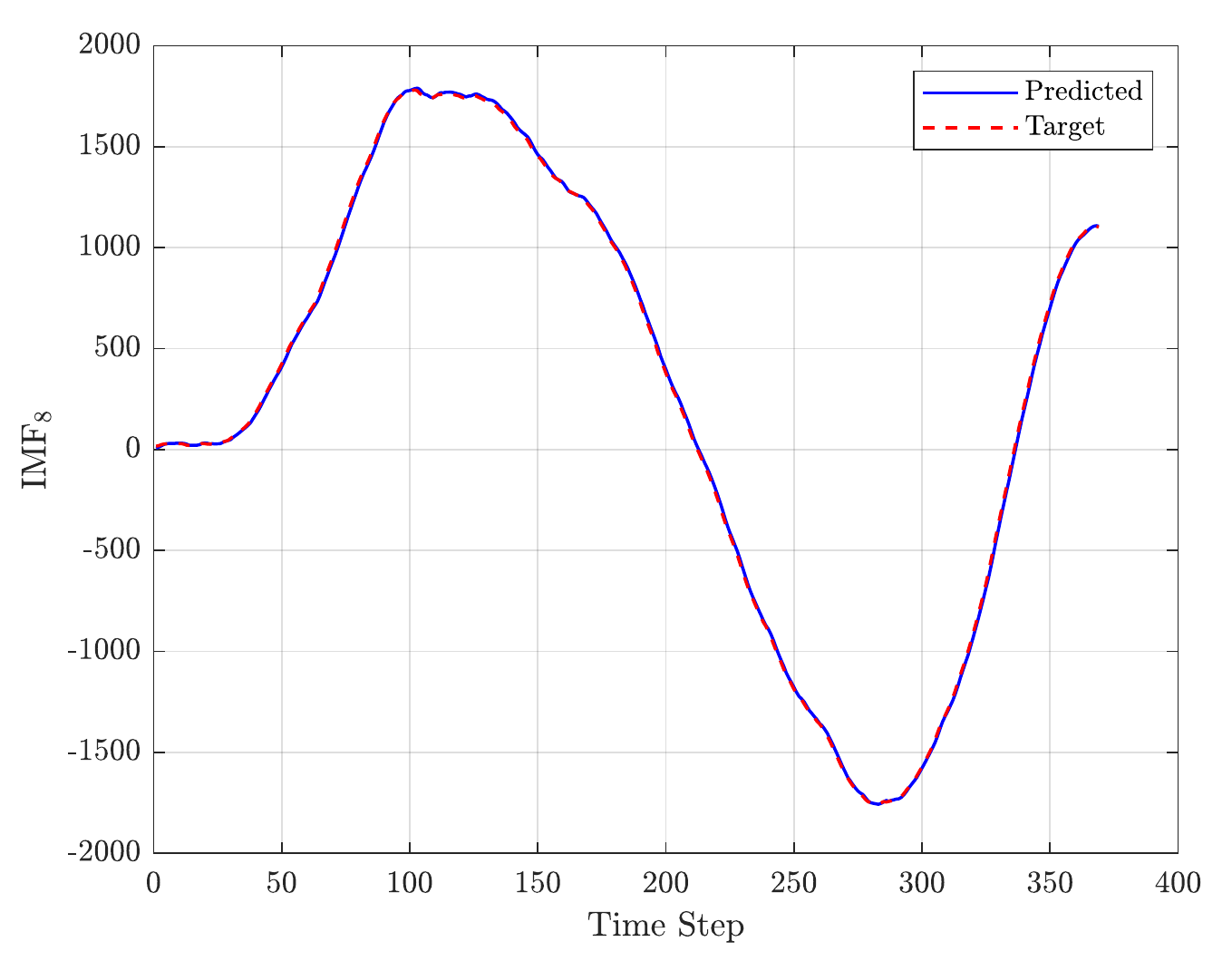}}
	\end{tabular}
	\begin{tabular}{c}
		\subfigure[][{\scriptsize IMF$_9$ Component}]{\includegraphics[width=5cm]{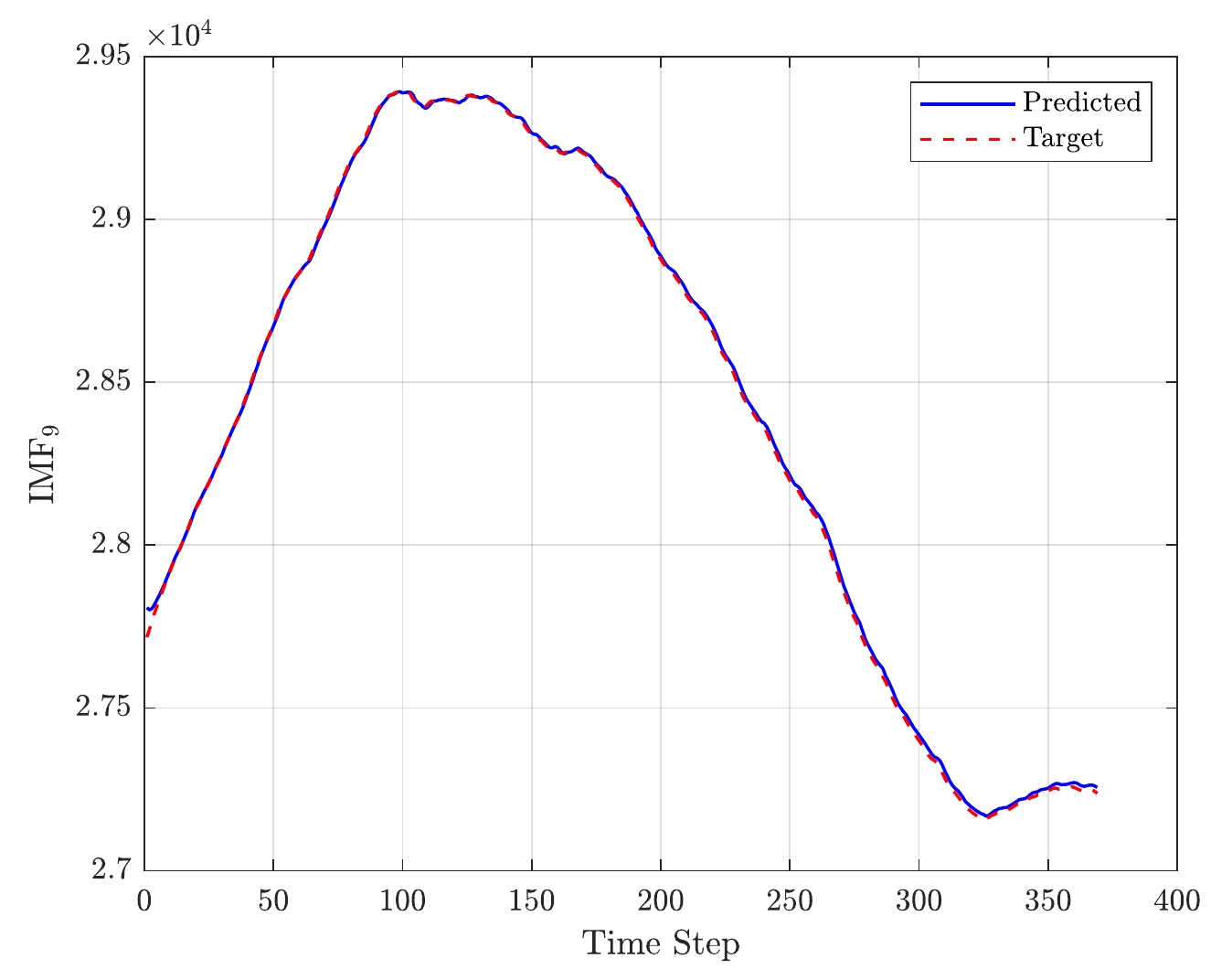}} 	
	\end{tabular}
	\caption{One-step-ahead prediction curves of VMD decomposition components for the HSI time series.}
	\label{fig:HSI_Prediction_VMD_Decomposition}	
\end{figure}

\begin{figure}[h]
	\centering
	\includegraphics[width=0.85\textwidth]{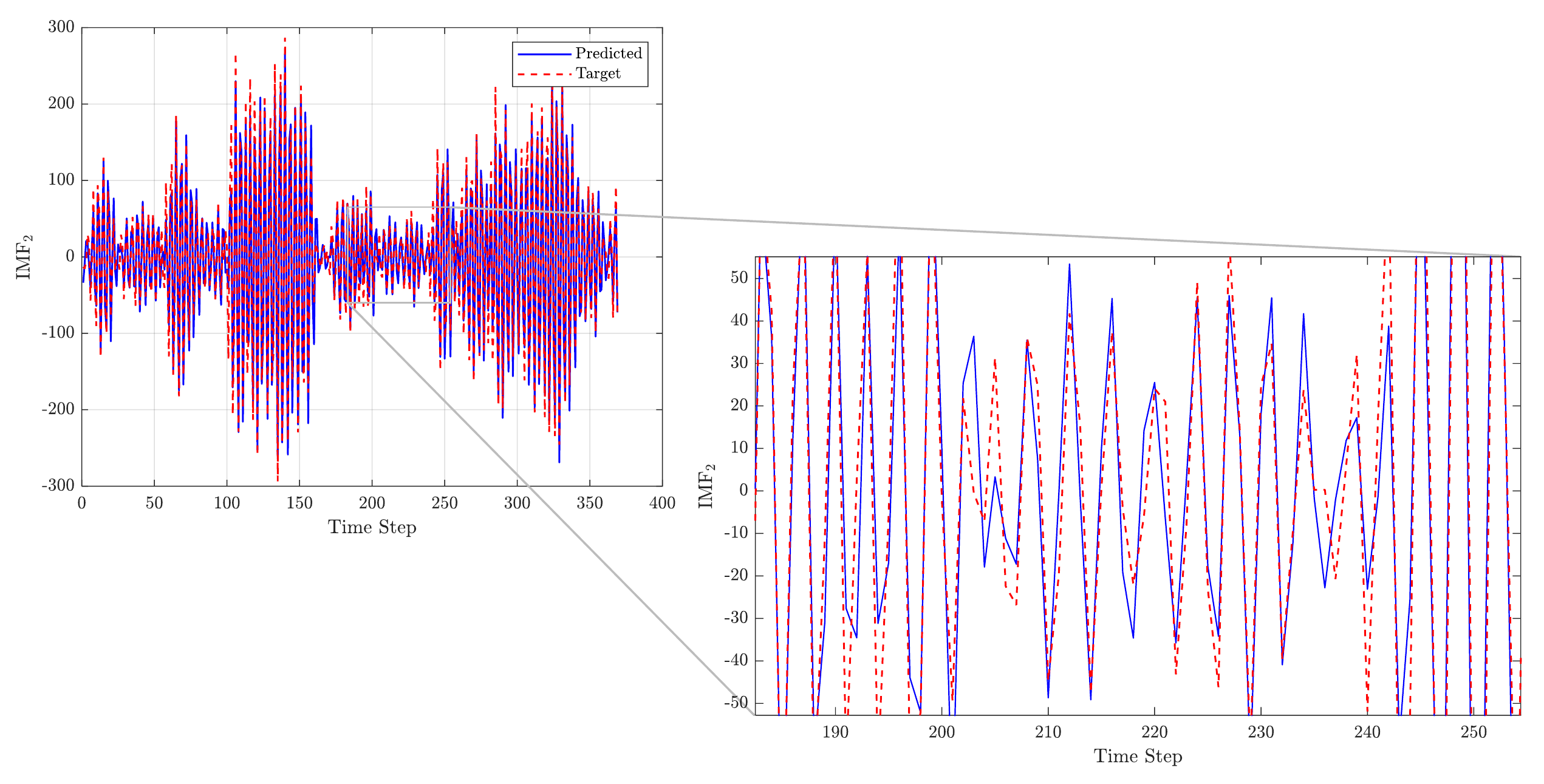}
	\caption{One-step-ahead prediction curve of IMF$_2$ for the HSI time series.}
	\label{fig:HSI_IMF2_Sep}
\end{figure}

\begin{figure}[!htbp]
	\centering
	\includegraphics[width=0.85\textwidth]{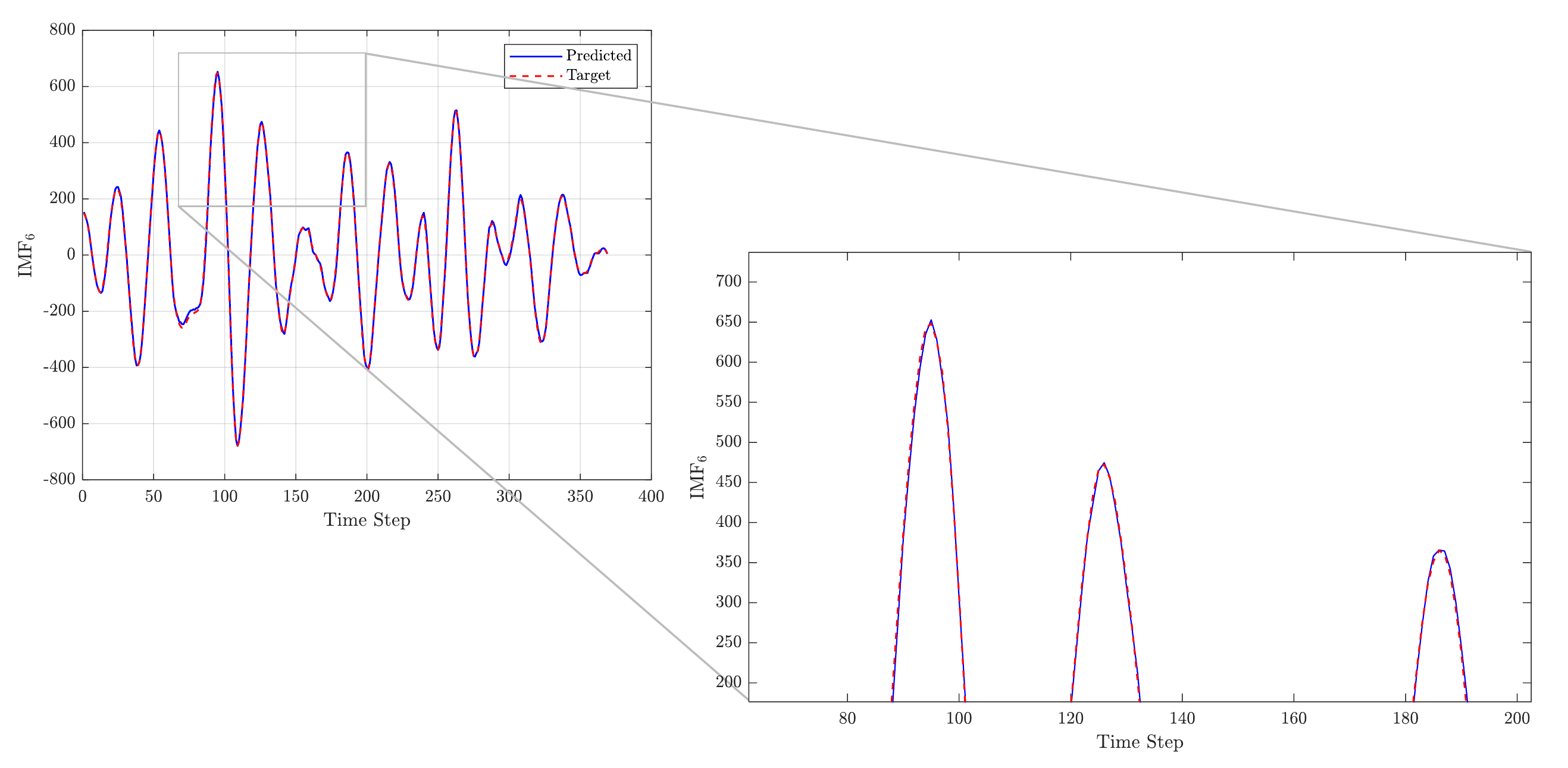}
	\caption{One-step-ahead prediction curve of IMF$_6$ for the HSI time series.}
	\label{fig:HSI_IMF6_Sep}
\end{figure}

\begin{figure}[htbp]
	\centering
	\begin{tabular}{ccc}
		\subfigure[][{\scriptsize One-step ahead prediction}]{\includegraphics[width=5cm]{HSI1Step}}	&  \subfigure[][{\scriptsize Three-step ahead prediction}]{\includegraphics[width=5cm]{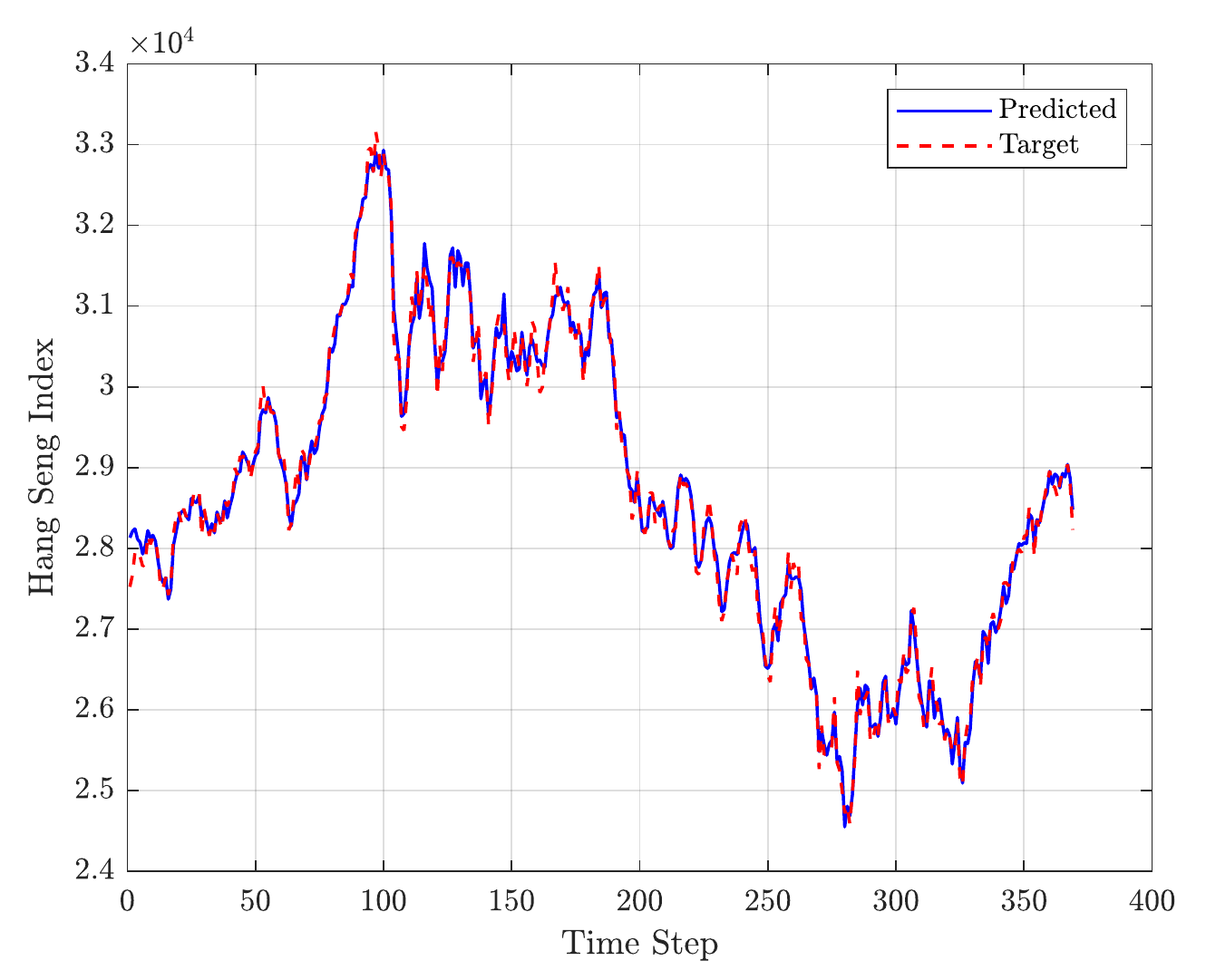}}   &
		\subfigure[][{\scriptsize Five-step ahead prediction}]{\includegraphics[width=5cm]{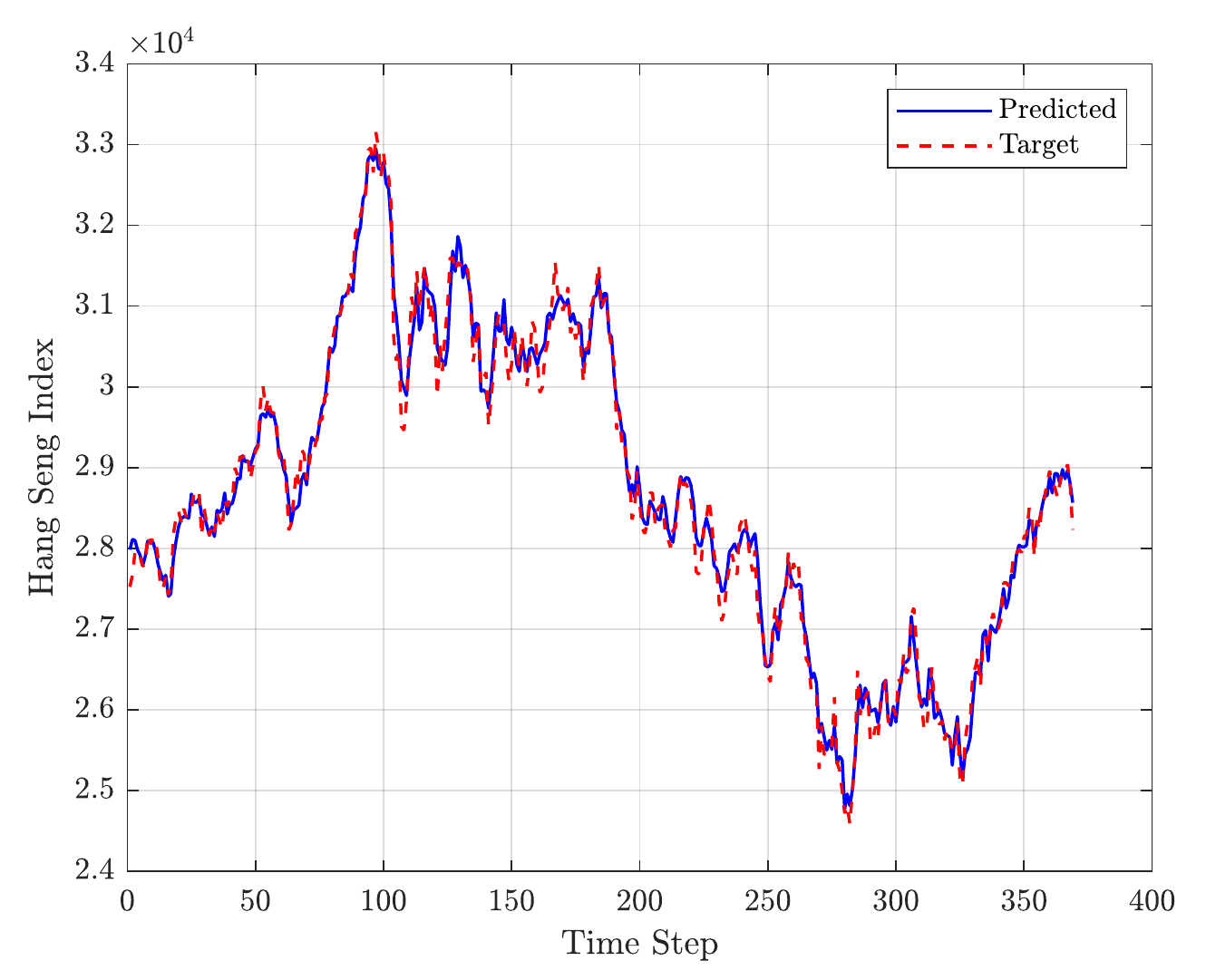}} 	\\   		\subfigure[][{\scriptsize Histogram of Errors (One-step ahead)}]{\includegraphics[width=5cm]{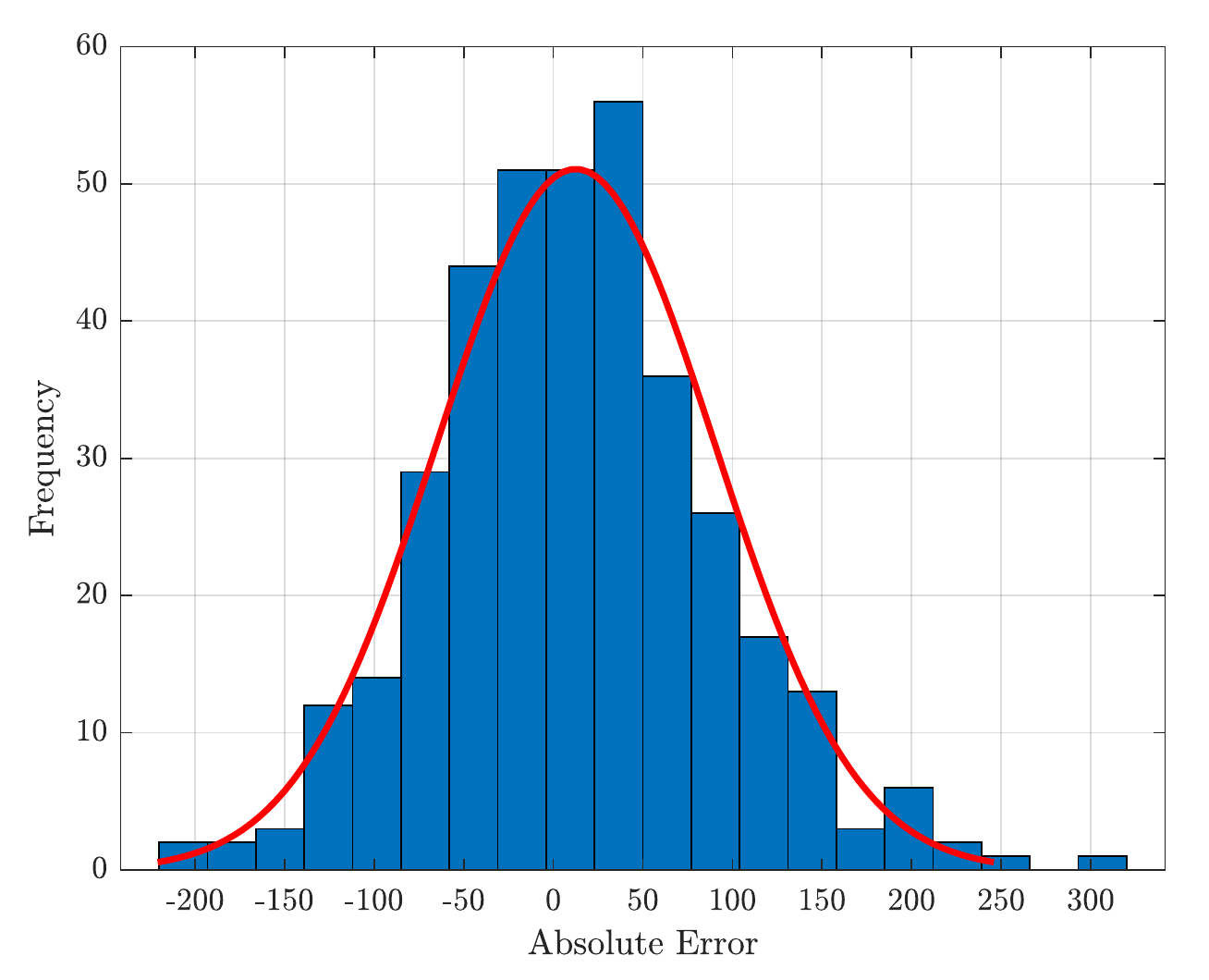}}  &
		\subfigure[][{\scriptsize Histogram of Errors (Three-step ahead)}]{\includegraphics[width=5cm]{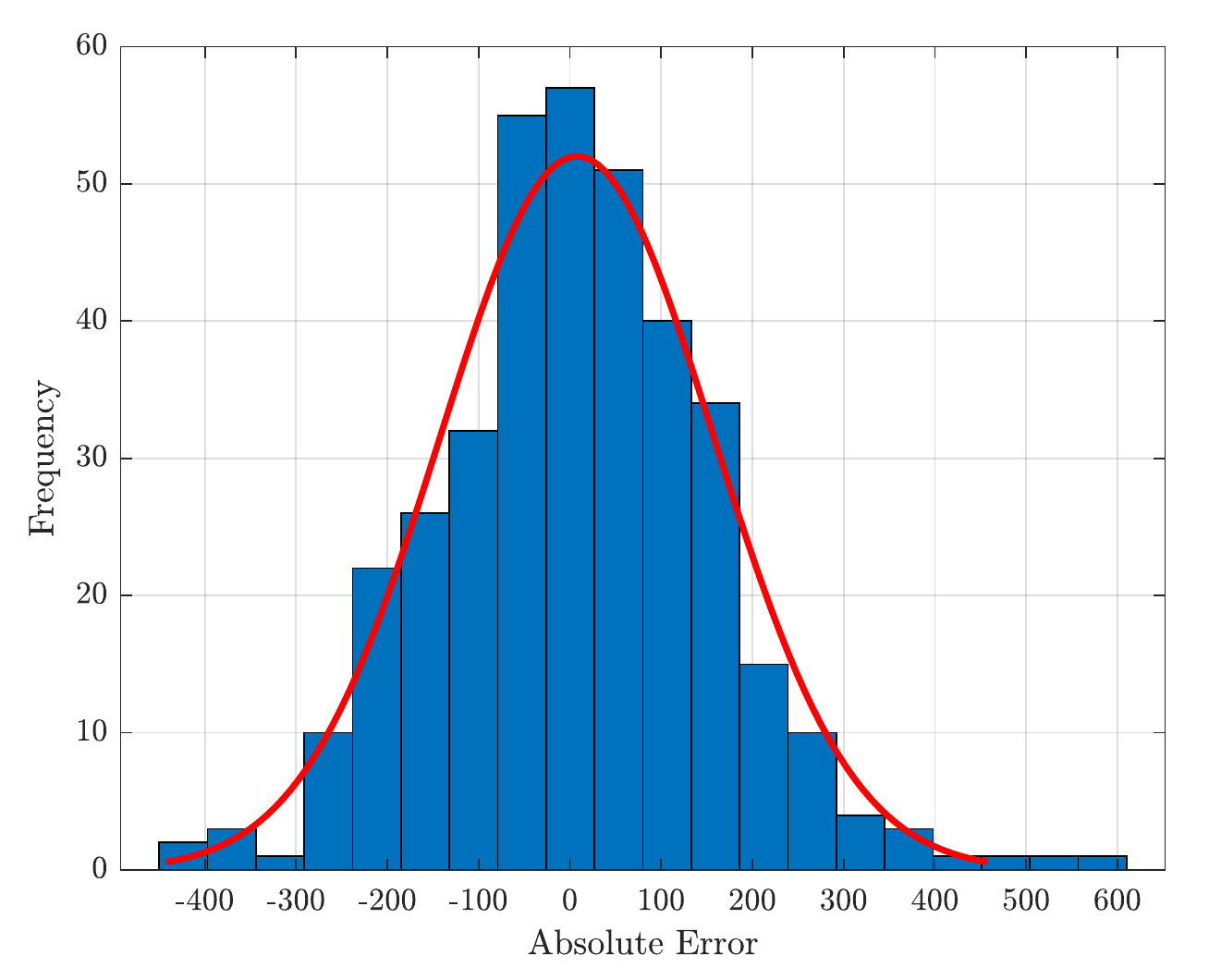}}  &   		\subfigure[][{\scriptsize Histogram of Errors (Five-step ahead)}]{\includegraphics[width=5cm]{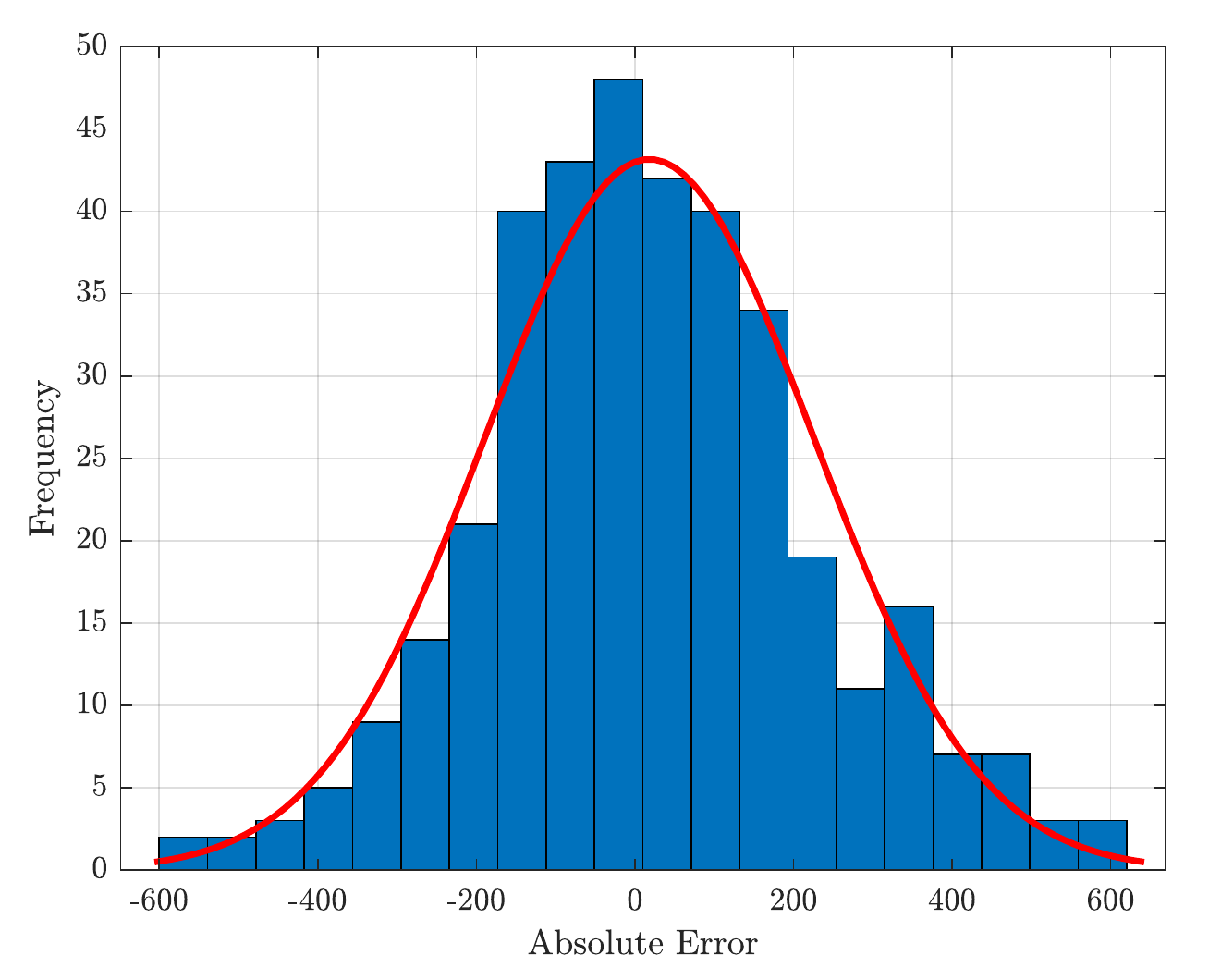}}  
	\end{tabular}
	
	\caption{Prediction curves and histograms of errors for the HSI time series.}
	\label{fig:HSI_Prediction_Plots}	
\end{figure}

Fig. \ref{fig:HSI_Prediction_VMD_Decomposition} demonstrates one-step-ahead prediction curves of VMD decomposition components for the HSI time series generated by VMD-MFRFNN. Fig. \ref{fig:HSI_IMF2_Sep} and Fig. \ref{fig:HSI_IMF6_Sep} illustrate the one-step-ahead prediction curves of IMF$_2$ and IMF$_6$, respectively. As can be seen, VMD-MFRFNN predicts low-frequency IMF components with high accuracy and high-frequency IMF components with low accuracy because the number of fuzzy rules is too small to model high-frequency IMF components accurately. Intuitively, the prediction of high-frequency IMF components is a bias for VMD-MFRFNN that adjust the final output. Fig. \ref{fig:HSI_Prediction_Plots} shows prediction curves and histograms of errors for the HSI time series produced by VMD-MFRFNN. The Gaussian distribution of errors indicates that VMD-MFRFNN effectively captured the dynamic properties of the time series. 

\subsection{Shanghai Stock Exchange (SSE)}
\label{sec:SSE}

The Shanghai Stock Exchange (SSE) is a stock exchange headquartered in Shanghai, China. According to market capitalization, the SSE is the third-largest stock market in the world and is representative of a developing market. The SSE time series data comprised 2218 samples obtained from Yahoo Finance during a nine-year period from 31-December-2010 to 21-February-2020. The first 1478 samples were used for the training, 370 samples for validation, and 370 samples for the test. 

The prediction results of the SSE time series are given in Table \ref{tab:SSE_Results}. As can be seen, the results showed better performance of VMD-MFRFNN compared to other methods in one-step-ahead prediction of SSE time series, closely followed by MEMD-LSTM. In three and five-step-ahead predictions based on MAPE, MEMD-LSTM outperformed other methods, and after that, VMD-MFRFNN had the lowest MAPE. Based on RMSE, VMD-MFRFNN had the highest accuracy in all prediction horizons, closely followed by MEMD-LSTM. Based on this metric, VMD-MFRFNN showed a decrease of 43.58\%, 22.96\%, and 8.10\% from the MEMD-LSTM in one, three, and five-step-ahead prediction tasks, respectively. 

\begin{table}[!t]
	\centering
	\caption{One, three, and five-step-ahead prediction results of the SSE}
	\label{tab:SSE_Results}
	\footnotesize
	\begin{tabular}{@{}lccccc@{}}
		\toprule
		Method     & Step & \multicolumn{1}{c}{MAPE (\%)}                                          & $p$-value & \multicolumn{1}{c}{RMSE}                                               & $p$-value \\ \midrule
		ARIMA      & 1    & 3.68E+00                                                               &    6.49E$-$43
		
		& 4.24E+02                                                               &   8.34E$-$46
		
		\\
		& 3    & 4.36E+00                                                               &   1.68E$-$54
		
		& 4.51E+02                                                               & 2.44E$-$65
		
		\\
		& 5    & 5.02E+00                                                               &    1.31E$-$39
		
		& 5.22E+02                                                               &   1.49E$-$47
		
		\\ \midrule
		FFNN       & 1    & 2.01E+00                                                               &    2.40E$-$37
		
		& 1.78E+02                                                               &   2.69E$-$38
		
		\\
		& 3    & 2.23E+00                                                               &  9.36E$-$48
		
		& 2.15E+02                                                               &   8.94E$-$59
		
		\\
		& 5    & 2.24E+00                                                               &   3.55E$-$31
		
		& 2.28E+02                                                               &   3.86E$-$40
		
		\\ \midrule
		LSTM       & 1    & 1.04E+00                                                               &   1.40E$-$30
		
		& 5.49E+01                                                               &  4.28E$-$27
		
		\\
		& 3    & 1.38E+00                                                               &  5.88E$-$42
		
		& 5.98E+01                                                               &  1.58E$-$45
		
		\\
		& 5    & 1.40E+00                                                               &  8.08E$-$25
		
		& 6.56E+01                                                               &  1.10E$-$26
		
		\\ \midrule
		EMD-LSTM   & 1    & 5.79E$-$01                                                               &   5.97E$-$23
		
		& 4.30E+01                                                               &   2.12E$-$24
		
		\\
		& 3    & 6.81E$-$01                                                               &  6.13E$-$27
		
		& 4.64E+01                                                               &  4.69E$-$42
		
		\\
		& 5    & 7.24E$-$01                                                               &  1.57E$-$04
		
		& 4.92E+01                                                               &   2.82E$-$22
		
		\\ \midrule
		MEMD-LSTM  & 1    & 3.81E$-$01                                                               &   2.61E$-$14
		
		& 2.18E+01                                                               &  7.34E$-$15
		
		\\
		& 3    &\textbf{4.22E$-$01}                                                              &  3.07E$-$26
		
		& 2.70E+01                                                               &   2.34E$-$30
		
		\\
		& 5    &\textbf{4.92E$-$01}                                                      &  8.14E$-$14
		
		& 2.84E+01                                                               &  3.39E$-$05
		
		\\ \midrule
		MFRFNN     & 1    & \begin{tabular}[c]{@{}c@{}}8.86E$-$01\\(7.69E$-$03)\end{tabular}                                                   &  3.03E$-$33
		
		& \begin{tabular}[c]{@{}c@{}}3.67E+01\\(2.38E$-$01)\end{tabular}                                                    &    5.91E$-$23
		
		\\
		& 3    & \begin{tabular}[c]{@{}c@{}}1.63E+00\\(6.45E$-$04)\end{tabular}                                                   &   7.02E$-$45
		
		& \begin{tabular}[c]{@{}c@{}}6.32E+01\\(9.95E$-$02)\end{tabular}                                                     &   9.80E$-$65
		
		\\
		& 5    & \begin{tabular}[c]{@{}c@{}}2.21E+00\\(6.38E$-$03)\end{tabular}                                                   &   5.21E$-$32
		
		& \begin{tabular}[c]{@{}c@{}}8.44E+01\\(8.37E$-$02)\end{tabular}                                                   &    5.58E$-$30
		\\ \midrule
		DCT-MFRFNN     & 1    & \begin{tabular}[c]{@{}c@{}}6.76E$-$01\\(9.72E$-$03)\end{tabular}                                                   &  5.62E$-$31
		
		& \begin{tabular}[c]{@{}c@{}}2.98E+01\\(1.00E+00)\end{tabular}                                                    &    5.14E$-$25
		
		\\
		& 3    & \begin{tabular}[c]{@{}c@{}}1.10E+00\\(1.86E$-$02)\end{tabular}                                                   &  1.37E$-$34
		
		& \begin{tabular}[c]{@{}c@{}}4.23E+01\\(6.64E$-$01)\end{tabular}                                                     &   4.34E$-$34
		
		\\
		& 5    & \begin{tabular}[c]{@{}c@{}}1.46E+00\\(2.56E$-$02)\end{tabular}                                                   &   3.70E$-$35
		
		& \begin{tabular}[c]{@{}c@{}}5.53E+01\\(1.13E+00)\end{tabular}                                                   &    2.98E$-$33
		
		\\ \midrule
		VMD-MFRFNN & 1    & \textbf{\begin{tabular}[c]{@{}c@{}}2.77E$-$01\\ (2.30E$-$02)\end{tabular}} &   $-$      & \textbf{\begin{tabular}[c]{@{}c@{}}1.23E+01\\ (1.96E+00)\end{tabular}} & $-$        \\
		& 3    & \begin{tabular}[c]{@{}c@{}}5.46E$-$01\\ (6.33E$-$03)\end{tabular} &  $-$       & \textbf{\begin{tabular}[c]{@{}c@{}}2.08E+01\\ (1.92E$-$01)\end{tabular}} &    $-$     \\
		& 5    & \begin{tabular}[c]{@{}c@{}}6.78E$-$01\\ (4.38E$-$02)\end{tabular}          &  $-$       & \textbf{\begin{tabular}[c]{@{}c@{}}2.61E+01\\ (1.91E+00)\end{tabular}} &  $-$       \\ \bottomrule
	\end{tabular}
\end{table}

\begin{figure}[!htbp]
	\centering
	\begin{tabular}{ccc}
		\subfigure[][{\scriptsize One-step ahead prediction}]{\includegraphics[width=5cm]{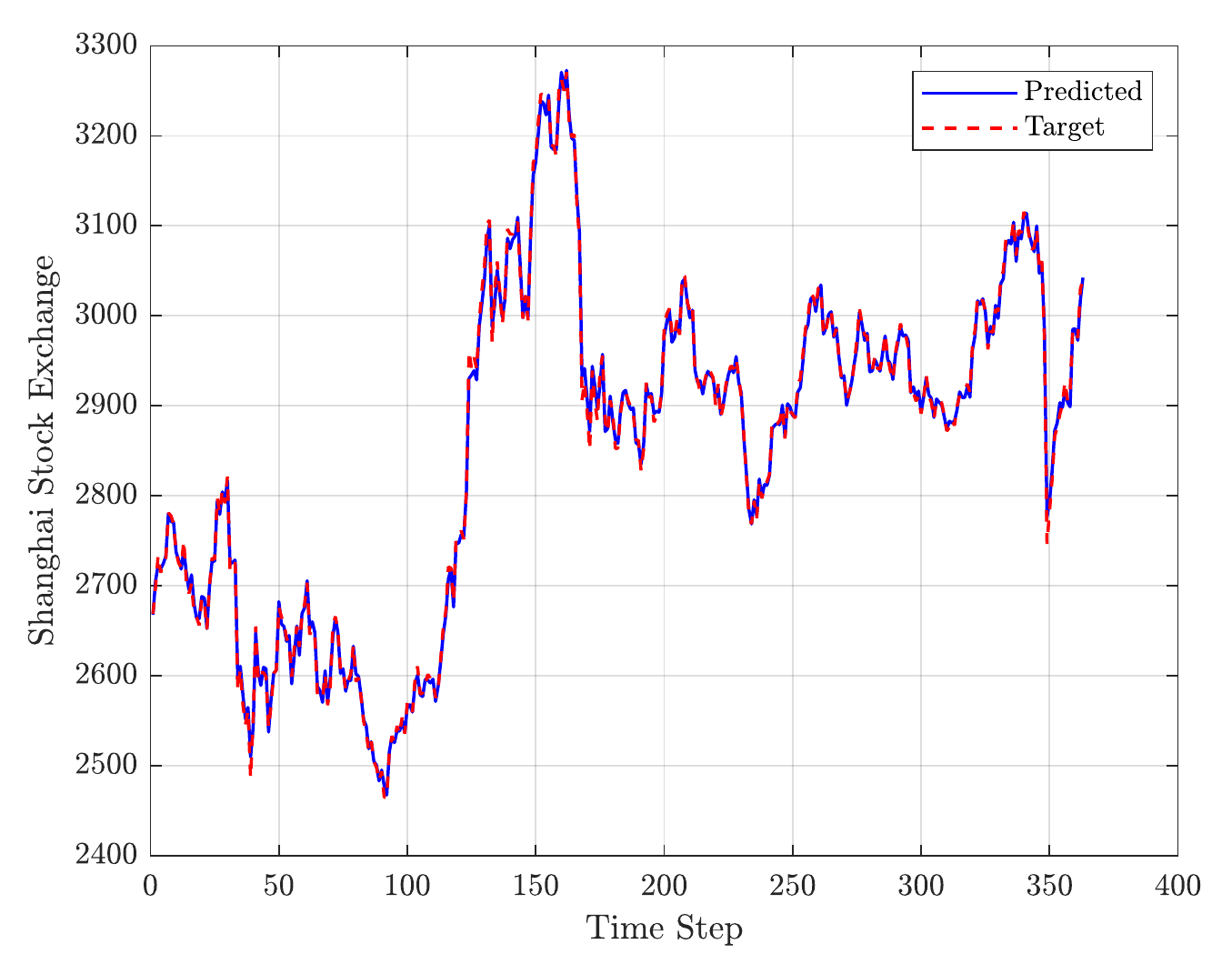}}	&  \subfigure[][{\scriptsize Three-step ahead prediction}]{\includegraphics[width=5cm]{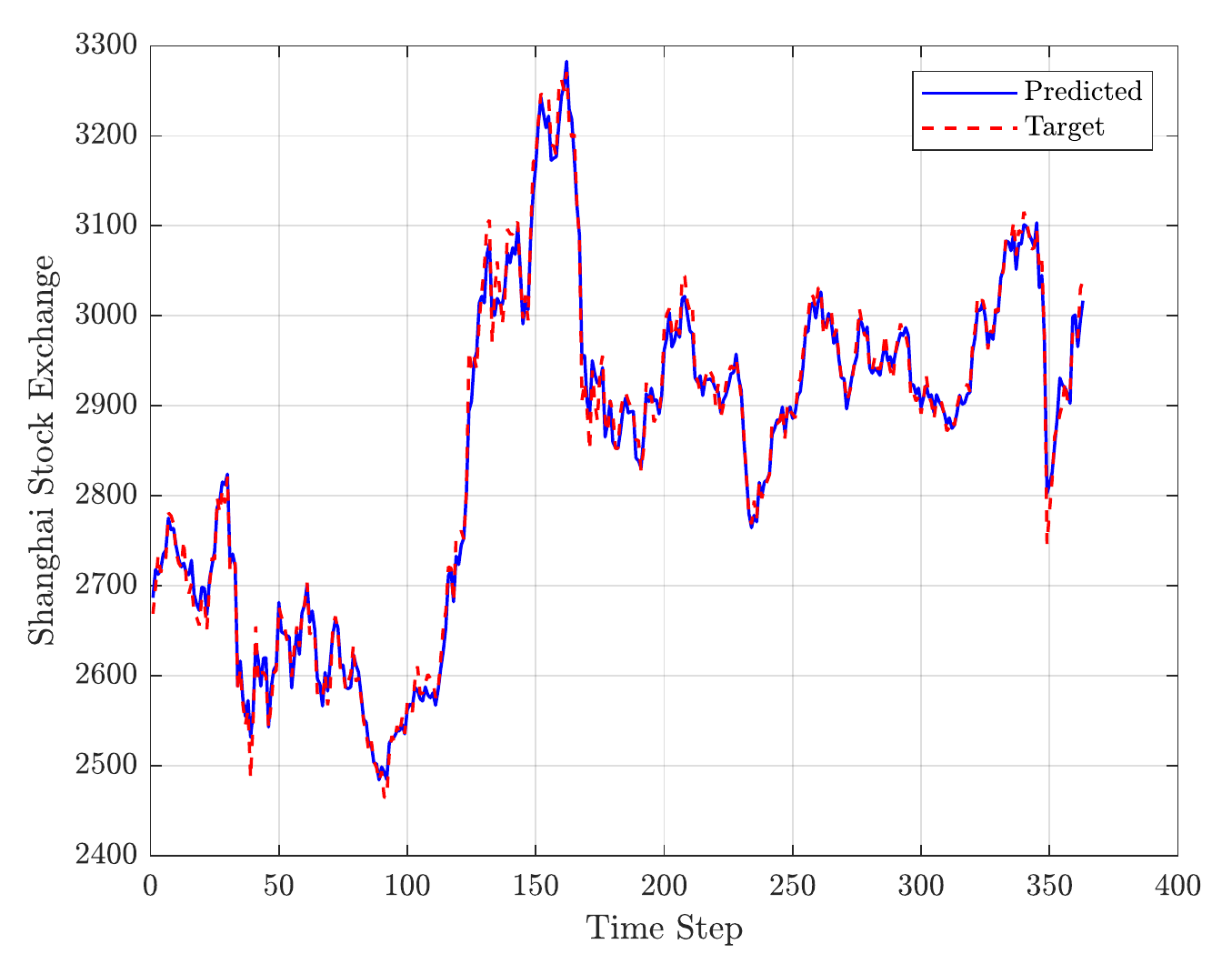}}   &
		\subfigure[][{\scriptsize Five-step ahead prediction}]{\includegraphics[width=5cm]{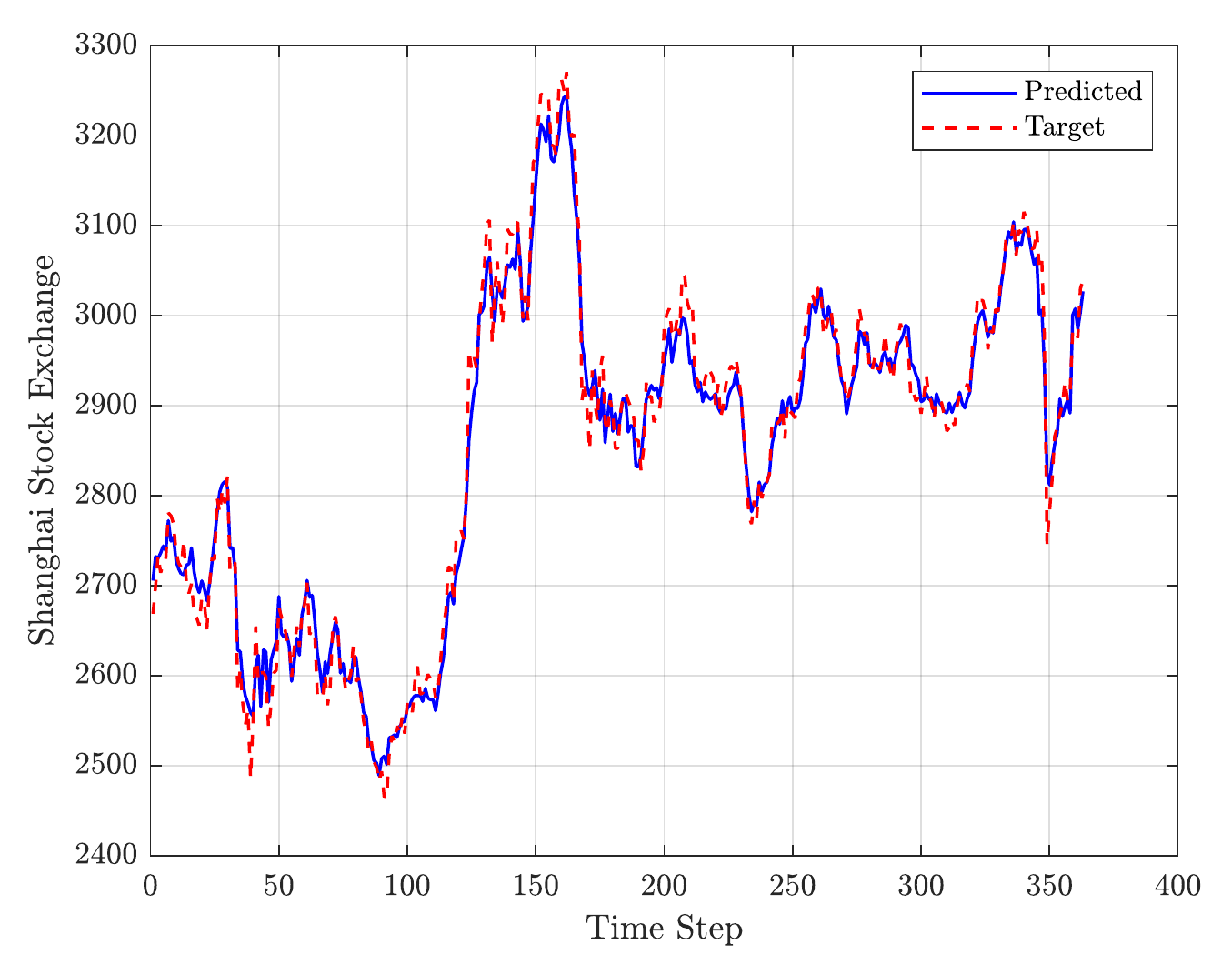}} 	\\   		\subfigure[][{\scriptsize Histogram of Errors (One-step ahead)}]{\includegraphics[width=5cm]{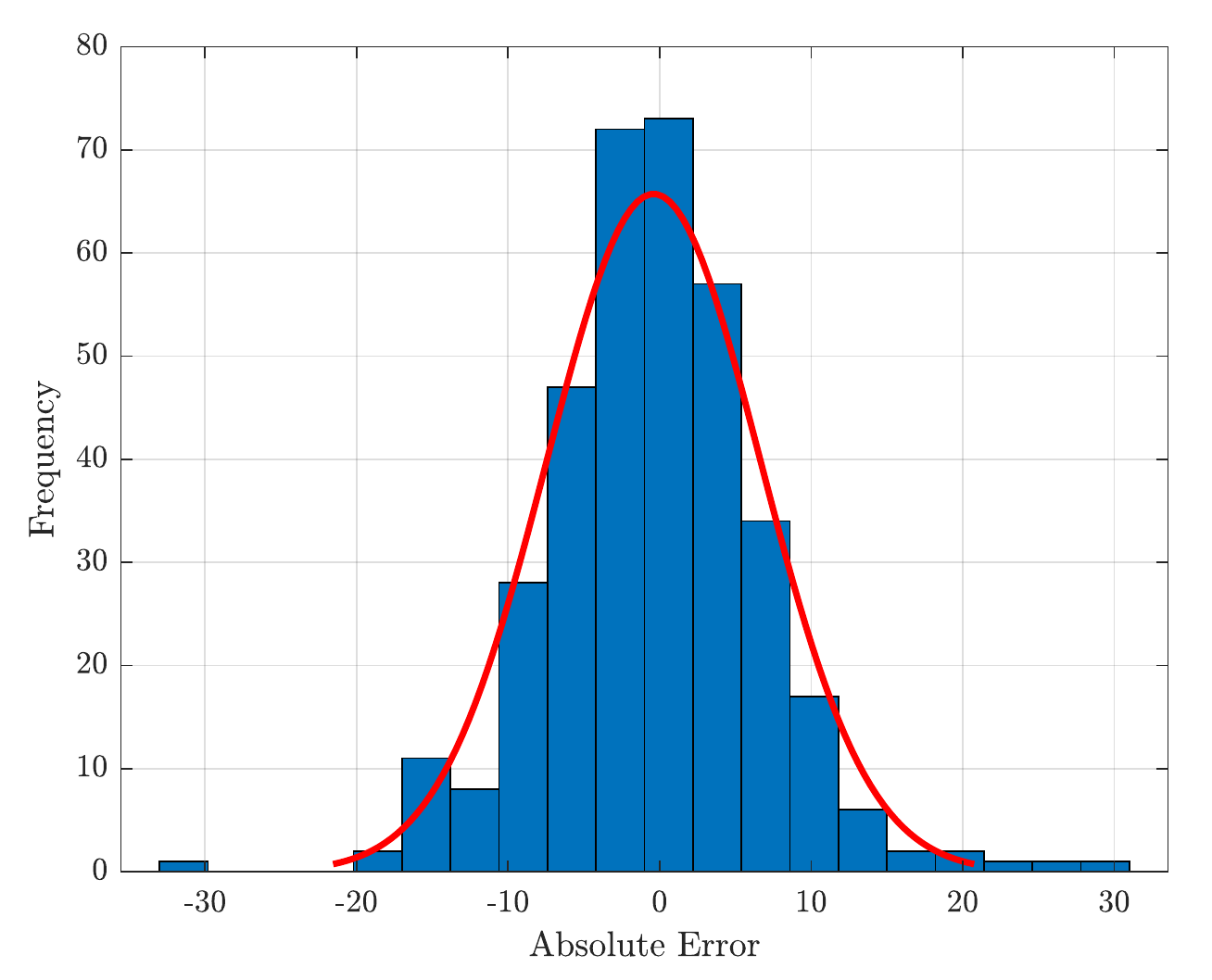}}  &
		\subfigure[][{\scriptsize Histogram of Errors (Three-step ahead)}]{\includegraphics[width=5cm]{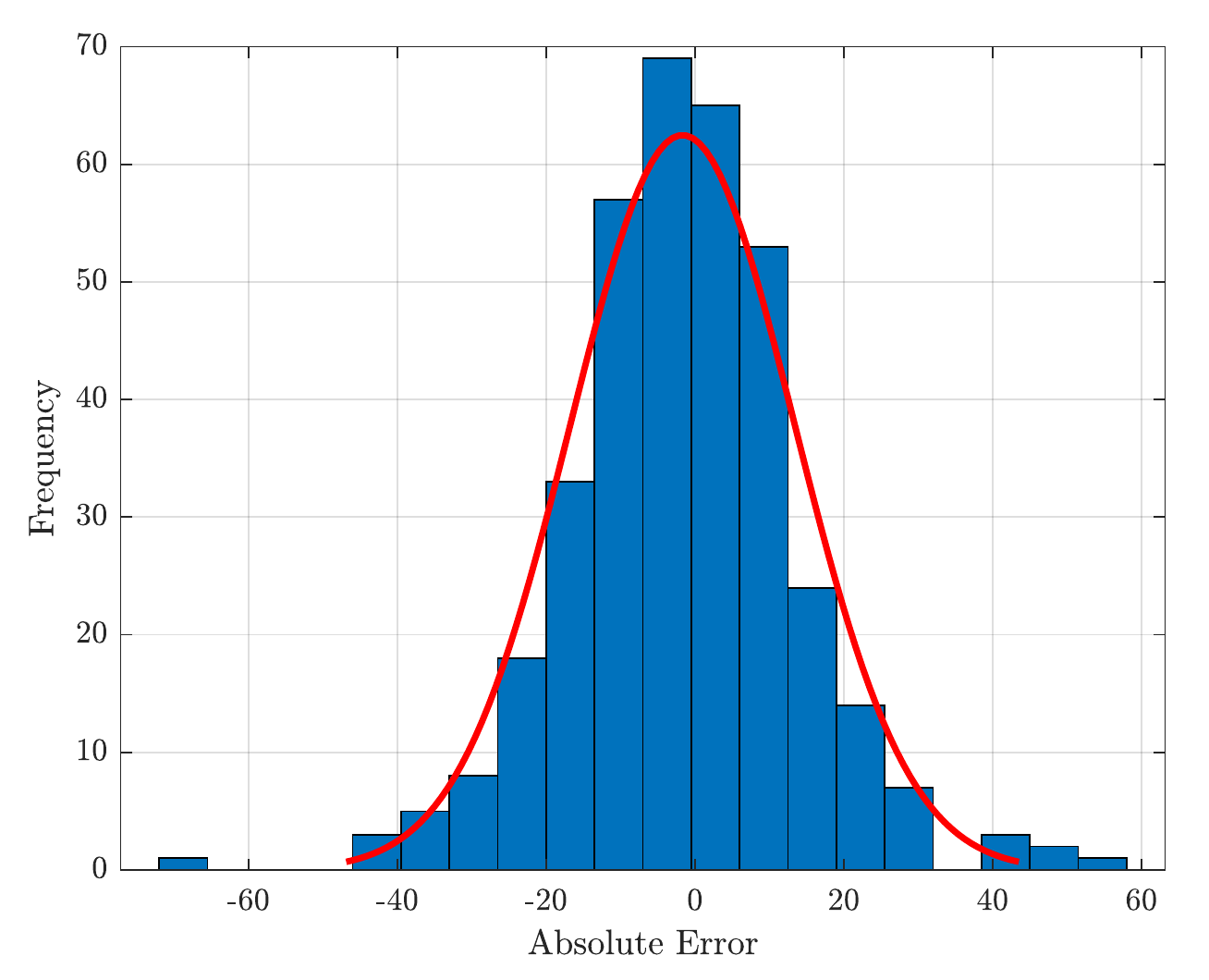}}  &   		\subfigure[][{\scriptsize Histogram of Errors (Five-step ahead)}]{\includegraphics[width=5cm]{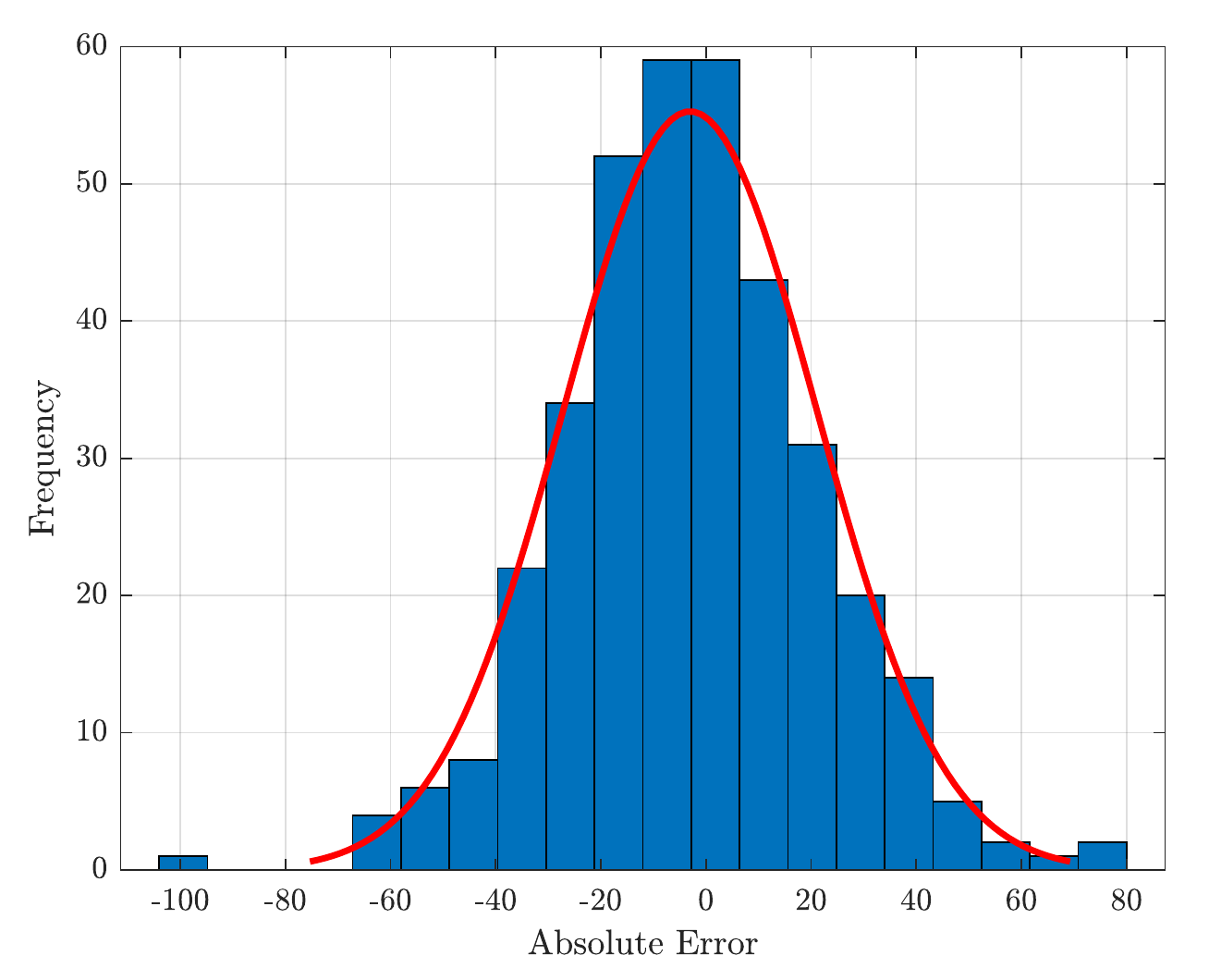}}  
	\end{tabular}
	
	\caption{Prediction curves and histograms of errors for the SSE time series.}
	\label{fig:SSE_Prediction_Plots}	
\end{figure}

From Table \ref{tab:SSE_Results}, it is evident that decomposition-based approaches (i.e., EMD-LSTM, MEMD-LSTM, and VMD-MFRFNN) obtained promising results compared to other methods. Again, ARIMA had the worst performance, and DCT-MFRFNN outperformed MFRFNN in all prediction horizons. Based on RMSE, DCT-MFRFNN showed a decrease of 18.80\%, 33.07\%, and 34.48\% from MFRFNN in one, three, and five-step-ahead forecasting tasks, respectively. The good performance of VMD-MFRFNN in this experiment can be explained by its potential to learn multiple functions simultaneously, along with employing a decomposition method to make the forecasting problem easier. Using two states allows the VMD-MFRFNN to learn and track dynamic behaviors of the SSE dataset over time. Moreover, VMD-MFRFNN, as an FNN, eliminates the time series noise. Note that, based on the statistical tests, the results of this experiment are regarded as statistically significant since all the reported p-values are less than the significance threshold (i.e., $\alpha = 0.05$). Prediction curves and histograms of errors produced by VMD-MFRFNN for the SSE time series are shown in Fig. \ref{fig:SSE_Prediction_Plots}.  

\subsection{Standard \& Poor's 500 Index (SPX)}
\label{sec:SPX}

The Standard \& Poor's 500 Index (SPX) tracks the stock price of 500 publicly traded domestic companies in the United States. SPX is typically recognized as one of the most developed markets in the world due to its mature market operation system \cite{deng2022multi}. The SPX data was also collected from Yahoo Finance during an eleven-year period from 4-January-2010 to 31-December-2020. From the entire time series data, 1845 samples were considered as the training set, 462 samples as the validation set, and the remaining 462 samples as the test set.

Table \ref{tab:SPX_Results} compares VMD-MFRFNN's performance in one, three, and five-step-ahead predictions of SPX with other methods. As shown in Table \ref{tab:SPX_Results}, VMD-MFRFNN obtained the smallest MAPE and RMSE in one and three-step-ahead predictions of the SPX time series. The MEMD-LSTM was the second-best method in these experiments. In five-step-ahead prediction based on MAPE, MEMD-LSTM outperformed other methods, and after that, VMD-MFRFNN had the lowest MAPE. The long-term forecasting capability of MEMD-LSTM can be attributed to the ability to capture important features and estimate the nonlinear dynamic function in the SPX time series.

\begin{table}[h]
	\centering
	\caption{One, three, and five-step-ahead prediction results of the SPX}
	\label{tab:SPX_Results}
	\footnotesize
	\begin{tabular}{@{}lccccc@{}}
		\toprule
		Method     & Step & \multicolumn{1}{c}{MAPE (\%)}                                          & $p$-value & \multicolumn{1}{c}{RMSE}                                               & $p$-value \\ \midrule
		ARIMA      & 1    & 3.84E+00                                                               &    1.31E$-$42
		
		     & 1.28E+02                                                               &  4.66E$-$40
		     
		            \\
		& 3    & 4.43E+00                                                               &  2.20E$-$54
		
		       & 1.58E+02                                                               &  1.70E$-$49
		       
		              \\
		& 5    & 5.59E+00                                                               &   3.86E$-$44
		
		      & 1.68E+02                                                               &  3.60E$-$46
		      
		             \\ \midrule
		FFNN       & 1    & 2.21E+00                                                               &   2.85E$-$37
		
		      & 1.11E+02                                                               & 1.06E$-$38
		      
		              \\
		& 3    & 2.29E+00                                                               &   6.43E$-$48
		
		      & 1.54E+02                                                               &   2.99E$-$49
		      
		           \\
		& 5    & 2.84E+00                                                               &  5.15E$-$37
		
		       & 1.66E+02                                                               &  4.83E$-$46
		       
		              \\ \midrule
		LSTM       & 1    & 1.54E+00                                                               &  2.08E$-$33
		
		       & 6.77E+01                                                               &  1.11E$-$33
		       
		              \\
		& 3    & 1.61E+00                                                               &   5.32E$-$44
		
		      & 7.00E+01                                                               &   7.18E$-$41
		      
		            \\
		& 5    & 1.72E+00                                                               & 2.86E$-$30
		
		       & 7.47E+01                                                               &  8.47E$-$36
		       
		              \\ \midrule
		EMD-LSTM   & 1    & 6.82E$-$01                                                               &   1.69E$-$21
		
		      & 5.65E+01                                                               &    1.15E$-$31
		      
		           \\
		& 3    & 7.29E$-$01                                                               &    3.13E$-$31
		
		     & 5.81E+01                                                               &  1.67E$-$38
		     
		            \\
		& 5    & 7.99E$-$01                                                               &  1.34E$-$05
		
		       & 6.09E+01                                                               &   5.82E$-$32
		       
		             \\ \midrule
		MEMD-LSTM  & 1    & 5.28E$-$01                                                               &    2.04E$-$14
		
		     & 3.60E+01                                                               &  7.47E$-$26
		     
		            \\
		& 3    & 6.51E$-$01                                                               &   5.88E$-$28
		
		      & 3.86E+01                                                               &  5.18E$-$32
		      
		             \\
		& 5    & \textbf{6.62E$-$01}                                                      &   6.84E$-$17
		
		      & 4.00E+01                                                   &   2.00E$-$13
		      
		            \\ \midrule
		MFRFNN     & 1    & \begin{tabular}[c]{@{}c@{}}9.95E$-$01\\(6.38E$-$03)\end{tabular}                                                  &   9.79E$-$31
		
		      & \begin{tabular}[c]{@{}c@{}}4.47E+01\\(1.41E$-$01)\end{tabular}                                                 &  1.45E$-$29
		      
		             \\
		& 3    & \begin{tabular}[c]{@{}c@{}}1.63E+00\\(1.10E$-$02)\end{tabular}                                                   &    2.50E$-$59
		
		     & \begin{tabular}[c]{@{}c@{}}7.31E+01\\(1.88E$-$01)\end{tabular}                                                   &  2.61E$-$54
		     
		           \\
		& 5    & \begin{tabular}[c]{@{}c@{}}2.13E+00\\(1.32E$-$02)\end{tabular}                                                   &   5.95E$-$44
		
		      & \begin{tabular}[c]{@{}c@{}}9.47E+01\\(1.82E$-$01)\end{tabular}                                                    &    3.61E$-$45
		            \\ \midrule
DCT-MFRFNN     & 1    & \begin{tabular}[c]{@{}c@{}}5.87E$-$01\\(5.40E$-$03)\end{tabular}                                                  &   6.87E$-$19

& \begin{tabular}[c]{@{}c@{}}2.84E+01\\(9.02E$-$01)\end{tabular}                                                 &  7.15E$-$32

\\
& 3    & \begin{tabular}[c]{@{}c@{}}9.43E$-$01\\(1.40E$-$02)\end{tabular}                                                   &   1.74E$-$39

& \begin{tabular}[c]{@{}c@{}}4.30E+01\\(1.02E+00)\end{tabular}                                                   &  2.91E$-$32

\\
& 5    & \begin{tabular}[c]{@{}c@{}}1.22E+00\\(1.09E$-$02)\end{tabular}                                                   &  6.25E$-$28

& \begin{tabular}[c]{@{}c@{}}5.57E+01\\(7.90E$-$01)\end{tabular}                                                    &   7.67E$-$42		      
		           \\ \midrule
		VMD-MFRFNN & 1    & \textbf{\begin{tabular}[c]{@{}c@{}}4.18E$-$01
				\\ (2.40E$-$02)\end{tabular}} &   $-$      & \textbf{\begin{tabular}[c]{@{}c@{}}1.60E+01
				\\ (1.07E+00)\end{tabular}} &  $-$       \\
		& 3    & \textbf{\begin{tabular}[c]{@{}c@{}}4.91E-01
				\\ (6.63E$-$03)\end{tabular}} &  $-$       & \textbf{\begin{tabular}[c]{@{}c@{}}2.23E+01
				\\ (4.13E$-$01)\end{tabular}} &   $-$      \\
		& 5    & \begin{tabular}[c]{@{}c@{}}8.35E$-$01
			\\ (2.77E$-$02)\end{tabular}          &  $-$        & \textbf{\begin{tabular}[c]{@{}c@{}}3.76E+01
			\\ (5.94E$-$01)\end{tabular}}          &    $-$     \\ \bottomrule
	\end{tabular}

\end{table}

Based on RMSE, VMD-MFRFNN had the highest accuracy in five-step-ahead prediction, closely followed by MEMD-LSTM. Based on this metric, VMD-MFRFNN showed a decrease of 55.56\%, 42.23\%, and 6.00\% from the MEMD-LSTM in one, three, and five-step-ahead predictions of the SPX dataset, respectively. In this time series, based on RMSE, DCT-MFRFNN showed a decrease of 36.47\%, 41.18\%, and 41.18\% from MFRFNN in prediction horizons of one, three, and five, respectively. 

\begin{figure}[!htbp]
	\centering
	\begin{tabular}{ccc}
		\subfigure[][{\scriptsize One-step ahead prediction}]{\includegraphics[width=5cm]{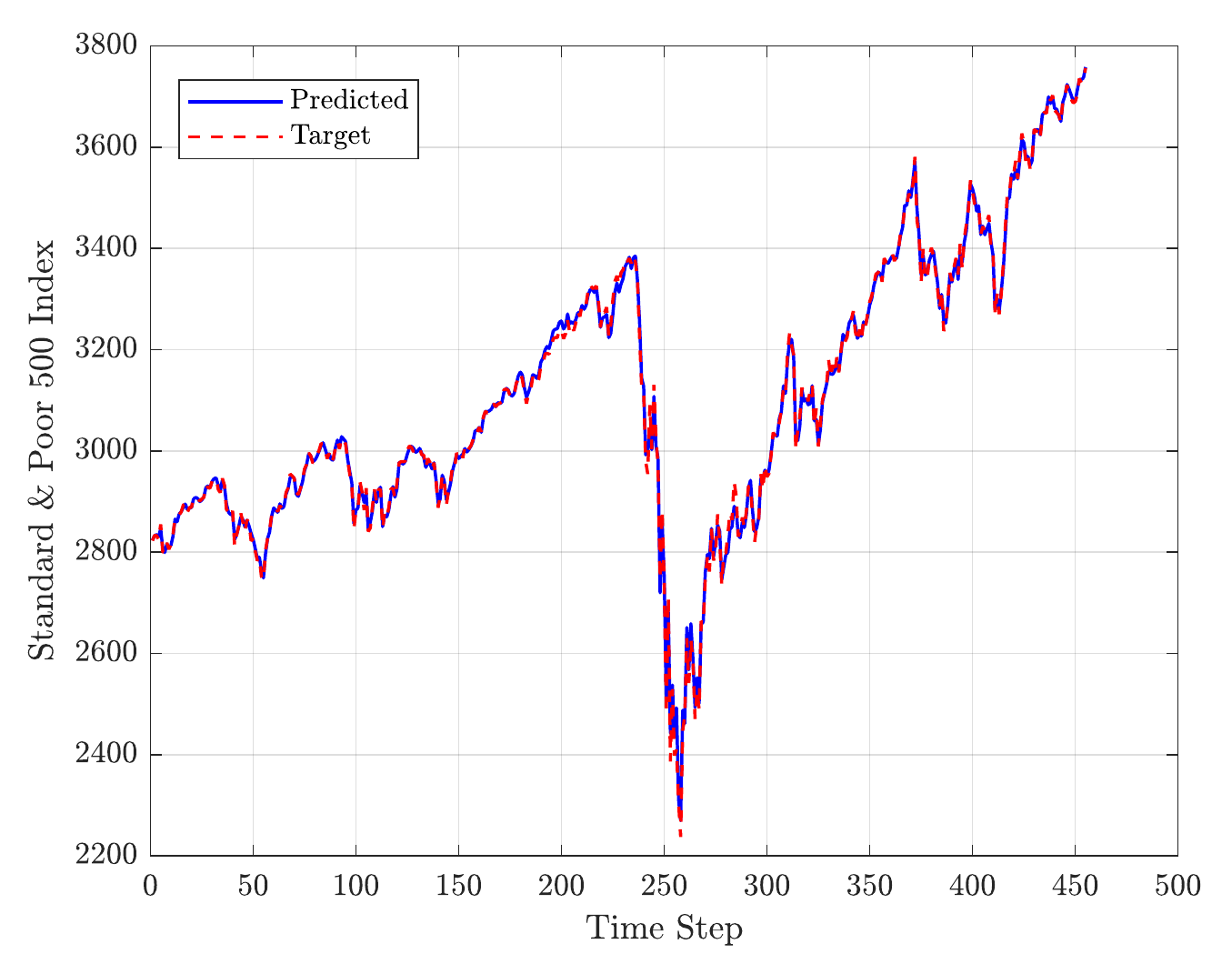}}	&  \subfigure[][{\scriptsize Three-step ahead prediction}]{\includegraphics[width=5cm]{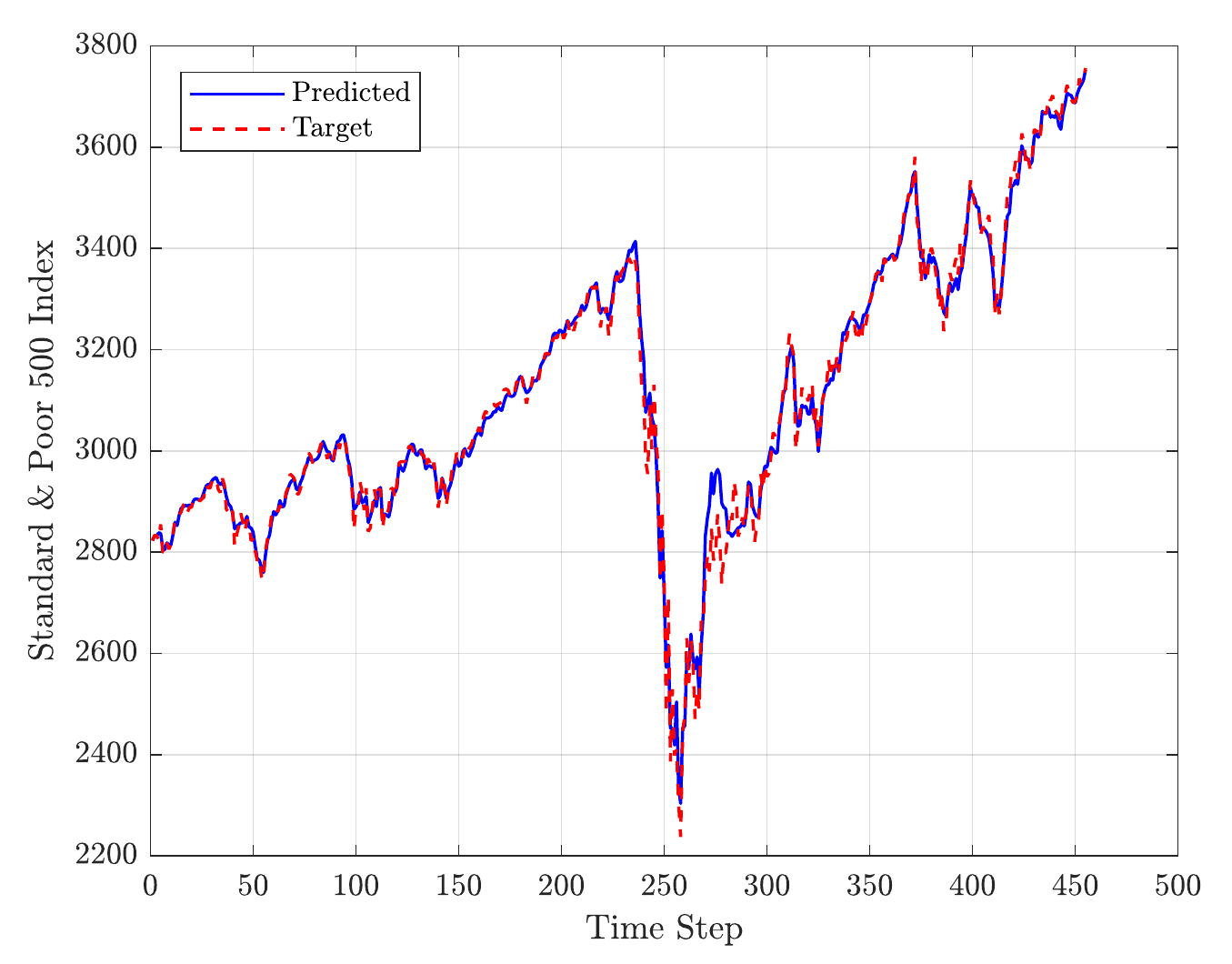}}   &
		\subfigure[][{\scriptsize Five-step ahead prediction}]{\includegraphics[width=5cm]{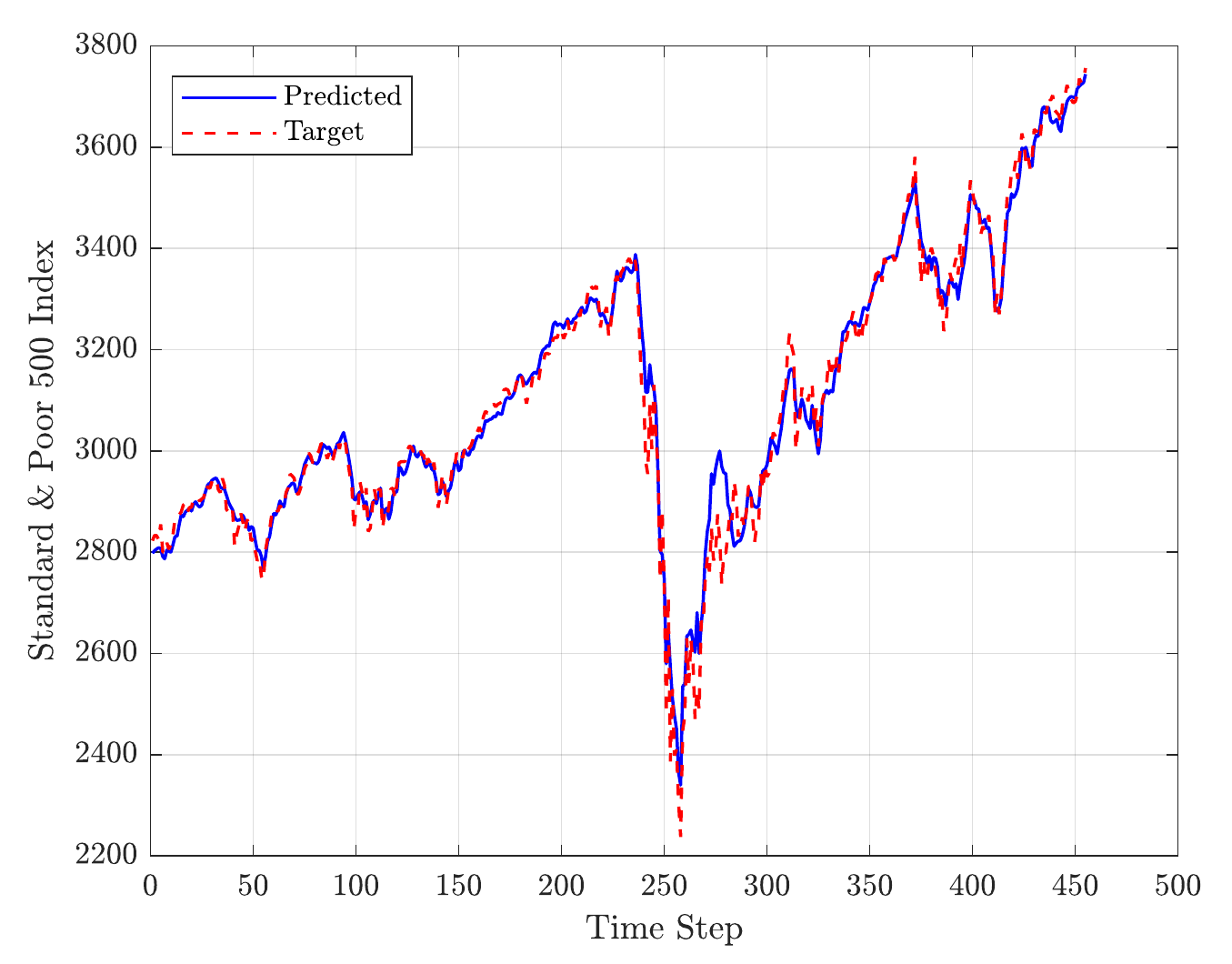}} 	\\   		\subfigure[][{\scriptsize Histogram of Errors (One-step ahead)}]{\includegraphics[width=5cm]{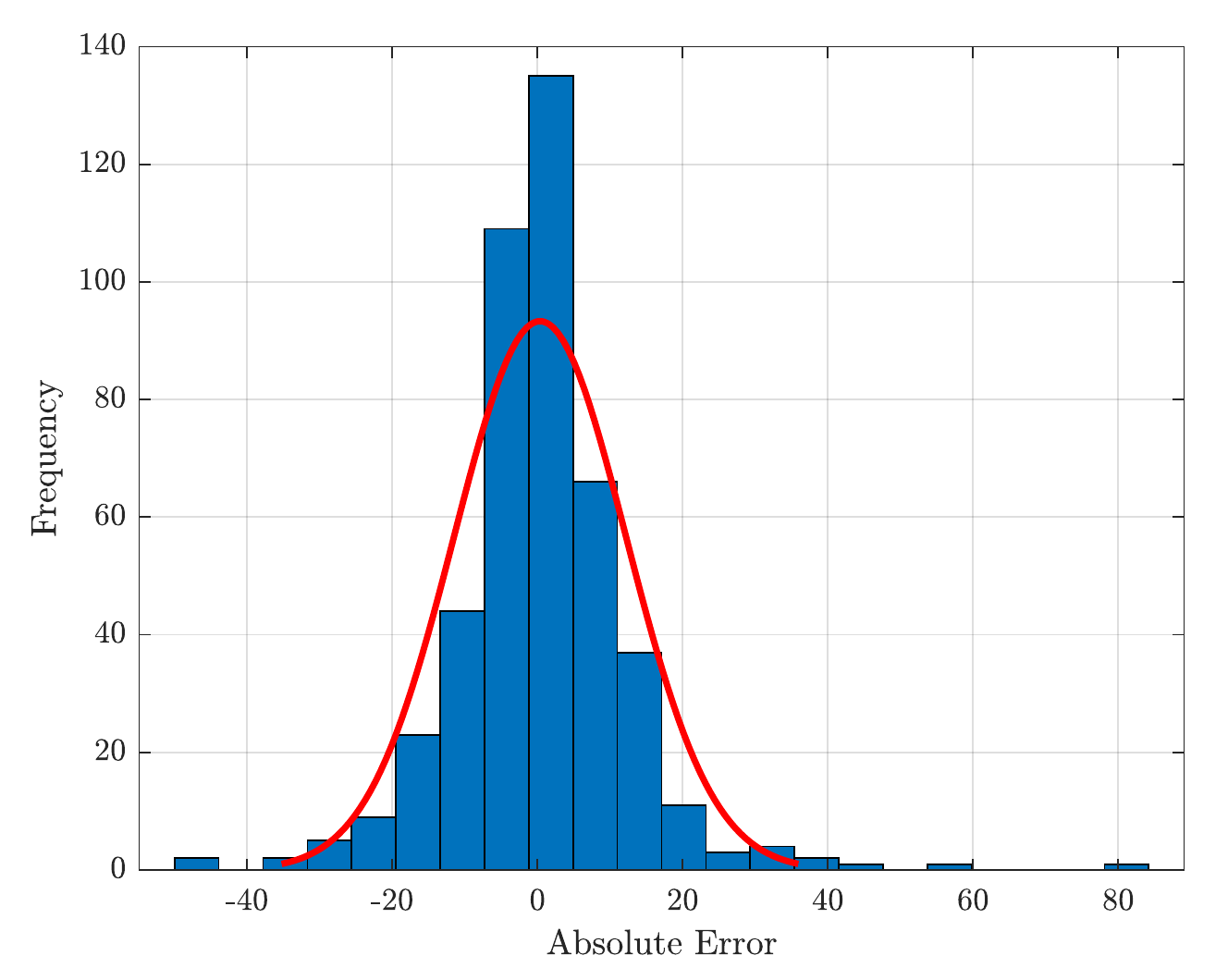}}  &
		\subfigure[][{\scriptsize Histogram of Errors (Three-step ahead)}]{\includegraphics[width=5cm]{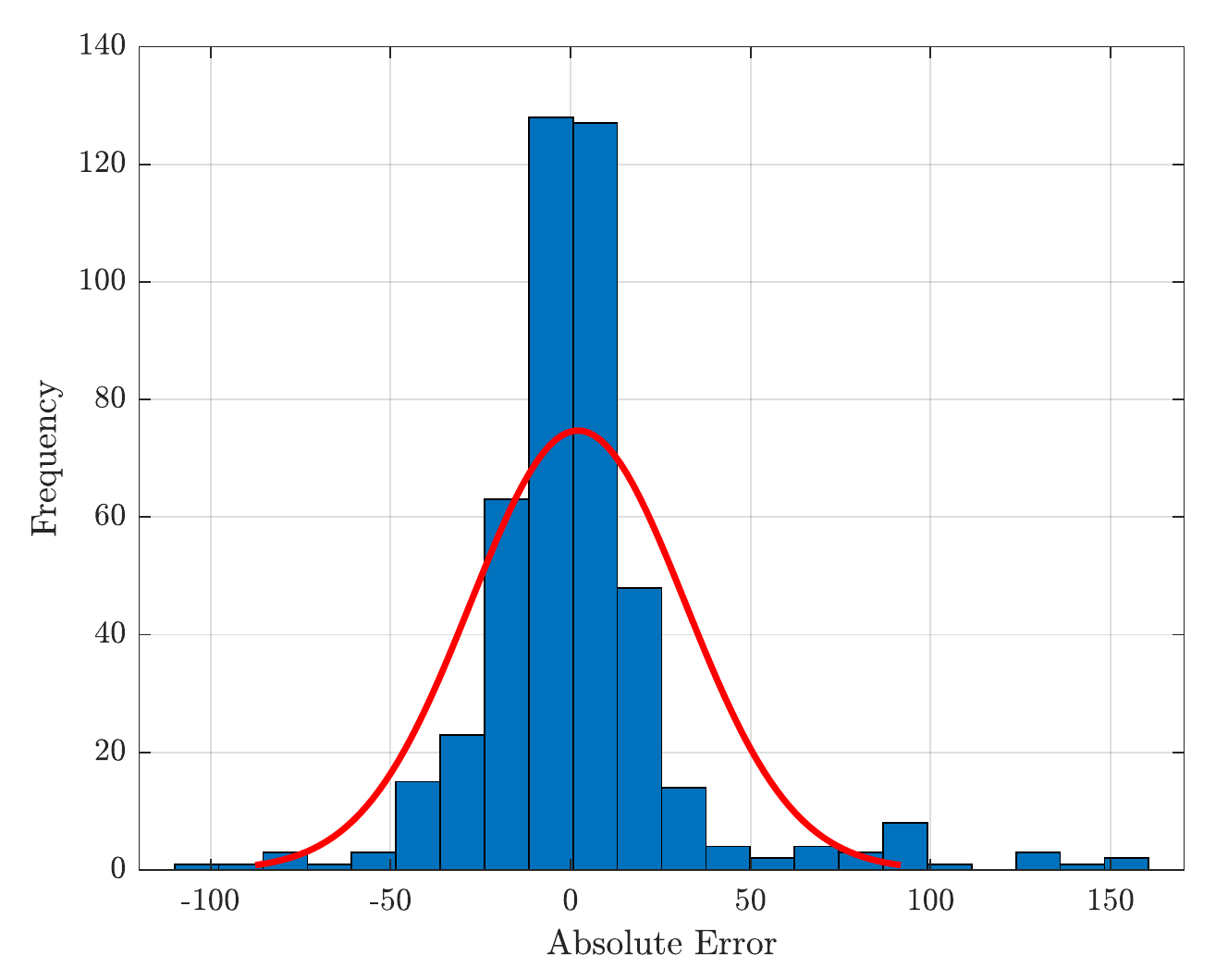}}  &   		\subfigure[][{\scriptsize Histogram of Errors (Five-step ahead)}]{\includegraphics[width=5cm]{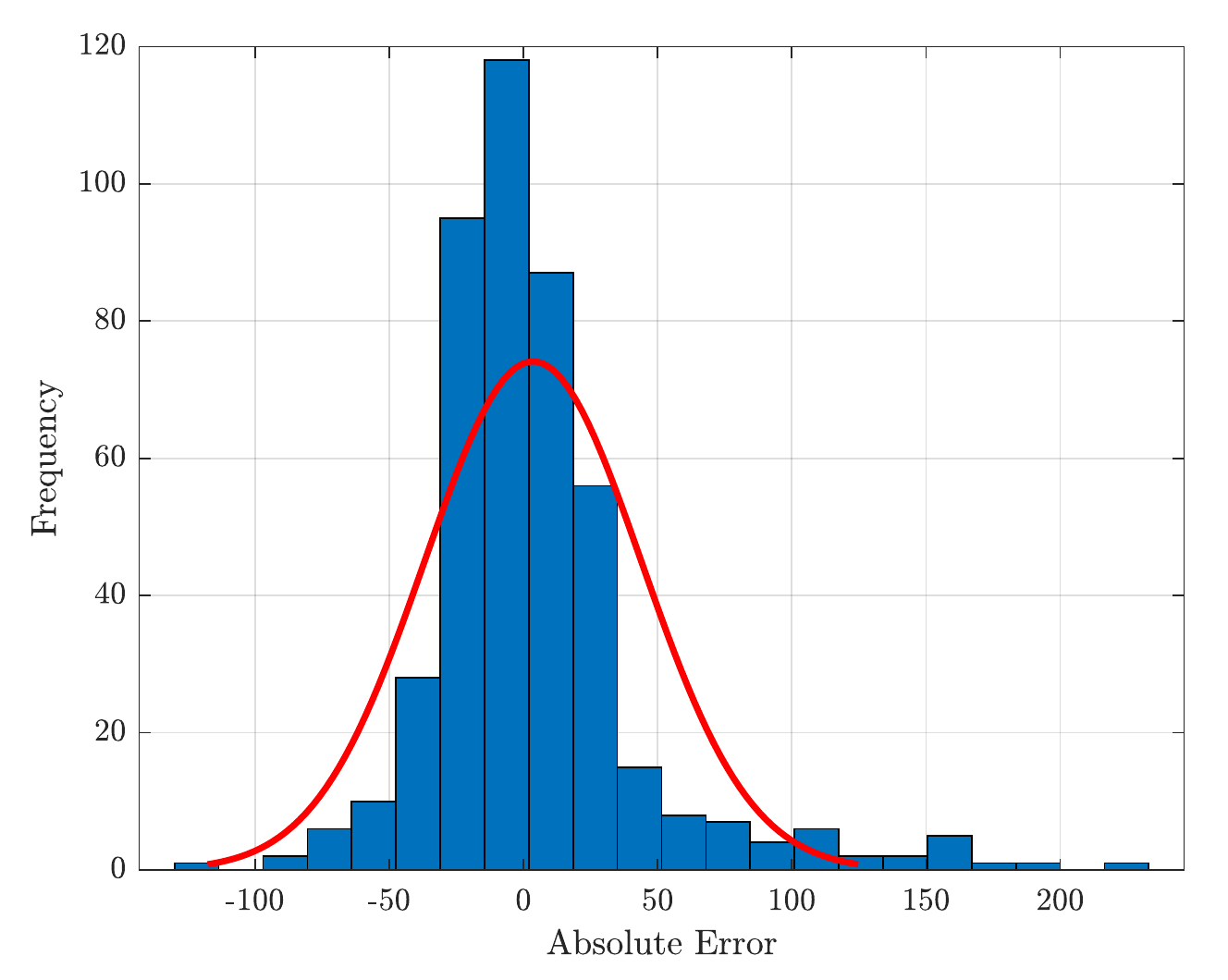}}  
	\end{tabular}
	
	\caption{Prediction curves and histograms of errors for the SPX time series.}
	\label{fig:SPX_Prediction_Plots}	
\end{figure}

Comparing the results of VMD-MFRFNN with MFRFNN revealed that VMD-MFRFNN had higher accuracy and standard deviation. Having a high standard deviation is a drawback of the proposed VMD-MFRFNN. Note that the variance of predictions increases as the standard deviation increases. Also, it is worth mentioning that LSTM outperformed FFNN in all experiments due to inherent problems of FFNN, such as overfitting and falling into local minima. The Welch's t-test performed on SPX prediction results confirmed that the results are statistically significant. Fig. \ref{fig:SPX_Prediction_Plots} shows prediction curves and histograms of errors produced by VMD-MFRFNN for the SPX dataset.

\subsection{Comparing Different Decomposition Methods}
This section compares various decomposition methods to strengthen this research further. We used different decomposition approaches in the decomposition phase of the proposed method, including VMD \cite{dragomiretskiy2014variational}, EMD \cite{huang1998empirical}, EEMD \cite{wu2009ensemble}, and Robust Empirical Mode Decomposition (REMD) \cite{liu2022improved}. REMD is an enhanced EMD that employs an adaptive sifting stop criterion to automatically terminate EMD's sifting process.

We set the prediction horizon equal to 5 in all experiments. Results of comparing decomposition methods are presented in Table \ref{tab:ComparingDecompositionMethods}. As seen in Table \ref{tab:ComparingDecompositionMethods}, the VMD method outperformed other methods in all financial time series, closely followed by EEMD. Based on the statistical tests, all the results are statistically significant. So, obviously, using VMD as a decomposition method in the structure of the proposed method is a good choice. Comparing the prediction results of EMD and EEMD demonstrated that EEMD achieved higher accuracy in all experiments. The good performance of EEMD can be attributed to the elimination of the mode mixing problem by this method.

\begin{table}[!htbp]
	\centering
	\caption{Comparison of different decomposition methods}
	\label{tab:ComparingDecompositionMethods}
	\footnotesize
	\begin{tabular}{cccccc}
		\toprule
		\multicolumn{2}{l}{Decomposition Method} & VMD                                                           & EMD                                                           & REMD & EEMD                                                   \\ \midrule
		\multirow{4}{*}{HSI}         & MAPE      & \begin{tabular}[c]{@{}c@{}}\textbf{8.99E$-$01}\\ \textbf{(2.24E$-$02)}\end{tabular} & \begin{tabular}[c]{@{}c@{}}1.08E+00\\ (2.58E$-$02)\end{tabular} & \begin{tabular}[c]{@{}c@{}}1.08E+00\\ (2.55E$-$02)\end{tabular} & \begin{tabular}[c]{@{}c@{}}9.81E$-$01\\ (1.33E$-$01)\end{tabular}  \\
		& $p$-value   &  $-$     &   4.57E$-$24
		                        &   3.25E$-$24
		                                                         &     1.32E$-$02
                                              \\
		& RMSE      & \begin{tabular}[c]{@{}c@{}}\textbf{3.27E+02}\\ \textbf{(6.53E+00)}\end{tabular} & \begin{tabular}[c]{@{}c@{}}4.03E+02\\ (7.51E+00)\end{tabular} & \begin{tabular}[c]{@{}c@{}}4.32E+02\\ (9.50E+00)\end{tabular} & \begin{tabular}[c]{@{}c@{}}3.53E+02\\ (4.20E+01)\end{tabular} \\
		& $p$-value   &     $-$                                               &         9.80E$-$30
		                                     &   3.10E$-$30
		                                                   &       1.28E$-$02
		                                                                                                           \\ \midrule
		\multirow{4}{*}{SSE}         & MAPE      & \begin{tabular}[c]{@{}c@{}}\textbf{6.78E$-$01}\\ \textbf{(4.38E$-$02)}\end{tabular} & \begin{tabular}[c]{@{}c@{}}1.19E+00\\ (2.59E$-$02)\end{tabular} & \begin{tabular}[c]{@{}c@{}}1.22E+00\\ (1.70E$-$02)\end{tabular} & \begin{tabular}[c]{@{}c@{}}9.28E$-$01\\ (1.51E$-$02)\end{tabular} \\
		& $p$-value   &     $-$                                                           &     1.08E$-$29
		                                  &   1.37E$-$26
		                                                       &      4.58E$-$18
		                                                                                                                \\
		& RMSE      & \begin{tabular}[c]{@{}c@{}}\textbf{2.61E+01}\\\textbf{(1.91E+00)}\end{tabular} & \begin{tabular}[c]{@{}c@{}}4.53E+01\\ (9.17E$-$01)\end{tabular} & \begin{tabular}[c]{@{}c@{}}4.87E+01\\ (6.55E$-$01)\end{tabular} & \begin{tabular}[c]{@{}c@{}}3.65E+01\\ (5.11E$-$01)\end{tabular}\\
		& $p$-value   &    $-$                                                            &         6.11E$-$26
		                  &    2.59E$-$25
		                                                  &     6.18E$-$17
		                                                                                                            \\ \midrule
		\multirow{4}{*}{SPX}    & MAPE      & \begin{tabular}[c]{@{}c@{}}\textbf{8.35E$-$01}\\ \textbf{(2.77E$-$02)}\end{tabular} & \begin{tabular}[c]{@{}c@{}}1.31E+00\\ (5.96E$-$02)\end{tabular} & \begin{tabular}[c]{@{}c@{}}1.56E+00\\ (7.25E$-$02)\end{tabular} & \begin{tabular}[c]{@{}c@{}}9.81E$-$01\\ (5.50E$-$02)\end{tabular}\\
		& $p$-value   &    $-$                                                            &       5.00E$-$23
		                              &     3.06E$-$24
		                                                   &       2.54E$-$11
		                                                                                                          \\
		& RMSE      & \textbf{\begin{tabular}[c]{@{}c@{}}3.76E+01\\(5.94E$-$01)\end{tabular}} & \begin{tabular}[c]{@{}c@{}}6.13E+01\\ (2.60E+00)\end{tabular} & \begin{tabular}[c]{@{}c@{}}8.09E+01\\ (5.28E+00)\end{tabular} & \begin{tabular}[c]{@{}c@{}}4.60E+01\\ (2.85E+00)\end{tabular}\\
		& $p$-value   &     $-$                                                           &       3.12E$-$21
		                            &     2.15E$-$19
		                                                  &       2.39E$-$11
		                                                                                                       \\ \bottomrule
	\end{tabular}	
\end{table}

\subsection{Sensitivity Analysis}
This section provides sensitivity analyses on the proposed methods to evaluate the sensitivity of VMD-MFRRFNN and DCT-MFRFNN to the number of IMFs and $\lambda$, respectively. Five and three-step-ahead prediction of the SSE time series were used as the benchmark for VMD-MFRFNN and DCT-MFRFNN, respectively. The number of IMFs ranged from 2 to 60, and $\lambda$ ranged from 0 to 99. Other parameters were the same as those provided in Table~\ref{tab:Parameters}. 

\begin{figure}[h]
	\centering
	\includegraphics[width=8cm]{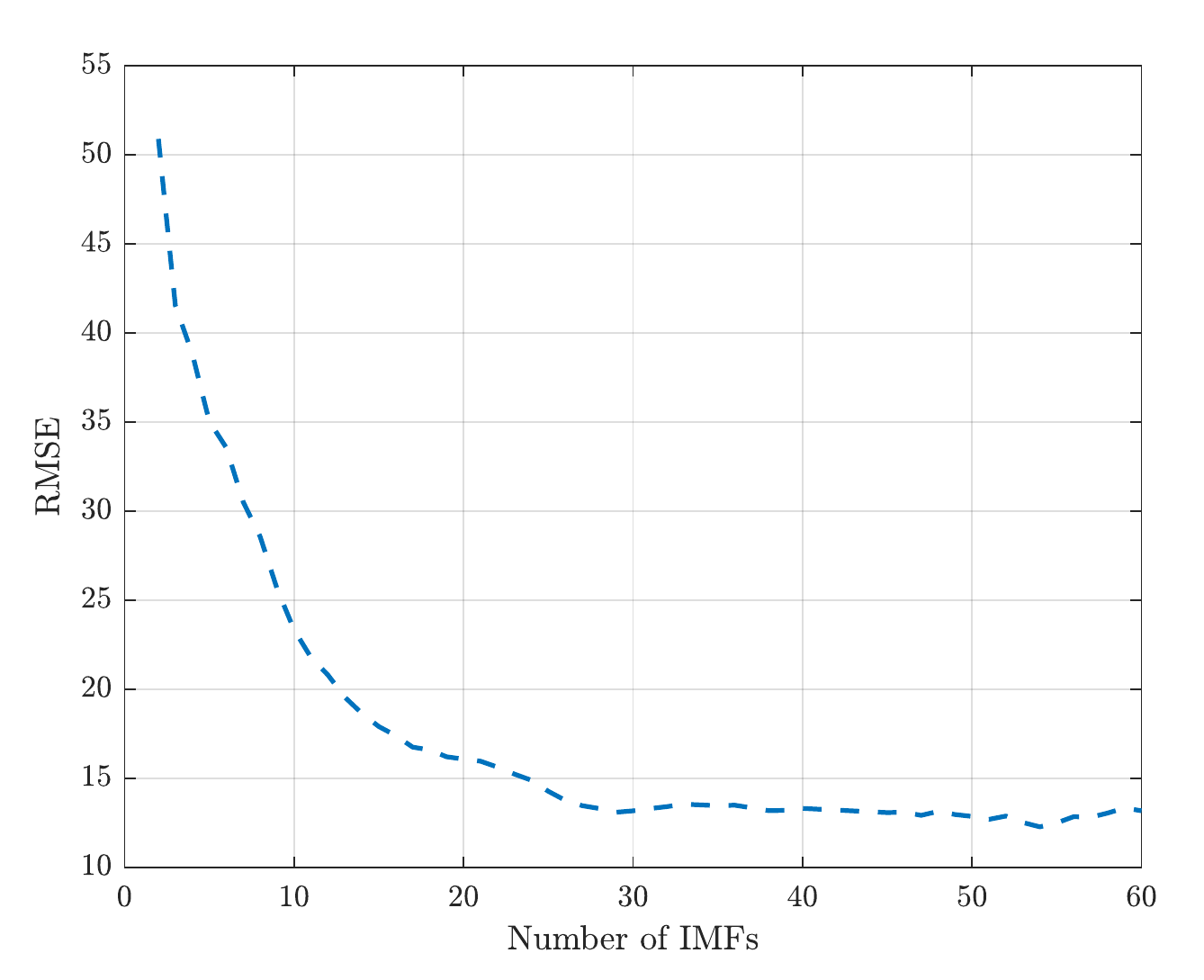}
	\caption{Relationship between RMSE and the number of IMFs in the five-step-ahead prediction of the SSE.}
	\label{fig:SensitivityAnalysis}
\end{figure}

The relationship between RMSE and the number of IMFs is shown in Fig.~\ref{fig:SensitivityAnalysis}. The results indicated that when the number of IMFs increased from 2 to 30, the RMSE was significantly reduced. Increasing the number of IMFs means predicting time series with simpler structures and more stationary trends, which is easier for the model. However, increasing the number of IMFs from 30 to 60 did not have much impact on RMSE. Note that increasing IMFs also increases the computational complexity of the model.

\begin{figure}[t]
	\centering
	\includegraphics[width=10cm]{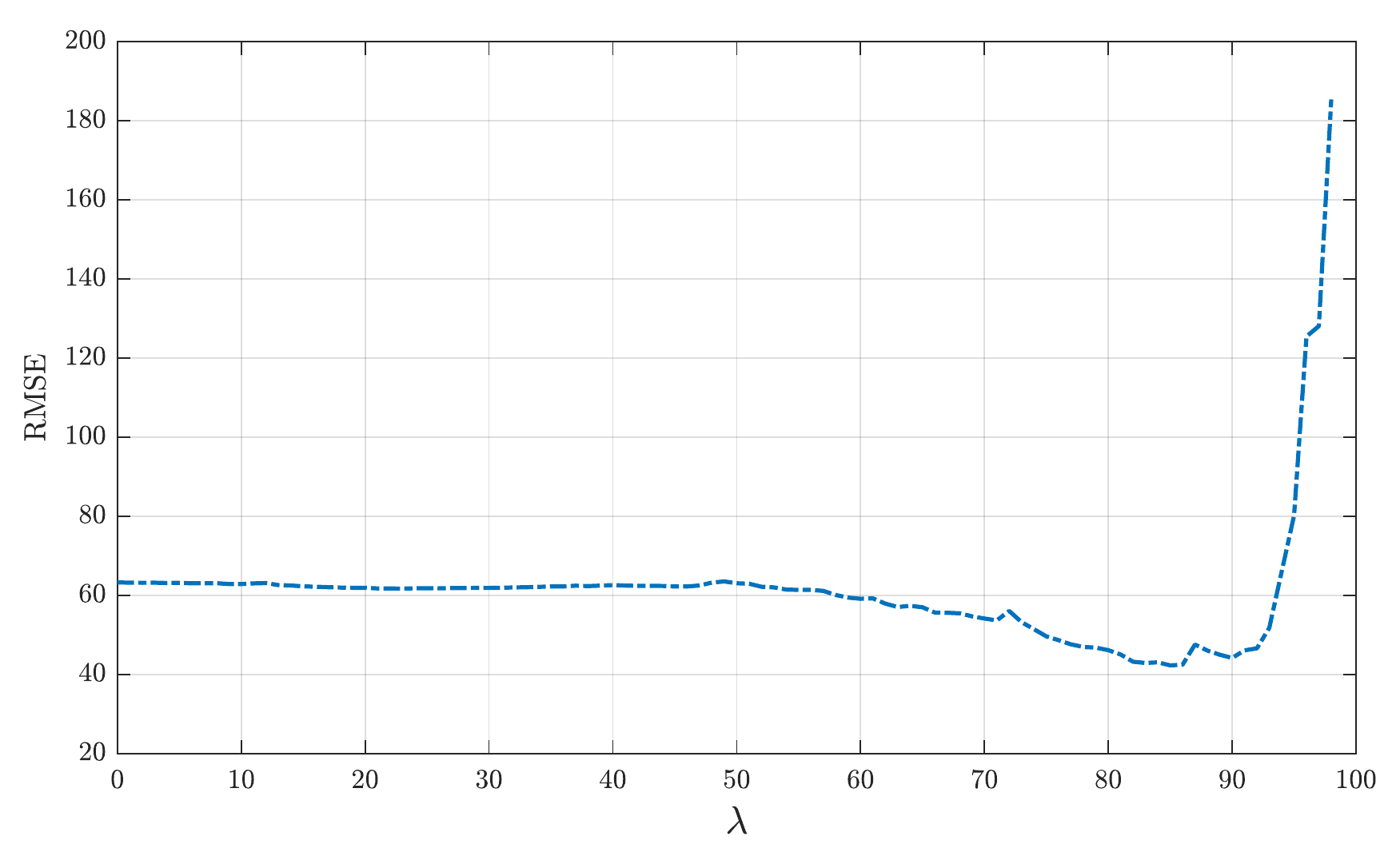}
	\caption{Relationship between RMSE and $\lambda$ in the three-step-ahead prediction of the SSE.}
	\label{fig:SensitivityAnalysis_DCT}
\end{figure}

\begin{figure}[!t]
	\centering
	\begin{tabular}{ccc}
		\subfigure[][{$\lambda = 0$}]{\includegraphics[width=5.2cm]{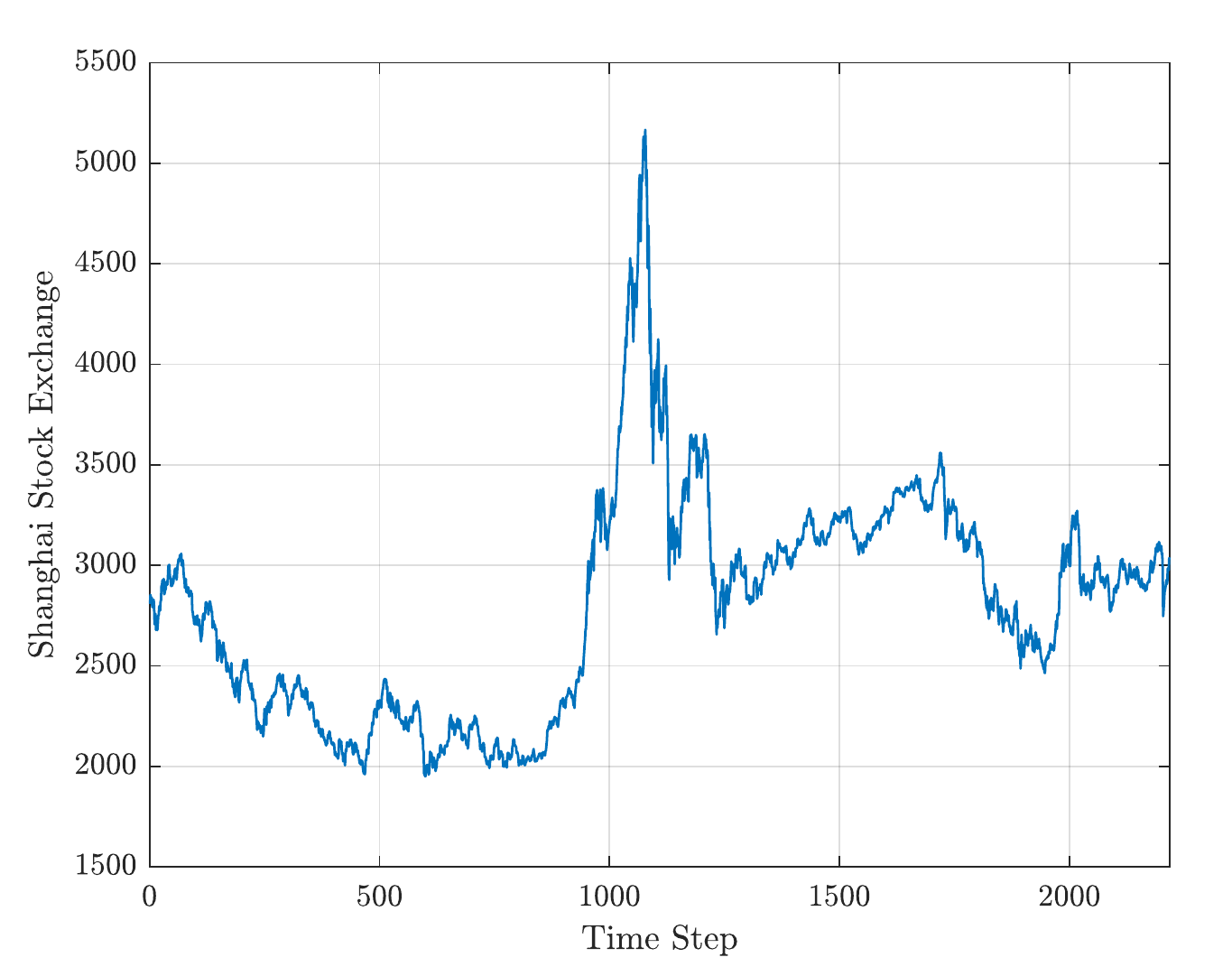}}	&  \subfigure[][{$\lambda = 85$}]{\includegraphics[width=5.2cm]{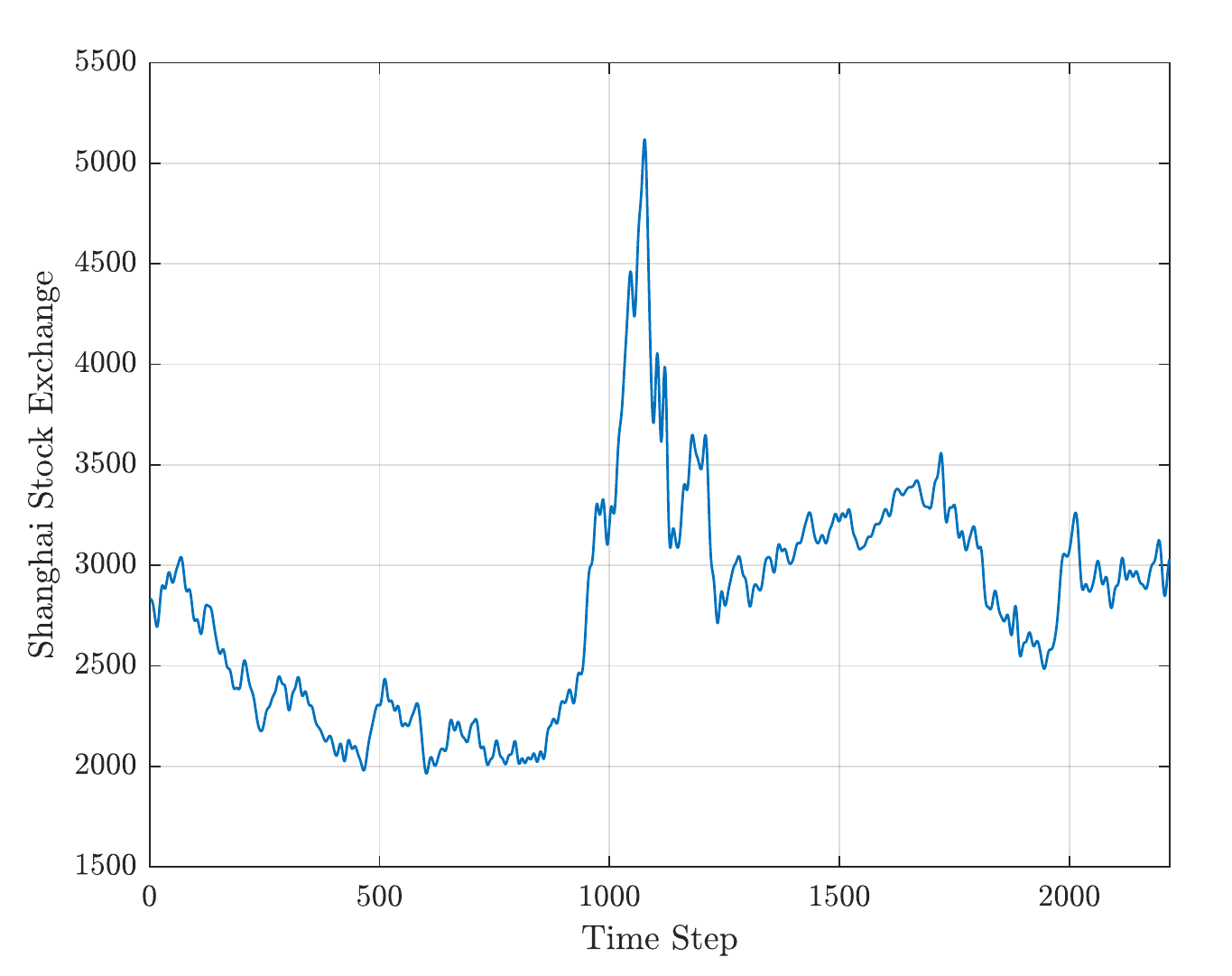}}   &
		\subfigure[][{$\lambda = 95$}]{\includegraphics[width=5.2cm]{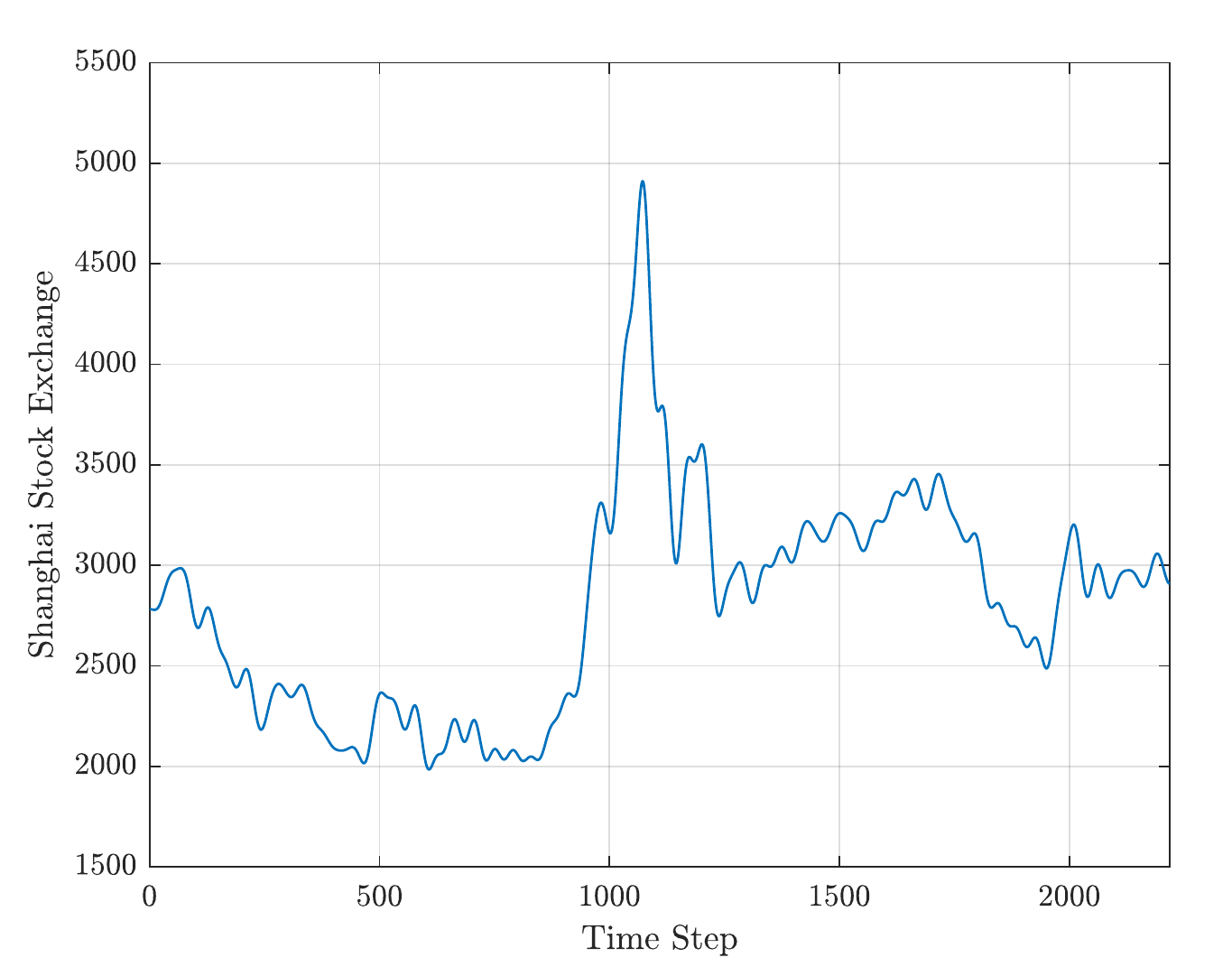}} 	 
	\end{tabular}
	
	\caption{Reconstructed signals of the SSE time series for different $\lambda$ values.}
	\label{fig:ReconstructedSignals}	
\end{figure}

The relationship between RMSE and $\lambda$ is illustrated in Fig. \ref{fig:SensitivityAnalysis_DCT}. Results indicated that when $\lambda$ increased from 50 to 85, the RMSE was reduced. Increasing $\lambda$ means reducing fluctuations and simplifying the time series structure, which leads to more accurate predictions. On the other hand, increasing $\lambda$ from 90 to 99 increases the RMSE. When $\lambda$ increases too much, a lot of meaningful information is discarded from the time series, and as a result, the predictions differ significantly from actual values. Fig. \ref{fig:ReconstructedSignals} shows reconstructed signals of the SSE time series for different $\lambda$ values.

\section{Conclusion}
\label{sec:Conclusion}

This study proposed two novel methods for multi-step-ahead stock price prediction. DCT-MFRFNN, a method based on DCT and MFRFNN, employed DCT to eliminate high frequencies, reduce fluctuations in the time series, and simplify its structure. It used MFRFNN to predict financial time series. VMD-MFRFNN, an approach based on VMD and MFRFNN, combined the benefits of VMD and MFRFNN. VMD-MFRFNN consisted of two phases: a) decomposition phase and b) prediction and reconstruction phase. The input time series was decomposed to several IMFs using VMD in the decomposition phase. Since these components had simpler structures and more stationary trends, their prediction was easier, and the expected accuracy was higher. In the prediction and reconstruction phase, each of IMFs was given to a separate MFRFNN model for prediction, and the predicted signals were summed to reconstruct the final output. Using the MFRFNN models enabled VMD-MFRFNN to learn and memorize historical data from previous samples, capture the dynamic characteristics of the financial time series, and approximate several functions simultaneously. 

The prediction performance of DCT-MFRFNN and VMD-MFRFNN was evaluated using three financial time series, including HSI, SSE, and SPX. The experimental results indicated that VMD-MFRFNN surpassed other state-of-the-art models and showed promising performance in multi-step-ahead prediction tasks. In more detail, based on RMSE, for the HSI time series, VMD-MFRFNN showed a decrease of 63.27\%, 41.27\%, and 3.25\% from the MEMD-LSTM in one, three, and five-step-ahead predictions, respectively. For the SSE time series, it demonstrated an RMSE decrease of 43.58\%, 22.96\%, and 8.10\% from the MEMD-LSTM in prediction horizons of one, three, and five, respectively. For the SPX time series, VMD-MFRFNN showed a decrease of 55.56\%, 42.23\%, and 6.00\% from the MEMD-LSTM in one, three, and five-step-ahead predictions, respectively, in terms of RMSE. Furthermore, DCT-MFRFNN outperformed MFRFNN in all experiments. DCT-MFRFNN, on average, showed 8.87\%, 28.78\%, and 39.61\% decreases in RMSE from MFRFNN for HSI, SSE, and SPX, respectively. The proposed methods in this study are definitely worth investigating further. One interesting area of future research is to apply DCT-MFRFNN and VMD-MFRFNN in other fields, such as meteorology and biomedical engineering.

\section{Code Availability}
The implementation of the proposed method and datasets needed to reproduce the results are provided on Github\footnote{\href{https://github.com/Hamid-Nasiri/VMD-MFRFNN}{https://github.com/Hamid-Nasiri/VMD-MFRFNN}}.

\bibliographystyle{elsarticle-num}
\bibliography{references}

\begin{thebibliography}{10}
\expandafter\ifx\csname url\endcsname\relax
  \def\url#1{\texttt{#1}}\fi
\expandafter\ifx\csname urlprefix\endcsname\relax\def\urlprefix{URL }\fi
\expandafter\ifx\csname href\endcsname\relax
  \def\href#1#2{#2} \def\path#1{#1}\fi

\bibitem{liu2022stock}
T.~Liu, X.~Ma, S.~Li, X.~Li, C.~Zhang, {A stock price prediction method based
  on meta-learning and variational mode decomposition}, Knowledge-Based Systems
  252 (2022) 109324.
\newblock \href {https://doi.org/10.1016/j.knosys.2022.109324}
  {\path{doi:10.1016/j.knosys.2022.109324}}.

\bibitem{deng2022multi}
C.~Deng, Y.~Huang, N.~Hasan, Y.~Bao, {Multi-step-ahead stock price index
  forecasting using long short-term memory model with multivariate empirical
  mode decomposition}, Information Sciences 607 (2022) 297--321.
\newblock \href {https://doi.org/10.1016/j.ins.2022.05.088}
  {\path{doi:10.1016/j.ins.2022.05.088}}.

\bibitem{farimani2022investigating}
S.~{Anbaee Farimani}, M.~{Vafaei Jahan}, A.~{Milani Fard}, S.~R.~K. Tabbakh,
  {Investigating the informativeness of technical indicators and news sentiment
  in financial market price prediction}, Knowledge-Based Systems 247 (2022)
  108742.
\newblock \href {https://doi.org/10.1016/j.knosys.2022.108742}
  {\path{doi:10.1016/j.knosys.2022.108742}}.

\bibitem{huang2021new}
Y.~Huang, Y.~Deng, {A new crude oil price forecasting model based on
  variational mode decomposition}, Knowledge-Based Systems 213 (2021) 106669.
\newblock \href {https://doi.org/10.1016/j.knosys.2020.106669}
  {\path{doi:10.1016/j.knosys.2020.106669}}.

\bibitem{seong2022forecasting}
N.~Seong, K.~Nam, {Forecasting price movements of global financial indexes
  using complex quantitative financial networks}, Knowledge-Based Systems 235
  (2022) 107608.
\newblock \href {https://doi.org/10.1016/j.knosys.2021.107608}
  {\path{doi:10.1016/j.knosys.2021.107608}}.

\bibitem{cao2019stock}
J.~Cao, J.~Wang, {Stock price forecasting model based on modified convolution
  neural network and financial time series analysis}, International Journal of
  Communication Systems 32~(12) (2019) e3987.
\newblock \href {https://doi.org/10.1002/dac.3987}
  {\path{doi:10.1002/dac.3987}}.

\bibitem{guo2022forecasts}
W.~Guo, Q.~Liu, Z.~Luo, Y.~Tse, {Forecasts for international financial series
  with VMD algorithms}, Journal of Asian Economics 80 (2022) 101458.
\newblock \href {https://doi.org/10.1016/j.asieco.2022.101458}
  {\path{doi:10.1016/j.asieco.2022.101458}}.

\bibitem{lin2021forecasting}
Y.~Lin, Y.~Yan, J.~Xu, Y.~Liao, F.~Ma, {Forecasting stock index price using the
  CEEMDAN-LSTM model}, The North American Journal of Economics and Finance 57
  (2021) 101421.
\newblock \href {https://doi.org/10.1016/j.najef.2021.101421}
  {\path{doi:10.1016/j.najef.2021.101421}}.

\bibitem{rezaei2021stock}
H.~Rezaei, H.~Faaljou, G.~Mansourfar, {Stock price prediction using deep
  learning and frequency decomposition}, Expert Systems with Applications 169
  (2021) 114332.
\newblock \href {https://doi.org/10.1016/j.eswa.2020.114332}
  {\path{doi:10.1016/j.eswa.2020.114332}}.

\bibitem{yujun2021research}
Y.~Yujun, Y.~Yimei, Z.~Wang, {Research on a hybrid prediction model for stock
  price based on long short-term memory and variational mode decomposition},
  Soft Computing 25~(21) (2021) 13513--13531.
\newblock \href {https://doi.org/10.1007/s00500-021-06122-4}
  {\path{doi:10.1007/s00500-021-06122-4}}.

\bibitem{xu2022self}
H.~Xu, D.~Cao, S.~Li, {A self-regulated generative adversarial network for
  stock price movement prediction based on the historical price and tweets},
  Knowledge-Based Systems 247 (2022) 108712.
\newblock \href {https://doi.org/10.1016/j.knosys.2022.108712}
  {\path{doi:10.1016/j.knosys.2022.108712}}.

\bibitem{nguyen2021ensemble}
H.-P. Nguyen, P.~Baraldi, E.~Zio, {Ensemble empirical mode decomposition and
  long short-term memory neural network for multi-step predictions of time
  series signals in nuclear power plants}, Applied Energy 283 (2021) 116346.
\newblock \href {https://doi.org/10.1016/j.apenergy.2020.116346}
  {\path{doi:10.1016/j.apenergy.2020.116346}}.

\bibitem{huang1998empirical}
N.~E. Huang, Z.~Shen, S.~R. Long, M.~C. Wu, H.~H. Shih, Q.~Zheng, N.-C. Yen,
  C.~C. Tung, H.~H. Liu, {The empirical mode decomposition and the Hilbert
  spectrum for nonlinear and non-stationary time series analysis}, Proceedings
  of the Royal Society of London. Series A: Mathematical, Physical and
  Engineering Sciences 454~(1971) (1998) 903--995.
\newblock \href {https://doi.org/10.1098/rspa.1998.0193}
  {\path{doi:10.1098/rspa.1998.0193}}.

\bibitem{liu2020robust}
Z.~Liu, J.~Liu, {A robust time series prediction method based on empirical mode
  decomposition and high-order fuzzy cognitive maps}, Knowledge-Based Systems
  203 (2020) 106105.
\newblock \href {https://doi.org/10.1016/j.knosys.2020.106105}
  {\path{doi:10.1016/j.knosys.2020.106105}}.

\bibitem{dragomiretskiy2014variational}
K.~Dragomiretskiy, D.~Zosso, {Variational Mode Decomposition}, IEEE
  Transactions on Signal Processing 62~(3) (2014) 531--544.
\newblock \href {https://doi.org/10.1109/TSP.2013.2288675}
  {\path{doi:10.1109/TSP.2013.2288675}}.

\bibitem{nasiri2022mfrfnn}
H.~Nasiri, M.~M. Ebadzadeh, {MFRFNN: Multi-Functional Recurrent Fuzzy Neural
  Network for Chaotic Time Series Prediction}, Neurocomputing 507 (2022)
  292--310.
\newblock \href {https://doi.org/10.1016/j.neucom.2022.08.032}
  {\path{doi:10.1016/j.neucom.2022.08.032}}.

\bibitem{zhou2019emd2fnn}
F.~Zhou, H.-m. Zhou, Z.~Yang, L.~Yang, {EMD2FNN: A strategy combining empirical
  mode decomposition and factorization machine based neural network for stock
  market trend prediction}, Expert Systems with Applications 115 (2019)
  136--151.
\newblock \href {https://doi.org/10.1016/j.eswa.2018.07.065}
  {\path{doi:10.1016/j.eswa.2018.07.065}}.

\bibitem{yang2022novel}
Y.~Yang, C.~Fan, H.~Xiong, {A novel general-purpose hybrid model for time
  series forecasting}, Applied Intelligence 52~(2) (2022) 2212--2223.
\newblock \href {https://doi.org/10.1007/s10489-021-02442-y}
  {\path{doi:10.1007/s10489-021-02442-y}}.

\bibitem{tang2020multistep}
Z.~Tang, T.~Zhang, J.~Wu, X.~Du, K.~Chen, {Multistep-Ahead Stock Price
  Forecasting Based on Secondary Decomposition Technique and Extreme Learning
  Machine Optimized by the Differential Evolution Algorithm}, Mathematical
  Problems in Engineering 2020 (2020) 1--13.
\newblock \href {https://doi.org/10.1155/2020/2604915}
  {\path{doi:10.1155/2020/2604915}}.

\bibitem{salimi2022novel}
A.~Salimi-Badr, M.~M. Ebadzadeh, {A novel learning algorithm based on computing
  the rules' desired outputs of a TSK fuzzy neural network with non-separable
  fuzzy rules}, Neurocomputing 470 (2022) 139--153.
\newblock \href {https://doi.org/10.1016/j.neucom.2021.10.103}
  {\path{doi:10.1016/j.neucom.2021.10.103}}.

\bibitem{liu2020non}
Y.~Liu, C.~Yang, K.~Huang, W.~Gui, {Non-ferrous metals price forecasting based
  on variational mode decomposition and LSTM network}, Knowledge-Based Systems
  188 (2020) 105006.
\newblock \href {https://doi.org/10.1016/j.knosys.2019.105006}
  {\path{doi:10.1016/j.knosys.2019.105006}}.

\bibitem{wu2009ensemble}
Z.~Wu, N.~E. Huang, {Ensemble Empirical Mode Decomposition: A Noise-Assisted
  Data Analysis Method}, Advances in Adaptive Data Analysis 01~(01) (2009)
  1--41.
\newblock \href {https://doi.org/10.1142/S1793536909000047}
  {\path{doi:10.1142/S1793536909000047}}.

\bibitem{rashida2020intelligent}
S.~Y. Rashida, M.~Sabaei, M.~M. Ebadzadeh, A.~M. Rahmani, {An intelligent
  approach for predicting resource usage by combining decomposition techniques
  with NFTS network}, Cluster Computing 23~(4) (2020) 3435--3460.
\newblock \href {https://doi.org/10.1007/s10586-020-03099-x}
  {\path{doi:10.1007/s10586-020-03099-x}}.

\bibitem{lahmiri2016variational}
S.~Lahmiri, {A variational mode decompoisition approach for analysis and
  forecasting of economic and financial time series}, Expert Systems with
  Applications 55 (2016) 268--273.
\newblock \href {https://doi.org/10.1016/j.eswa.2016.02.025}
  {\path{doi:10.1016/j.eswa.2016.02.025}}.

\bibitem{niu2020hybrid}
H.~Niu, K.~Xu, W.~Wang, {A hybrid stock price index forecasting model based on
  variational mode decomposition and LSTM network}, Applied Intelligence
  50~(12) (2020) 4296--4309.
\newblock \href {https://doi.org/10.1007/s10489-020-01814-0}
  {\path{doi:10.1007/s10489-020-01814-0}}.

\bibitem{wang2022asian}
J.~Wang, Q.~Cui, X.~Sun, M.~He, {Asian stock markets closing index forecast
  based on secondary decomposition, multi-factor analysis and attention-based
  LSTM model}, Engineering Applications of Artificial Intelligence 113 (2022)
  104908.
\newblock \href {https://doi.org/10.1016/j.engappai.2022.104908}
  {\path{doi:10.1016/j.engappai.2022.104908}}.

\bibitem{vinciguerra2021discrete}
A.~Vinciguerra, M.~Aleardi, A.~Hojat, M.~H. Loke, E.~Stucchi, {Discrete cosine
  transform for parameter space reduction in Bayesian electrical resistivity
  tomography}, Geophysical Prospecting 70~(1) (2022) 193--209.
\newblock \href {https://doi.org/10.1111/1365-2478.13148}
  {\path{doi:10.1111/1365-2478.13148}}.

\bibitem{begum2022hybrid}
M.~Begum, J.~Ferdush, M.~S. Uddin, {A Hybrid robust watermarking system based
  on discrete cosine transform, discrete wavelet transform, and singular value
  decomposition}, Journal of King Saud University - Computer and Information
  Sciences 34~(8) (2022) 5856--5867.
\newblock \href {https://doi.org/10.1016/j.jksuci.2021.07.012}
  {\path{doi:10.1016/j.jksuci.2021.07.012}}.

\bibitem{zhang2021application}
Z.~Zhang, W.-C. Hong, {Application of variational mode decomposition and
  chaotic grey wolf optimizer with support vector regression for forecasting
  electric loads}, Knowledge-Based Systems 228 (2021) 107297.
\newblock \href {https://doi.org/10.1016/j.knosys.2021.107297}
  {\path{doi:10.1016/j.knosys.2021.107297}}.

\bibitem{he2022fault}
D.~He, C.~Liu, Z.~Jin, R.~Ma, Y.~Chen, S.~Shan, {Fault diagnosis of flywheel
  bearing based on parameter optimization variational mode decomposition energy
  entropy and deep learning}, Energy 239 (2022) 122108.
\newblock \href {https://doi.org/10.1016/j.energy.2021.122108}
  {\path{doi:10.1016/j.energy.2021.122108}}.

\bibitem{yang2021denoising}
H.~Yang, Y.~Cheng, G.~Li, {A denoising method for ship radiated noise based on
  Spearman variational mode decomposition, spatial-dependence recurrence sample
  entropy, improved wavelet threshold denoising, and Savitzky-Golay filter},
  Alexandria Engineering Journal 60~(3) (2021) 3379--3400.
\newblock \href {https://doi.org/10.1016/j.aej.2021.01.055}
  {\path{doi:10.1016/j.aej.2021.01.055}}.

\bibitem{lian2018adaptive}
J.~Lian, Z.~Liu, H.~Wang, X.~Dong, {Adaptive variational mode decomposition
  method for signal processing based on mode characteristic}, Mechanical
  Systems and Signal Processing 107 (2018) 53--77.
\newblock \href {https://doi.org/10.1016/j.ymssp.2018.01.019}
  {\path{doi:10.1016/j.ymssp.2018.01.019}}.

\bibitem{zojaji2022semantic}
Z.~Zojaji, M.~M. Ebadzadeh, H.~Nasiri, {Semantic schema based genetic
  programming for symbolic regression}, Applied Soft Computing 122 (2022)
  108825.
\newblock \href {https://doi.org/10.1016/j.asoc.2022.108825}
  {\path{doi:10.1016/j.asoc.2022.108825}}.

\bibitem{liu2022improved}
Z.~Liu, D.~Peng, M.~J. Zuo, J.~Xia, Y.~Qin, {Improved Hilbert–Huang transform
  with soft sifting stopping criterion and its application to fault diagnosis
  of wheelset bearings}, ISA Transactions 125 (2022) 426--444.
\newblock \href {https://doi.org/10.1016/j.isatra.2021.07.011}
  {\path{doi:10.1016/j.isatra.2021.07.011}}.

\end{thebibliography}

\vspace{0.5cm}

\parpic{\includegraphics[width=0.95in,clip,keepaspectratio]{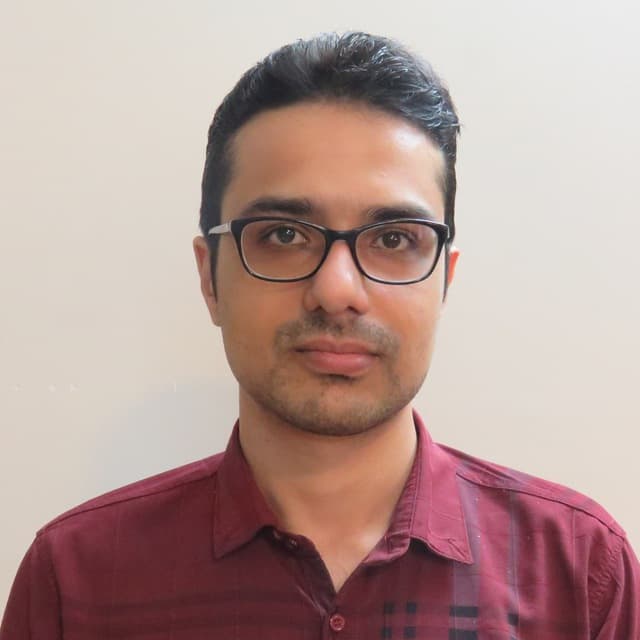}}
\noindent {\bf Hamid Nasiri} received the B.Sc. degree in Computer Engineering from the Semnan University, Semnan, Iran, in 2014, and the M.Sc. degree in Computer Engineering from the Amirkabir University of Technology, Tehran, Iran, in 2016. He is currently a Ph.D. candidate at the Department of Computer Engineering, Amirkabir University of Technology, under the supervision of Dr. Mohammad Mehdi Ebadzadeh. His research interests include machine learning, deep learning, time series forecasting, and fuzzy systems.

\parpic{\includegraphics[width=0.95in,clip,keepaspectratio]{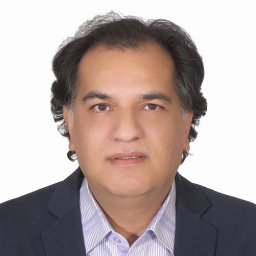}}
\noindent {\bf Mohammad Mehdi Ebadzadeh} received the B.Sc. degree in Electrical Engineering from Sharif University of Technology, and his M.Sc. degree in Computer Engineering from Amirkabir University of Technology, Tehran, Iran, in 1991 and 1995, respectively. He received the Ph.D. degree in Image and Signal Processing from TELECOM ParisTech, Paris, France, in 2004. Currently, he is a Professor in Department of Computer Engineering of Amirkabir University of Technology, Tehran, Iran. His research interests include deep reinforcement learning, computational intelligence, and computational neuroscience.

\end{document}